\documentclass[twocolumn,runningheads]{svjour3}

\usepackage[latin1]{inputenc}
\usepackage{times}

\usepackage{psfrag,amsmath, amssymb, amscd, amsfonts, latexsym, epsf, graphicx,amscd}
\usepackage{natbib}

\def\bbbr{{\mathbb R}} 
 
\def\diag{\operatorname{diag}}

\newcommand{\orth}{\bot}
\newcommand{\erf}{\operatorname{erf}}

\def\norm{\scriptsize\mbox{norm}}
\def\simple{\scriptsize\mbox{simple}}
\def\spacespace{\scriptsize\mbox{space}}
\def\time{\scriptsize\mbox{time}}
\def\spat{\scriptsize\mbox{spat}}
\def\sep{\scriptsize\mbox{sep}}
\def\vel{\scriptsize\mbox{vel}}
\def\even{\scriptsize\mbox{even}}
\def\odd{\scriptsize\mbox{odd}}
\def\complex{\scriptsize\mbox{complex}}

\journalname{arXiv preprint}

\begin{document}

\titlerunning{Orientation selectivity properties for the affine Gaussian derivative
  and the affine Gabor models for visual receptive fields}
\title{\bf Orientation selectivity properties for the affine Gaussian derivative
  and the affine Gabor models for visual receptive fields%
\thanks{The support from the Swedish Research Council 
              (contract 2022-02969) is gratefully acknowledged. }}
\author{Tony Lindeberg}

\institute{Computational Brain Science Lab,
        Division of Computational Science and Technology,
        KTH Royal Institute of Technology,
        SE-100 44 Stockholm, Sweden.
        \email{tony@kth.se}.
      ORCID: 0000-0002-9081-2170.}

\date{}

\maketitle

\begin{abstract}
\noindent
This paper presents an in-depth theoretical analysis of the orientation
selectivity properties of simple cells and complex cells, that
can be well modelled by
the generalized Gaussian derivative model for visual receptive fields,
with the purely spatial component of the receptive fields determined by
oriented affine Gaussian derivatives for different orders of spatial
differentiation.

A detailed mathematical analysis is presented for the three different
cases of either: (i) purely spatial receptive fields, (ii) space-time
separable spatio-temporal receptive fields and (iii) velocity-adapted
spatio-temporal receptive fields. Closed-form theoretical expressions for
the orientation selectivity curves for idealized models of
simple and complex cells are derived
for all these main cases, and it is shown that the 
orientation selectivity of the receptive fields becomes more narrow, as a scale
parameter ratio $\kappa$, defined as the ratio between the scale
parameters in the directions perpendicular to {\em vs.\/}\ parallel with the
preferred orientation of the receptive field, increases. It is also shown that
the orientation selectivity becomes more narrow with increasing order of
spatial differentiation in the underlying affine Gaussian derivative
operators over the spatial domain.

A corresponding theoretical orientation selectivity analysis is also
presented for
purely spatial receptive fields according to an affine Gabor model, showing that: (i)~the orientation selectivity becomes
more narrow when making the receptive fields wider in the direction
perpendicular to the preferred orientation of the receptive field;
while (ii)~an additional degree of freedom in the affine Gabor model
does, however, also strongly affect the orientation selectivity properties.

\keywords{Receptive field \and Orientation selectivity \and
                  Affine covariance \and Simple cell \and Complex cell \and Vision}


\end{abstract}

\section{Introduction}
\label{sec-intro}

The receptive fields%
\footnote{Notes concerning the terminology used in this paper:
  While the notion of a visual receptive field originally referred
  to the region in the visual field that may invoke a stimulus
  response for a visual neuron
  (Hubel and Wiesel \citeyear{HubWie59-Phys,HubWie05-book}),
  we here adopt a somewhat different functional definition
  of the notion of a receptive field, as the visual operator that computes the output
  from a visual neuron for a (non-spiking) model of the computational
  function of the visual neuron. From such a viewpoint, the above more traditional notion of
  a visual receptive field is according to the definition used in this
  paper instead referred to as the (effective) support region of the receptive
  field.

  
  Concerning simple cells, we throughout this paper make use of idealized
  purely feed-forward computational models of such receptive fields in terms of linear
  convolution operations, whereas our idealized models of complex
  cells do instead combine the output from such linear simple cells in
  a non-linear, also purely feed-forward, manner. These highly idealized
  simplifications are made here to make theoretical analysis feasible
  to handle in closed form, thereby leading to a set of closed-form
  characteristics regarding the orientation selectivity properties 
  for the two main paradigms of modelling simple and complex cells in
  terms of either Gaussian derivative operators or Gabor functions.}
in the primary visual cortex (V1) capture
properties of the visual patterns, which are then passed on to higher
layers in the visual hierarchy. Being able to understand the
computational function of these receptive fields is, hence, essential
for understanding the computational function of the visual system.

The task of understanding the visual system has been addressed both
neurophysiologically, by measuring the response properties of neurons to
visual stimuli, and by formulating mathematical models, that aim at
both explaining the computational function of visual neurons, as well
as enabling computational simulation of neural functions in terms of
biologically plausible computer vision algorithms, or aiming at making
theoretical predictions, which can then be investigated experimentally.

If we want to build computational models of the vision system, then
the receptive fields of the visual neurons can be seen as the primary
objects with regard to the theoretical modelling step. In terms of
neurophysiological measurements, it does, however, usually imply quite
a complex procedure reconstruct the receptive fields of individual
neurons. First of all, the problem of deriving a receptive field model
for a visual neuron constitutes an inverse problem, which may require
both an explicit model for the computational function of the neuron,
as well as the need for performing a substantial number of
measurements for different visual stimuli, to span a sufficiently
large subspace of the variability in the possible types of image input
that the neuron may be exposed to
(DeAngelis {\em et al.\/}\
\citeyear{DeAngOhzFre95-TINS,deAngAnz04-VisNeuroSci},
Ringach \citeyear{Rin04-JPhys},
Sharpee \citeyear{Sha13-AnnRevNeurSci},
Walker {\em et al.\/}
\citeyear{WalSinCobMuhFroFahEckReiPitTol19-NatNeurSci}).

Mapping the orientation selectivity
properties of a visual neuron, can, on the other hand, often be
performed by a comparably much more straightforward procedure, by just
using a probing stimulus, often a sine wave pattern, and then
measuring the variations in the output of the neuron when the
orientation of the stimulus is varied in the image domain
(Ringach {\em et al.\/} \citeyear{RinShaHaw03-JNeurSci},
Nauhaus {\em et al.\/} \citeyear{NauBenCarRin09-Neuron},
Scholl {\em et al.\/} \citeyear{SchTanCorPri13-JNeurSci},
Mazurek {\em et al.\/} \citeyear{MazKagHoo14-FrontNeurCirc}).
For this
reason, there is a substantially larger amount of available
neurophysiological measurements in terms of orientation selectivity,
than in terms of actually reconstructed receptive fields of visual
neurons.

The purpose of this article is to present an in-depth theoretical
analysis, that aims at bridging the gap between these two conceptually
different ways of characterizing the properties of visual neurons, by
deriving closed-form expressions for links between inherent properties of the
receptive fields for simple and complex cells in the primary visual
cortex and their orientation selectivity properties. In particular, we
are interested in characterizing how the degree of elongation,
or the eccentricity, for the receptive fields of simple cells is related to
the orientation selectivity properties, for the commonly used models
of simple cells in terms of either (i)~an extension of the regular
Gaussian derivative model
(Koenderink and van Doorn \citeyear{KoeDoo87-BC,KoeDoo92-PAMI};
 Young \citeyear{You87-SV})
into the affine Gaussian derivative
model for visual receptive fields
(Lindeberg \citeyear{Lin13-BICY,Lin21-Heliyon}),
or (ii)~the affine Gabor model for
visual receptive fields
(Marcelja \citeyear{Mar80-JOSA};
Jones and Palmer \citeyear{JonPal87a,JonPal87b};
Ringach \citeyear{Rin01-JNeuroPhys}).

The motivation for studying these two models for visual receptive
fields are that:
(i)~the generalized Gaussian derivative model can be regarded as the
theoretically most principled model, by being derived by necessity in
an axiomatic way from principled symmetry requirements, while
(ii)~the Gabor model can be regarded as the most commonly used model
in the field.
In addition to such idealized linear models for simple cells, we will
also, assuming that complex cells can be
computationally modelled as combining the output from multiple simple
cells in a non-linear manner, derive corresponding
relationships between properties of the receptive fields for the underlying simple cells and
the orientation selectivity properties for idealized models of complex cells.

Specifically, we will focus on quantifying how
the degree of orientation selectivity of a receptive field depends
on how anisotropic or elongated that the
underlying affine Gaussian derivatives are. We will derive explicit
expressions for how the degree of orientation selectivity increases
with the ratio between the scale parameters in the orientations
perpendicular to {\em vs.\/}\ parallel with the preferred orientation of
the receptive fields. In this way, we also demonstrate how closed-form
theoretical analysis is possible for the generalized Gaussian derivative
model for visual receptive fields, which arises from the normative
theory of visual receptive fields in (Lindeberg \citeyear{Lin21-Heliyon}).
Affine Gaussian smoothing has also been earlier used for computing more accurate cues to local surface orientation, than is possible if basing the spatial smoothing operations on rotationally symmetric Gaussian kernels only
(Lindeberg and G{\aa}rding \citeyear{LG96-IVC}).

We propose that these theoretical connections, between the orientation
selectivity and the degree of elongation of the receptive fields, can
be useful, when to relate the results from neurophysiological
recordings of the orientation selectivity properties of visual neurons
to other modelling-based functional characteristics of biological
receptive fields.
Specifically, in a companion paper (Lindeberg
\citeyear{Lin24-arXiv-HypoElongVarRF}),  theoretical results derived
in this paper are, in combination with properties of biological
measurements of the orientation selectivity of receptive fields in the
primary visual cortex by Nauhaus {\em et al.\/}
(\citeyear{NauBenCarRin09-Neuron}) and
Goris {\em et al.\/}\ (\citeyear{GorSimMov15-Neuron}), used to provide
potential indirect support for a previously formulated hypothesis
(Hypothesis~1 in Lindeberg (\citeyear{Lin23-FrontCompNeuroSci}) Section~3.2.1).
That hypothesis states that the
the receptive field shapes in the primary visual cortex of higher
mammals may span a variability over the
eccentricity, or the degree of elongation, of the receptive fields, to
support affine covariance over the population of biological receptive
fields, to in turn enable the computation of more accurate cues to 3-D
scene geometry, when observing the same local surface patch from
different slant angles.

More generally, we propose that the theoretical results to be
presented, regarding observable properties of the affine Gaussian
derivative model for visual receptive fields, as well as for the
affine Gabor model also studied for comparison, are important for
understanding the properties of these models for visual receptive
fields in the primary visual cortex. These
theoretical results are also important when relating
biological measurements of orientation selectivity, which have been
extensively performed in the area of biological vision, to
mathematical models of functional properties of the visual neurons.

Compared to the results of purely numerical simulations of properties of
networks of visual neurons, the closed-form nature of the results
derived in this paper for the affine Gaussian derivative model
specifically makes it possible to use these results as
first-class results, which can be used as primitives for inferring further properties of receptive
fields based on closed-form mathematical analysis.
See also (Lindeberg \citeyear{Lin24-arXiv-HypoElongVarRF}) for
other closed-form mathematical characterizations of
properties of idealized models for visual receptive fields, based on
results derived in this paper.

By performing such idealized
theoretical analysis of functional properties and characteristics of visual neurons,
we argue that it should be possible to reveal both inherent
possibilities and limitations with regard to the computational
functions of these functional primitives in the visual pathway.
See also Geisler (\citeyear{Gei11-VisRes}) and
Burge (\citeyear{Bur20-AnnRevVisSci}) for over\-views
of the structurally related notion of idealized observers for visual perception.
The main difference in relation to our approach, pursued according to
our normative conceptual framework for visual receptive fields,
is that we instead base the foundations of our analysis on symmetry
properties of the environment, in a structurally similar way as
symmetry properties can constitute the foundations for
formulating fundamental theories in theoretical physics.

Additionally, when using a computational modelling approach to understand the
function of the vision system, it is essential to also understand the theoretical
properties of the model. The analysis presented in this paper shows
that it is possible to fully understand the orientation selectivity
properties for the affine Gaussian derivative model for simple cells
in terms of the degree of elongation of the receptive fields,
combined with the order of spatial differentiation.
As will be shown in the paper,
there is, however, not such a direct relationship between the
orientation selectivity properties and the degree of elongation
for the affine Gabor model as for the affine Gaussian derivative
model, which is important to consider when modelling the functional
properties of visual receptive fields in terms of idealized models.

Furthermore, with regard to theory developments, the results derived in this paper extend
the axiomatically determined computational and normative theory for
visual receptive fields in terms of the generalized Gaussian
derivative model
(Lindeberg \citeyear{Lin13-BICY,Lin21-Heliyon})
from provable covariance properties under natural image transformations
(Lindeberg \citeyear{Lin23-FrontCompNeuroSci,Lin24-arXiv-UnifiedJointCovProps})
to explicit formulations of orientation selectivity properties in closed form, for the
idealized receptive fields according to this theoretically principled
model for visual receptive fields.

\subsection{Structure of this article}

This paper is organized as follows:
After an overview of related work in Section~\ref{sec-rel-work},
Section~\ref{sec-theor-bg} provides the
theoretical background to this work,
by describing the generalized Gaussian derivative model for visual
receptive fields, both in the cases of a purely spatial domain, where
the receptive fields are pure affine Gaussian derivatives, and for a
joint spatio-temporal domain, where the affine Gaussian derivatives
are complemented by temporal derivatives of either a non-causal
Gaussian kernel over the temporal domain, or a genuine time-causal kernel,
referred to as the time-causal limit kernel, as well as complemented
with possible velocity adaptation, to enable Galilean covariance.

Beyond these models of
simple cells, we do both review a previously formulated purely spatial
model for complex cells, based on a Euclidean combination of
scale-normalized affine Gaussian derivatives of orders 1 and~2, as
well as propose two new spatio-temporal models for complex cells, based
on image measurements in terms of affine Gaussian derivatives
over the spatial domain, complemented by
explicit temporal processing operations. Two special cases are treated,
in terms of either space-time separable spatio-temporal receptive fields,
or velocity-adapted spatio-temporal receptive fields, with the latter tuned to 
particular motion directions and motion velocities in joint space-time.

Section~\ref{sec-anal-ori-sel} then performs a detailed mathematical
analysis of the orientation selectivity of these models, over the three
main cases of either (i)~a purely spatial domain, (ii)~a space-time
separable spatio-temporal domain, or (iii)~a velocity-adapted
spatio-temporal domain. For each one of these main cases, we analyze the
properties of simple cells of orders 1 and~2, corresponding to first-
or second-order Gaussian derivatives, as well as the orientation
selectivity properties for idealized models of complex cells, defined
in terms of quasi-quadrature measures, constituting Euclidean
combinations of the underlying idealized models of simple cells.

Concerning simple cells, it is shown that both the pure spatial
receptive fields and the joint spatio-temporal receptive fields have
similar orientation selectivity properties, which only depend on the
order of spatial differentiation and the degree of anisotropy of the
receptive field. For complex cells, a different dependency is,
however, derived for the model based on space-time separable receptive
fields, as opposed to the models for either a purely spatial domain or
a joint spatio-temporal domain based on velocity-adapted
spatio-temporal receptive fields.

To widen the scope of the treatment in the paper, beyond a
theoretical analysis of properties of idealized visual receptive field
models according to the generalized Gaussian derivative model for
visual receptive fields, we will also in Section~\ref{app-ori-sel-anal-gabor}
perform a corresponding orientation selectivity analysis for an affine
Gabor model of receptive fields, as well as give a derivation of affine
covariant properties of a further generalized affine Gabor model, with the details of that
analysis in Appendix~\ref{app-aff-transf-prop-gabor}.

As a service for the reader, who may want to use the results from the
theoretical analysis in this paper for comparing biological
measurements of the orientation selectivity of visual neurons to
characteristic properties of idealized models of visual receptive
fields, we do then in Section~\ref{sec-rel-comp-quant-measures}
compute two explicit compact measures of the degree of orientation
selectivity for the idealized models of receptive fields in terms of
affine Gaussian derivatives, in terms of the resultant of the circular
distribution of the orientation selectivity curves as well as the
orientation bandwidth, as functions of the scale parameter ratio
$\kappa$ of the receptive fields.

For the purpose of relating results of biological measurements
performed under conditions when the angular frequency of the sine wave
probe is not adapted to each stimulus orientation,
as done in the main theoretical analysis in Section~\ref{sec-anal-ori-sel},
we do additionally
in Section~\ref{sec-perturb-anal} present a complementary analysis of how
different variations in the choice of the angular frequency of the
sine wave probe, and for spatio-temporal receptive fields also
variations in the image velocity of the sine wave stimulus,
affect the shapes of the resulting idealized
orientation selectivity curves.

In this complementary analysis, it is specifically shown that the
shapes of resulting orientation selectivity curves may be strongly
dependent on the relationships between the parameters of the probing
sine wave stimulus and the parameters of the receptive field, unless
adapting the parameters of the sine wave probe to maximize the 
response of the receptive field, as done in
the theoretical analysis in Section~\ref{sec-anal-ori-sel}.

Finally, Section~\ref{sec-summ-disc} gives a summary and discussion
about some of the main results, including suggestions concerning
possible extensions of the presented work.

\section{Related work}
\label{sec-rel-work}

Hubel and Wiesel
(\citeyear{HubWie59-Phys,HubWie62-Phys,HubWie68-JPhys,HubWie05-book})
pioneered the study of visual neurons in the primary visual cortex,
and introduced the taxonomy of simple and complex cells.
Simple cells were simple characterized by their properties of:
(i)~having distinct excitatory and inhibitory regions,
(ii)~obeying roughly linear summation properties,
(iii)~the excitatory and inhibitory regions balance each other in
diffuse lighting.
Visual neurons that did not obey these properties were referred to as
complex cells, The response of a complex cell to a visual stimulus was
also reported to be much less sensitive to the position of the stimulus in the visual
field than for a simple cell.

More detailed characterizations of the receptive fields of simple cells
have then been performed by
DeAngelis {\em et al.\/}\
(\citeyear{DeAngOhzFre95-TINS,deAngAnz04-VisNeuroSci}),
Ringach (\citeyear{Rin01-JNeuroPhys,Rin04-JPhys}),
Conway and Livingstone (\citeyear{ConLiv06-JNeurSci}),
Johnson {\em et al.\/}\ (\citeyear{JohHawSha08-JNeuroSci})
and
De and Horwitz (\citeyear{DeHor21-JNPhys}),
where specifically the use of multiple white noise stimuli permit a
reconstruction of the full receptive field of a visual neuron, 
based on theoretical results in system
identification theory, assuming linearity of the neuron.
More specialized methodologies to characterize the response properties of
possibly non-linear visual
neurons, based on using deep predictive models to generate tailored
stimuli for probing and modelling the neurons, have also been been more recently developed by
Walker {\em et al.\/}
(\citeyear{WalSinCobMuhFroFahEckReiPitTol19-NatNeurSci}).

Furthermore, more detailed characterizations of the orientation selectivity of
visual neurons have been performed by
Watkins and Berkley (\citeyear{WatBer73-ExpBrainRes}),  
Rose and Blakemore (\citeyear{RosBla74-ExpBrainRes}),
Schiller {\em et al.\/} (\citeyear{SchFinVol76-JNeuroPhys}),
Albright (\citeyear{Alb84-JNeuroPhys}),
Ringach {\em et al.\/} (\citeyear{RinShaHaw03-JNeurSci}),
Nauhaus {\em et al.\/}\ (\citeyear{NauBenCarRin09-Neuron}),
Scholl {\em et al.\/} (\citeyear{SchTanCorPri13-JNeurSci}),
Sadeh and Rotter (\citeyear{SadRot14-BICY}),
Goris {\em et al.\/}\ (\citeyear{GorSimMov15-Neuron}),
Li {\em et al.\/}\ (\citeyear{LiLiuChoZhaTao15-JNeuroSci})
and Almasi {\em et al.\/}\ (\citeyear{AlmMefCloWonYunIbb20-CerCort}).

While several works have been concerned with biological mechanisms for achieving orientation selectivity in the visual neurons
(Somers {\em et al.\/}\ \citeyear{SomNelSur95-JNeuroSci},
Sompolinsky and Shapley \citeyear{SomSha97-CurrOpNeuroBio},
Carandini and Ringach \citeyear{CarRin97-VisRes},
Lampl {\em et al.\/}\ \citeyear{LamAndGilFer01-Neuron},
Ferster and Miller \citeyear{FerMil00-AnnRevNeuroSci},
Shapley {\em et al.\/}\ \citeyear{ShaHawRin03-Neuron},
Seri{\`e}s {\em et al.\/}\ \citeyear{SerLatPou04-NatNeuroSci},
Hansel and van~Vreeswijk \citeyear{HanVre12-JNeuroSci},
Moldakarimov {\em et al.\/}\ \citeyear{MolBazSej14-PLOSCompBiol},
Gonzalo Cogno and Mato \citeyear{GonMat15-FrontNeurCirc},
Priebe \citeyear{Pri16-AnnRevVisSci},
Pattadkal {\em et al.\/} \citeyear{PatMatVrePriHan18-CellRep},
Nguyen and Freeman \citeyear{NguFre19-PLOSCompBiol},
Merkt {\em et al.\/} \citeyear{MerSchRot19-PLOSCompBiol},
Wei {\em et al.\/} \citeyear{WeiMerRot22-bioRxiv},
Wang {\em et al.\/} \citeyear{WanDeyLagBehCalSta24-CellRep})
as well as characterizing properties of cortical maps
(Bonhoeffer and Grinvald \citeyear{BonGri91-Nature},
Blasdel \citeyear{Bla92-JNeuroSci},
Maldonado {\em et al.\/} \citeyear{MalGodGraBon97-Science},
Koch {\em et al.\/} \citeyear{KocJinAloZai16-NatComm},
Kremkow {\em et al.\/}\ \citeyear{KreJinWanAlo16-Nature},
Najafian {\em et al.\/}\ \citeyear{NajKocTehJinRahZaiKreAlo22-NatureComm},
Jung {\em et al.\/}\ \citeyear{JunAlmSunYunCloBauRenMefIbb22-SciAdv},
Fang {\em et al.\/}\ \citeyear{FanCaiLu-PNAS},
Vita  {\em et al.\/}\ \citeyear{BitOrsStaClaTir24-bioRxiv}),
the focus of this work is, however, not on such specific
biological implementation mechanisms, but instead on purely
{\em functional properties of individual neurons\/}, that will constitute the effective results from such underlying computational mechanisms between different neurons.

A detailed study of how the orientation selective
of neurons in the primary visual cortex can be modulated for
cats that wear permanently mounted googles, that alter
the directional distribution of the incoming visual stimuli,
has been presented by Sasaki {\em et al.\/}
(\citeyear{SakKimNimTabTanFukAsaAraInaNakBabDaiNisSanTanImaTanOhz15-SciRep}),
showing that long exposure to such stimuli affects the elongation of
receptive fields, and how such elongation affects the orientational selectivity.

Mathematical models of simple cells have been formulated, in
terms of Gabor filters 
(Marcelja \citeyear{Mar80-JOSA};
Jones and Palmer \citeyear{JonPal87a,JonPal87b};
Porat and Zeevi \citeyear{PorZee88-PAMI})
or Gaussian derivatives
(Koenderink and van Doorn \citeyear{Koe84,KoeDoo87-BC,KoeDoo92-PAMI};
 Young and his co-workers \citeyear{You87-SV,YouLesMey01-SV,YouLes01-SV},
 Lindeberg \citeyear{Lin13-BICY,Lin21-Heliyon}).
Specifically, theoretical models of early visual processes in terms of
Gaussian derivatives have been formulated or used in
Lowe (\citeyear{Low00-BIO}),
May and Georgeson (\citeyear{MayGeo05-VisRes}),
Hesse and Georgeson (\citeyear{HesGeo05-VisRes}),
Georgeson  {\em et al.\/}\ (\citeyear{GeoMayFreHes07-JVis}),
Hansen and Neumann (\citeyear{HanNeu09-JVis}),
Wallis and Georgeson (\citeyear{WalGeo09-VisRes}),  
Wang and Spratling (\citeyear{WanSpra16-CognComp}),
Pei {\em et al.\/}\ (\citeyear{PeiGaoHaoQiaAi16-NeurRegen}),
Ghodrati {\em et al.\/}\ (\citeyear{GhoKhaLeh17-ProNeurobiol}),
Kristensen and Sandberg (\citeyear{KriSan21-SciRep}),
Abballe and Asari (\citeyear{AbbAsa22-PONE}),
Ruslim {\em et al.\/}\ (\citeyear{RusBurLia23-bioRxiv}) and
Wendt and Faul (\citeyear{WenFay24-JVis}).

Beyond the work by Hubel and Wiesel, the properties of complex cells
have been further characterized by
Movshon {\em et al.\/}\ (\citeyear{MovThoTol78-JPhys}), 
Emerson {\em et al.\/}\ (\citeyear{EmeCitVauKle87-JNeuroPhys}),
Martinez and Alonso (\citeyear{MarAlo01-Neuron}),
Touryan {\em et al.\/}\ (\citeyear{TouLauDan02-JNeuroSci,TouFelDan05-Neuron}),
Rust {\em et al.\/}\ (\citeyear{RusSchMovSim05-Neuron}),
van~Kleef {\em et al.\/}\ (\citeyear{KleCloIbb10-JPhys}),
Goris {\em et al.\/}\ (\citeyear{GorSimMov15-Neuron}),
Li {\em et al.\/}\ (\citeyear{LiLiuChoZhaTao15-JNeuroSci})
and Almasi {\em et al.\/}\ (\citeyear{AlmMefCloWonYunIbb20-CerCort}),
as well as modelled computationally by Adelson and Bergen
(\citeyear{AdeBer85-JOSA}), Heeger (\citeyear{Hee92-VisNeuroSci}),
Serre and Riesenhuber (\citeyear{SerRie04-AIMemo}),
Einh{\"a}user {\em et  al.\/} (\citeyear{EinKayKoeKoe02-EurJNeurSci}),
Kording {\em et al.\/}\ (\citeyear{KorKayWinKon04-JNeuroPhys}),
Merolla and Boahen (\citeyear{MerBoa04-NIPS}),
Berkes and Wiscott (\citeyear{BerWis05-JVis}),
Carandini (\citeyear{Car06-JPhys}),
Hansard and Horaud (\citeyear{HanHor11-NeurComp}),
Franciosini {\em et al.\/}\ (\citeyear{FraBouPer19-AnnCompNeurSciMeet}),
Lian {\em et  al.\/} (\citeyear{LiaAlmGraKamBurMef21-PLOSCompBiol}),
Oleskiw {\em et al.\/}\ (\citeyear{OleLieSimMov23-bioRxiv})
and Yedjour and Yedjour (\citeyear{YedYed24-CognNeurDyn}).

The notion of affine Gaussian smoothing, with its
associated notion of affine Gaussian derivatives, was derived
axiomatically in (Lindeberg \citeyear{Lin10-JMIV}) and was proposed
as a spatial model of simple cells in 
(Lindeberg \citeyear{Lin13-BICY,Lin21-Heliyon}).
This model has also been extended to complex cells in
(Lindeberg \citeyear{Lin20-JMIV} Section~5).
Parallel extensions of the affine Gaussian derivative model for simple
cells to spatio-temporal receptive fields
have been performed in (Lindeberg
\citeyear{Lin16-JMIV,Lin21-Heliyon}).

\section{Receptive field models based on affine Gaussian derivatives}
\label{sec-theor-bg}

For modelling the response properties of simple cells, we will in this
paper make use of the affine Gaussian derivative model for linear
receptive fields, which has been theoretically derived in a
principled axiomatic manner in 
(Lindeberg \citeyear{Lin10-JMIV}), and then been demonstrated
in (Lindeberg \citeyear{Lin13-BICY,Lin21-Heliyon}) to well model the
spatial properties of simple cells in the primary cortex, as established
by neurophysiological measurements by DeAngelis {\em et al.\/}
(\citeyear{DeAngOhzFre95-TINS,deAngAnz04-VisNeuroSci})
and Johnson {\em et al.\/} (\citeyear{JohHawSha08-JNeuroSci});
see Figures~16--17 in (Lindeberg \citeyear{Lin21-Heliyon}) for
comparisons between biological receptive fields and computational models in terms of
affine Gaussian derivatives, as will be used as a basis for the
analysis in this paper.

In this section, we shall review main components of this theory, which
we shall then build upon in the theoretical analysis in
Section~\ref{sec-anal-ori-sel}, concerning the orientation selectivity properties of these
receptive field models, based on the affine Gaussian derivative model
for simple cells over a purely spatial domain, as well as a corresponding generalized
affine Gaussian derivative model for simple cells over a joint
spatio-temporal domain. Based on these models of simple cells, we
shall also define models for complex cells, both over a purely spatial
image domain and over a joint spatio-temporal domain, where the joint
spatio-temporal models are new.
Since the analysis to be performed in Section~\ref{sec-anal-ori-sel}
will make use of very specific properties of this receptive field
model, while this model may be less known than the Gabor model for visual
receptive fields, this review will be rather detailed,
so as to be sufficiently self-contained, to support the theoretical
analysis that will be later presented in Section~\ref{sec-anal-ori-sel}.

\subsection{Purely spatial model of simple cells}
\label{sec-spat-simple-cells}

If we initially disregard the temporal dependencies of the simple
cells, we can formulate a purely spatial model of linear receptive
fields with orientation preference according to
(Lindeberg \citeyear{Lin21-Heliyon} Equation~(23)),
see (Lindeberg \citeyear{Lin21-Heliyon} Figure~7) for illustrations of
such receptive fields;
\begin{multline}
  \label{eq-spat-RF-model}
  T_{\simple}(x_1, x_2;\; \sigma_{\varphi}, \varphi, \Sigma_{\varphi}, m) \\
  = T_{\varphi^m,\norm}(x_1, x_2;\; \sigma_{\varphi}, \Sigma_{\varphi})
  = \sigma_{\varphi}^{m} \, \partial_{\varphi}^{m} \left( g(x_1, x_2;\; \Sigma_{\varphi}) \right),
\end{multline}
where
\begin{itemize}
\item
   $\varphi$ represents the preferred orientation of the receptive
   field,
\item
  $\sigma_{\varphi}$ represents the amount of spatial smoothing%
\footnote{If we let $e_{\varphi}$ denote the unit vector in the
  direction of $\varphi$, according to
  $e_{\varphi} = (\cos \varphi, \sin \varphi)^T$, then we can determine
  $\sigma_{\varphi}$ from the spatial covariance matrix $\Sigma_{\varphi}$
  according to $\sigma_{\varphi} = \sqrt{e_{\varphi}^T \Sigma_{\varphi} \, e_{\varphi}}$.}
  in the direction $\varphi$ (in units of the spatial standard deviation%
\footnote{In this paper, we parameterize both the spatial and the
  temporal scale parameters in units of the standard deviations of the
  corresponding spatial or temporal smoothing kernels. These
  parameters are related to the corresponding variance-based
  parameterizations in the earlier work, that we build upon, according to
  $\sigma_{\spacespace} = \sqrt{s}$ and $\sigma_{\time} = \sqrt{\tau}$.
  By this change of notation, some equations in this paper will have somewhat
  different appearance, compared to the previous work that we
  cite. This change of notation will, however, lead to easier notation
in the mathematical analysis that will follow in Section~\ref{sec-anal-ori-sel}}),
\item
  $\partial_{\varphi}^m =
  (\cos \varphi \, \partial_{x_1} + \sin  \varphi \, \partial_{x_2})^m$
  is an $m$:th-order directional derivative operator%
\footnote{Concerning the notation, we write derivatives either in
  operator form $\partial_{x_1}$, which constitutes short notation for
  $\frac{\partial}{\partial x_1}$, or as subscripts, $L_{x_1} = \partial_{x_1} L$,
  which is also short notation for $\frac{\partial L}{\partial x_1}$.
  Directional derivatives of functions are also written as subscripts
  $L_{\varphi} = \partial_{\varphi} L$ and can, of course, also be
  applied repeatedly $L_{\varphi\varphi} = \partial_{\varphi}^2 L$.
  Observe, however, that subscripts are also used with a different
  meaning in connections constants. The notation $\sigma_{\varphi}$ means the
  $\sigma$-value used when computing directional derivatives, as
  opposed to $\sigma_t$, to be introduced later, which means the
  $\sigma$-value that is used when computing temporal derivatives.}
   in the direction $\varphi$,
 \item
   $\Sigma_{\varphi}$ is a symmetric positive definite covariance matrix, with
   one of its eigenvectors aligned with the direction of $\varphi$, and
 \item
   $g(x;\; \Sigma_{\varphi})$ is a 2-D affine Gaussian kernel with its shape
   determined by the covariance matrix $\Sigma_{\varphi}$
   \begin{equation}
     g(x;\; \Sigma_{\varphi})
     = \frac{1}{2 \pi \sqrt{\det \Sigma_{\varphi}}}
         e^{-x^T \Sigma_{\varphi}^{-1} x/2}
    \end{equation}
    for $x = (x_1, x_2)^T$, and with one of the eigenvectors of
    $\Sigma_{\varphi}$ parallel to the orientation $\varphi$.
\end{itemize}
For $m = 1$ and $m = 2$, differentiation of the affine
Gaussian kernel and introducing the following parameterization of
the spatial covariance matrix
\begin{equation}
  \label{def-spat-cov-mat}
  \Sigma_{\varphi}
  = \left(
        \begin{array}{cc}
          C_{11} & C_{12} \\
          C_{12} & C_{22}
        \end{array}
      \right)
\end{equation}
with
\begin{align}
    \begin{split}
       \label{eq-expl-par-Cxx}
       C_{11} & = \sigma_1^2 \, \cos^2 \varphi + \sigma_2^2 \, \sin^2 \varphi,
    \end{split}\\
    \begin{split}
          \label{eq-expl-par-Cxy}
        C_{12} & = (\sigma_1^2 - \sigma_2^2)  \, \cos \varphi  \, \sin \varphi,
    \end{split}\\
    \begin{split}
       \label{eq-expl-par-Cyy}
        C_{22} & = \sigma_1^2 \, \sin^2 \varphi + \sigma_2^2  \, \cos^2 \varphi,
   \end{split}
\end{align}
as well as reformulating the arguments with respect to this more
explicit parameterization, leads to the following explicit
expressions, for an arbitrary preferred orientation $\varphi$:
\begin{align}
  \begin{split}
     T_{\simple}(x_1, x_2;\; \sigma_1, \sigma_2, \varphi, 1) =
  \end{split}\nonumber\\
  \begin{split}
    = \sigma_1 \, (\cos (\varphi) \, \partial_{x_1} + \sin(\varphi) \, \partial_{x_2}) \, 
        g(x_1, x_2;\; \Sigma_{\varphi})
  \end{split}\nonumber\\
  \begin{split}
     =
     -\frac{(x_1 \cos (\varphi )+x_2 \sin (\varphi ))}
               {2 \pi  \, \sigma_1^2 \, \sigma_2} \times
   \end{split}\nonumber\\
  \begin{split}
    \label{eq-1dirder-affgauss}
    e^{
   -\frac{\left(\sigma_1^2+\sigma_2^2\right)
   \left(x_1^2+x_2^2\right)-(\sigma_1-\sigma_2) (\sigma_1+\sigma_2) (2 x_1
   x_2 \sin (2 \varphi )+\cos (2 \varphi ) (x_1-x_2) (x_1+x_2))}{4 \sigma_1^2
   \sigma_2^2}}
  \end{split}\\
  \begin{split}
     T_{\simple}(x_1, x_2;\; \sigma_1, \sigma_2, \varphi, 2) =
  \end{split}\nonumber\\
 \begin{split}
   = \sigma_1^2 \, (\cos^2 (\varphi) \, \partial_{x_1 x_1} +
   2 \cos(\varphi) \sin(\varphi) \, \partial_{x_1 x_2} +
    \sin^2(\varphi) \, \partial_{x_2 x_2}) \, 
 \end{split}\nonumber\\
  \begin{split}
         \quad\quad g(x_1, x_2;\; \Sigma_{\varphi})
  \end{split}\nonumber\\
  \begin{split}
     =
     \frac{
       \left(
         x_1^2+x_2^2 -2 \sigma_1^2
         + \cos (2 \varphi ) \left(x_1^2-x_2^2\right)
         +2 x_1 x_2 \sin (2 \varphi )
        \right)}
            {4 \pi  \, \sigma_1^3 \, \sigma_2} \times
  \end{split}\nonumber\\
  \begin{split}
    \label{eq-2dirder-affgauss}    
      e^{
   -\frac{\left(\sigma_1^2+\sigma_2^2\right)
   \left(x_1^2+x_2^2\right)-(\sigma_1-\sigma_2) (\sigma_1+\sigma_2) (2 x_1
   x_2 \sin (2 \varphi )+\cos (2 \varphi ) (x_1-x_2) (x_1+x_2))}{4 \sigma_1^2
   \sigma_2^2}}.
  \end{split}
\end{align}
In the above expressions for the spatial receptive field model
$T_{\simple}$, the multiplication of the
$m$:th-order directional derivative operator in the direction
$\varphi$ by the spatial scale
parameter $\sigma_{\varphi}$ in the same direction, implements
scale-normalized derivatives (Lindeberg \citeyear{Lin97-IJCV})
according to
\begin{equation}
  \label{sec-spat-sc-norm-ders}
  \partial_{x_1^{\alpha_1} x_2^{\alpha_2},\norm}
  = \sigma^{\gamma (\alpha_1 + \alpha_2)} \, \partial_{x_1^{\alpha_1}  x_2^{\alpha_2}},
\end{equation}
where we here, for simplicity, choose the scale normalization power
$\gamma = 1$ to simplify the following calculations.

This notion of spatial scale-normalized
derivatives implies that we measure the amplitude of local spatial
variations with respect to a given scale level, and makes it possible
to define scale-invariant feature responses, that assume the same
magnitude for input image structures of different spatial extent,
provided that the spatial scale levels are adapted to the
characteristic length of spatial variations in the image data.
Specifically, this spatial scale normalization of the spatial
receptive field response implies scale selective properties in the
sense that the receptive field will produce its maximum response over
spatial scales at a spatial scale proportional to a characteristic
length in the image data.

Note that this model for the spatial dependency of simple cells, in
terms of affine Gaussian derivatives, goes
beyond the previous biological modelling results by Young (\citeyear{You87-SV}),
in turn with very close relations to theoretical modelling results by
Koenderink and van Doorn (\citeyear{KoeDoo87-BC,KoeDoo92-PAMI}),
in that the spatial smoothing part of the receptive field is here spatially
anisotropic, and thereby allows for higher orientation selectivity
compared to defining (regular) Gaussian derivatives from partial
derivatives or directional derivatives of rotationally symmetric
Gaussian kernels, as used by Young and Koenderink and van Doorn.
Direct comparisons with biological receptive fields,
see Figures~16--17 in (Lindeberg \citeyear{Lin21-Heliyon}), also show
that biological simple cells are more aniso\-tropic than can be well modelled
by directional derivatives of rotationally symmetric Gaussian kernels.

\subsection{Purely spatial model of complex cells}
\label{sec-spat-model-comp-cells}

In (Lindeberg \citeyear{Lin20-JMIV} Section~5) it was proposed that
some of the qualitative properties of complex cells, of being
both (i)~polarity-independent, (ii)~approximately phase-independ\-ent and (ii)~not obeying
a superposition principle, as opposed to
polarity-dependent as well as strongly phase-dependent, as simple cells
are,
can modelled by
combining first- and second-order directional affine Gaussian
derivative responses of the form%
\footnote{In the expressions below, the symbol $L$ denotes the result
  of pure spatial smoothing of the input image $f(x_1, x_2)$ with an affine
  Gaussian kernel $g(x_1, x_2;\; \Sigma)$, {\em i.e.\/},
  $L(x_1, x_2;\; \Sigma) = g(x_1, x_2;\; \Sigma) * f(x_1, x_2)$.
  In the area of scale-space theory, this representation is referred
  to as the (affine) spatial scale-space representation of $f$.}
\begin{equation}
  \label{eq-quasi-quad-dir}
  {\cal Q}_{\varphi,\spat,\norm} L
  = \sqrt{\frac{L_{\varphi,\norm}^2
               + C_{\varphi} \, L_{\varphi\varphi,\norm}^2}{\sigma_{\varphi}^{2\Gamma}}},
\end{equation}
where
\begin{itemize}
\item
  $L_{\varphi,\norm}$ and $L_{\varphi\varphi,\norm}$ represent the results of
  applying scale-normalized directional affine Gaussian derivative operators of orders 1 and~2,
  respectively, according to (\ref{eq-spat-RF-model}),
  to the input image $f$:
  \begin{multline}
       L_{\varphi,\norm}(x_1, x_2;\;  \sigma_{\varphi}, \Sigma_{\varphi}) =\\
       = T_{\varphi,\norm}(x_1, x_2;\; \sigma_{\varphi}, \Sigma_{\varphi}) *
       f(x_1, x_2),
    \end{multline}
    \begin{multline}
       L_{\varphi\varphi,\norm}(x_1, x_2;\;  \sigma_{\varphi}, \Sigma_{\varphi}) =\\
        = T_{\varphi\varphi,\norm}(x_1, x_2;\; \sigma_{\varphi}, \Sigma_{\varphi}) *
       f(x_1, x_2),
    \end{multline}
\item
  $C_{\varphi} > 0$ is a weighting factor between first and second-order
  information, and
\item
  $\Gamma \geq 0$ is a complementary scale
normalization parameter, that we, however, henceforth will set to zero,
to simplify the following treatment.
\end{itemize}
This model can be seen as an affine Gaussian derivative analogue of
the energy model of complex cells proposed by
Adelson and Bergen (\citeyear{AdeBer85-JOSA}) and
Heeger (\citeyear{Hee92-VisNeuroSci}), specifically the fact that
receptive fields similar to
first- {\em vs.\/}\ second-order derivatives are reported to occur in
pairs (De~Valois {\em et al.\/}\ \citeyear{ValCotMahElfWil00-VR}),
resembling properties of approximate quadrature pairs, as related by a
Hilbert transform  (Bracewell \citeyear{Bra99}, pp.~267--272).
The model is also closely related to the proposal by 
Koenderink and van Doorn (\citeyear{KoeDoo90-BC}) to  sum up the squares of
first- and second-order derivative response, in a corresponding way as
cosine wave and sine wave responses are combined in a Euclidean manner,
to get a more phase-independent response.

For a perfect quadrature pair of filters, the sum of the squares of
the filter responses will be spatially constant for any sine wave of
any frequency and phase. The quasi-quadra\-ture entity will instead have the
property that the response will be spatially constant, or
alternatively have only relatively moderate ripples, at or near the
spatial scale level at which the quasi-quadrature measure assumes its
maximum value over spatial scales, provided that the value of the
weighting parameter $C_{\varphi}$ is properly chosen
(see Lindeberg (\citeyear{Lin97-IJCV}) Figure~21 and
Lindeberg (\citeyear{Lin18-SIIMS}) Equation~(26)).

In this way, the quasi-quadrature
measure combines the responses of the first- and second-order affine
Gaussian derivative operators in a complementary manner, where the
first-order derivatives correspond to odd filters (antisymmetric
under reflection) and the second-order
derivatives to even filters (symmetric under reflection).
  
\subsection{Joint spatio-temporal models of simple cells}
\label{sec-spat-temp-simpl-cells}

For modelling the joint spatio-temporal behaviour of linear receptive fields with
orientation preference, in (Lindeberg \citeyear{Lin21-Heliyon} Section~3.2)
the following model is derived from theoretical arguments,
see (Lindeberg \citeyear{Lin21-Heliyon} Figures~10-11 for illustrations)
\begin{align}
  \begin{split}
    \label{eq-spat-temp-RF-model-der-norm-caus}
    T_{\simple}(x_1, x_2, t;\; \sigma_{\varphi}, \sigma_t, \varphi, v, \Sigma_{\varphi}, m, n) 
  \end{split}\nonumber\\
  \begin{split}
   & = T_{{\varphi}^m, {\bar t}^n,\norm}(x_1, x_2, t;\; \sigma_{\varphi}, \sigma_t, v, \Sigma_{\varphi})
  \end{split}\nonumber\\
  \begin{split}
   &  = \sigma_{\varphi}^{m} \, 
          \sigma_t^{n} \, 
         \partial_{\varphi}^{m} \,\partial_{\bar t}^n 
          \left( g(x_1 - v_1 t, x_2 - v_2 t;\; \Sigma_{\varphi}) \,
           h(t;\; \sigma_t) \right),
  \end{split}
\end{align}
where (for symbols not previously defined in connection
with Equation~(\ref{eq-spat-RF-model}))
\begin{itemize}
\item
  $\sigma_t$ represents the amount of temporal smoothing (in units of
  the temporal standard deviation),
\item
  $v = (v_1, v_2)^T$ represents a local motion vector, in the
  direction $\varphi$ of the spatial orientation of the receptive field,
\item
  $\partial_{\bar t}^n = (\partial_t + v_1 \, \partial_{x_1} + v_2 \, \partial_{x_2})^n$
  represents an $n$:th-order velocity-adapted temporal derivative
  operator,
\item
  $h(t;\; \sigma_t)$ represents a temporal smoothing kernel with temporal
  standard deviation $\sigma_t$.
\end{itemize}
In the case of non-causal time (where the future can be accessed), the
temporal kernel can be determined to be a 1-D Gaussian kernel
\begin{equation}
  h(t;\; \sigma_t) = \frac{1}{\sqrt{2 \pi} \sigma_t} e^{-t^2/2\sigma_t^2},
\end{equation}
whereas in the case of time-causal time (where the future cannot be
accessed), the temporal kernel can instead be chosen as the
time-causal limit kernel
(Lindeberg \citeyear{Lin16-JMIV} Section~5,
 \citeyear{Lin23-BICY} Section~3)
\begin{equation}
  \label{eq-time-caus-lim-kern}
  h(t;\; \sigma_t) = \psi(t;\; \sigma_t, c),
\end{equation}
defined by having a Fourier transform of the form
\begin{equation}
  \label{eq-FT-comp-kern-log-distr-limit}
     \hat{\Psi}(\omega;\; \sigma_t, c) 
     = \prod_{k=1}^{\infty} \frac{1}{1 + i \, c^{-k} \sqrt{c^2-1} \, \sigma_t \, \omega},
\end{equation}
and corresponding to an infinite set of first-order integrators coupled in
cascade with specifically chosen time constants to enable temporal
scale covariance, where the distribution parameter $c > 1$ describes the
ratio between adjacent discrete temporal scale levels in this temporal
scale-space model.

In analogy with the spatial scale normalization in the previous
purely spatial model, multiplication of
the $n$:th-order velocity-adapted temporal derivative operator
$\partial_{\bar t}^n$ by the temporal standard deviation $\sigma_t$
raised to the power of $n$ implements scale-normalized
velocity-adapted temporal derivatives according to
  \begin{equation}
    \partial_{\bar t,\norm}^n = \sigma_t^{\gamma n} \, \partial_{\bar t}^n,
  \end{equation}
as an
extension of Equation~(\ref{sec-spat-sc-norm-ders}) from the spatial
to the temporal domain, see (Lindeberg \citeyear{Lin17-JMIV}). 
Also here, for simplicity, we restrict ourselves to the specific choice of the
scale normalization parameter $\gamma = 1$.
By this temporal scale normalization, the temporal component of the
spatio-temporal receptive field will have scale selective properties,
implying that it will produce its maximum response over temporal
scales at a temporal scale proportional to a characteristic temporal
duration of the temporal structures in the video data.
  
In Figure~18 in (Lindeberg \citeyear{Lin21-Heliyon}), it is
demonstrated that this model well captures the qualitative properties of simple
cells in the primary visual cortex, as established by
neurophysiological cell recordings by DeAngelis {\em et al.\/}
(\citeyear{DeAngOhzFre95-TINS,deAngAnz04-VisNeuroSci}),
regarding both space-time separable receptive fields and
velocity-adapted receptive fields, tuned to particular motion
directions in joint space-time.

Note that these spatio-temporal models of simple cells go beyond the
previous biological modelling results by Young and his co-workers
(\citeyear{YouLesMey01-SV,YouLes01-SV}) in that:
(i)~the purely spatial smoothing component of the receptive field model is based
on aniso\-tropic Gaussian kernels as opposed to rotationally symmetric
Gaussian kernels,
(ii)~this model also incorporates a truly time-causal model, that takes
into explicit account that the future cannot be accessed in a
real-world situation, and
(iii)~the parameterization of the spatio-temporal filter shapes is
different, and more closely aligned to the inherent geometry of the
imaging situation.

\subsection{Joint spatio-temporal models of complex cells}
\label{sec-spat-temp-compl-cells}

\subsubsection{Model based on space-time separable receptive fields}

For modelling qualitative properties of complex cells over the joint
spatio-temporal domain, we can extend the spatial quasi-quadrature measure in
(\ref{eq-quasi-quad-dir}) to the following spatio-temporal
quasi-quadrature measure that operates on a combination of spatial directional
derivatives and temporal derivatives according to
\begin{multline}
  \label{eq-quasi-quad-dir-spat-temp}
  ({\cal Q}_{\varphi,\sep,\norm} L)^2
  = \left(
         \left( L_{\varphi,t,\norm}^2
                 + C_t \, L_{\varphi,tt,\norm}^2
         \right)
       \right. \\
       \left.      
         + \, C_{\varphi} 
             \left( L_{\varphi\varphi, t,\norm}^2
                      + C_t \, L_{\varphi\varphi,tt,\norm}^2
             \right)
       \right)/(\sigma_{\varphi}^{2\Gamma_{\varphi}} \sigma_{t}^{2 \Gamma_t}),
\end{multline}
where the individual components in this expression are defined
from space-time separable receptive fields according to
\begin{multline}
  \label{eq-spat-temp-quasi-comp1}
    L_{\varphi,t,\norm}(x_1, x_2, t;\;  \sigma_{\varphi}, \sigma_t, 0, \Sigma_{\varphi}) =\\
    = T_{\varphi,t,\norm}(x_1, x_2, t;\; \sigma_{\varphi}, \sigma_t, 0, \Sigma_{\varphi}) *
        f(x_1, x_2, t),
\end{multline}
\begin{multline}
  \label{eq-spat-temp-quasi-comp2}
    L_{\varphi,tt,\norm}(x_1, x_2, t;\;  \sigma_{\varphi}, \sigma_t, 0, \Sigma_{\varphi}) =\\
    = T_{\varphi,tt,\norm}(x_1, x_2, t;\; \sigma_{\varphi}, \sigma_t, 0, \Sigma_{\varphi}) *
        f(x_1, x_2, t),
\end{multline}
\begin{multline}
    \label{eq-spat-temp-quasi-comp3}
    L_{\varphi\varphi,t,\norm}(x_1, x_2, t;\;  \sigma_{\varphi}, \sigma_t, 0, \Sigma_{\varphi}) =\\
    = T_{\varphi\varphi,t,\norm}(x_1, x_2, t;\; \sigma_{\varphi}, \sigma_t, 0, \Sigma_{\varphi}) *
        f(x_1, x_2, t),
\end{multline}
\begin{multline}
    \label{eq-spat-temp-quasi-comp4}
    L_{\varphi\varphi,tt,\norm}(x_1, x_2, t;\;  \sigma_{\varphi}, \sigma_t, 0, \Sigma_{\varphi}) =\\
    = T_{\varphi\varphi,tt,\norm}(x_1, x_2, t;\; \sigma_{\varphi}, \sigma_t, 0, \Sigma_{\varphi}) *
        f(x_1, x_2, t),
\end{multline}
with the underlying space-time separable spatio-temporal receptive fields
$T_{\varphi^m, t^n,\norm}(x_1, x_2, t;\; \sigma_{\varphi}, \sigma_t, 0, \Sigma_{\varphi}) $
according to (\ref{eq-spat-temp-RF-model-der-norm-caus}) for $v = 0$.

In the expression (\ref{eq-quasi-quad-dir-spat-temp}),
the quasi-quadrature measure operates on both pairs of
first- and second-order directional derivates as well as pairs of
first- and second-order velocity-adapted derivatives simultaneously,
aimed at balancing the responses of odd {\em vs.\/}\ even filter
responses over both space or time in parallel.

This (new) spatio-temporal quasi-quadrature measure constitutes an extension
of the spatio-temporal quasi-quadrature measure in
(Lindeberg \citeyear{Lin18-SIIMS} Section~4.2), by being additionally adapted to be selective
to particular image orientations over the spatial domain, as opposed
to being rotationally isotropic, as well as being
based on affine Gaussian derivative operators, as opposed to partial
derivatives of rotationally symmetric Gaussian kernels over the
spatial domain.

This oriented spatio-temporal quasi-quadrature measure will
specifically inherit the qualitative properties of the previously
presented purely spatial oriented quasi-quadrature measure
(\ref{eq-quasi-quad-dir-pure-spat-anal}), in that it will be
polarity-independent as well as much less
sensitive to the phase in the input video data compared to the above
models of simple cells.

\subsubsection{Model based on velocity-adapted receptive fields}

If we would compute the above spatio-temporal quasi-quadra\-ture measure
based on velocity-adapted spatio-temporal receptive fields, then the
quasi-quadrature measure would be zero if the velocity vector of the
velocity-adapted receptive fields is equal to the velocity-value of
the moving sine wave pattern.
To define a quasi-quadrature measure that instead will give a
maximally strong response if the velocity vector is adapted to the
velocity vector of a moving pattern, we do therefore instead define a
quasi-quadrature measure for velocity-adapted receptive fields by
extending the purely spatial quasi-quadrature measure (\ref{eq-quasi-quad-dir}) to operate on
spatial receptive fields complemented by a temporal smoothing stage,
{\em i.e.\/}, velocity-adapted receptive fields for zero order of
temporal differentiation ($n = 0$ in  (\ref{eq-spat-temp-RF-model-der-norm-caus})):
\begin{multline}
  \label{eq-quasi-quad-dir-vel-adapt-spat-temp}
  ({\cal Q}_{\varphi,\vel,\norm} L)
  = \sqrt{\frac{L_{\varphi,\norm}^2 
              + \, C_{\varphi} \, L_{\varphi\varphi,\norm}^2}{\sigma_{\varphi}^{2\Gamma}}},
\end{multline}
where the individual components in this expression are defined
from space-time separable receptive fields according to
\begin{multline}
  \label{eq-spat-temp-quasi-vel-adapt-comp1}
    L_{\varphi,\norm}(x_1, x_2, t;\;  \sigma_{\varphi}, \sigma_t, v, \Sigma_{\varphi}) =\\
    = T_{\varphi,\norm}(x_1, x_2, t;\; \sigma_{\varphi}, \sigma_t, v, \Sigma_{\varphi}) *
        f(x_1, x_2, t),
 \end{multline}
\begin{multline}
    \label{eq-spat-temp-quasi-vel-adapt-comp2}
    L_{\varphi\varphi,\norm}(x_1, x_2, t;\;  \sigma_{\varphi}, \sigma_t, v, \Sigma_{\varphi}) =\\
    = T_{\varphi\varphi,\norm}(x_1, x_2, t;\; \sigma_{\varphi}, \sigma_t, v, \Sigma_{\varphi}) *
        f(x_1, x_2, t),
\end{multline}
with the underlying space-time separable spatio-temporal receptive fields
$T_{\varphi^m, t^n,\norm}(x_1, x_2, t;\; \sigma_{\varphi}, \sigma_t, v, \Sigma_{\varphi}) $
according to (\ref{eq-spat-temp-RF-model-der-norm-caus}) for $n = 0$.

Again, the intention behind this (also new) definition is that also this
spatio-temporal quasi-quadrature measure should be both
polarity-independent with respect to the input and much less sensitive
to the phase in the input compared to the velocity-adapted
spatio-temporal models of simple cells, while also in conceptual
agreement with the energy models of complex cells.

\section{Orientation selectivity properties for models of simple cells
  and complex cells in terms of affine Gaussian derivatives}
\label{sec-anal-ori-sel}

In this section, we will theoretically analyze the orientation
selectivity properties of the above described purely spatial as well as joint
spatio-temporal models for receptive fields, when exposed to sine wave
patterns with different orientations in relation to the preferred
orientation of the receptive field.

\subsection{Motivation for separate analyses for the different classes
  of purely spatial or joint spatio-temporal receptive field models,
  as well as motivation for studying the orientation selectivity
  curves for models of both simple cells and complex cells}

A main reason for performing separate analyses for both purely
spatial and joint spatio-temporal models, is that, {\em a priori\/},
it may not be clear how the results from an analysis of the orientation
selectivity of a purely spatial model would relate to the results from
an orientation selectivity analysis of a joint spatio-temporal model.
Furthermore, within the domain of joint spatio-temporal models, it is not
{\em a priori\/} clear how the results from an orientation selectivity
analysis of a space-time separable spatio-temporal model would relate to the results
for a velocity-adapted spatio-temporal model. For this reason, we will
perform separate individual analyses for these three main classes of
either purely spatial or joint spatio-temporal receptive field models.

Additionally, it may
furthermore not be {\em a priori\/} clear how the results 
from orientation selectivity analyses of models of simple cells that have
different numbers of dominant spatial or spatio-temporal lobes would
relate to each other, nor how the results from an orientation
selectivity analysis of a model of a complex cell would relate to the analysis of
orientation selectivity of models of simple cells. For this reason, we will also
perform separate analyses for models of simple and complex cells, for each one of the above
three main classes of receptive field models, which leads to a total
number of $(2 + 1) + (2 \times 2 + 1) + (2 + 1) = 11$ separate
analyses, in the cases of either purely spatial, space-time separable or
velocity-adapted receptive spatio-temporal fields, respectively,
given that we restrict
ourselves to simple cells that can be modelled in terms of
either first- or second-order spatial and/or temporal derivatives of
affine Gaussian kernels.
The reason why there are $2 \times 2 + 1 = 5$ subcases for the
space-time separable spatio-temporal model is that we consider all the possible combinations of
first- and second-order spatial derivatives with all possible
combinations of first- and second-order temporal derivatives.

As will be demonstrated from the results, to be summarized in
Table~\ref{tab-summ-ori-sel-diff-models}, the
results from the orientation selectivity of the simple cells will
turn out be similar within the class of purely first-order simple cells
between the three classes of receptive field models
(purely spatial, space-time separable or velocity-adapted
spatio-temporal), as well as also similar
within the class of purely second-order simple cells between the three
classes of receptive field models. Regarding the models of complex
cells, the results of the orientation selectivity analysis will be
similar between the purely spatial and the velocity-adapted
spatio-temporal receptive field models, whereas the results from the
space-time separable spatio-temporal model of complex cells
will differ from the other
two. The resulting orientation selectivity curves will,
however, differ between the sets of (i)~first-order simple cells,
(ii)~second-order simple cells and (iii)~complex cells.

The resulting special handling of the different cases that will arise from this
analysis, with their different response characteristics, is
particularly important, if one wants to fit parameterized models of
orientation selectivity curves to actual neurophysiological data.
If we want to relate neurophysiological findings measured for neurons,
that are exposed to time-varying visual stimuli, or for neurons that have a
strong dependency on temporal variations in the visual stimuli, then
our motivation for performing a genuine spatio-temporal analysis for
each one of the main classes of spatio-temporal models for simple
cells, is to reduce the explanatory gap between the models and
actual biological neurons.

\subsection{Modelling scenario}

\begin{figure}[hbtp]
  \begin{center}
    \includegraphics[width=0.40\textwidth]{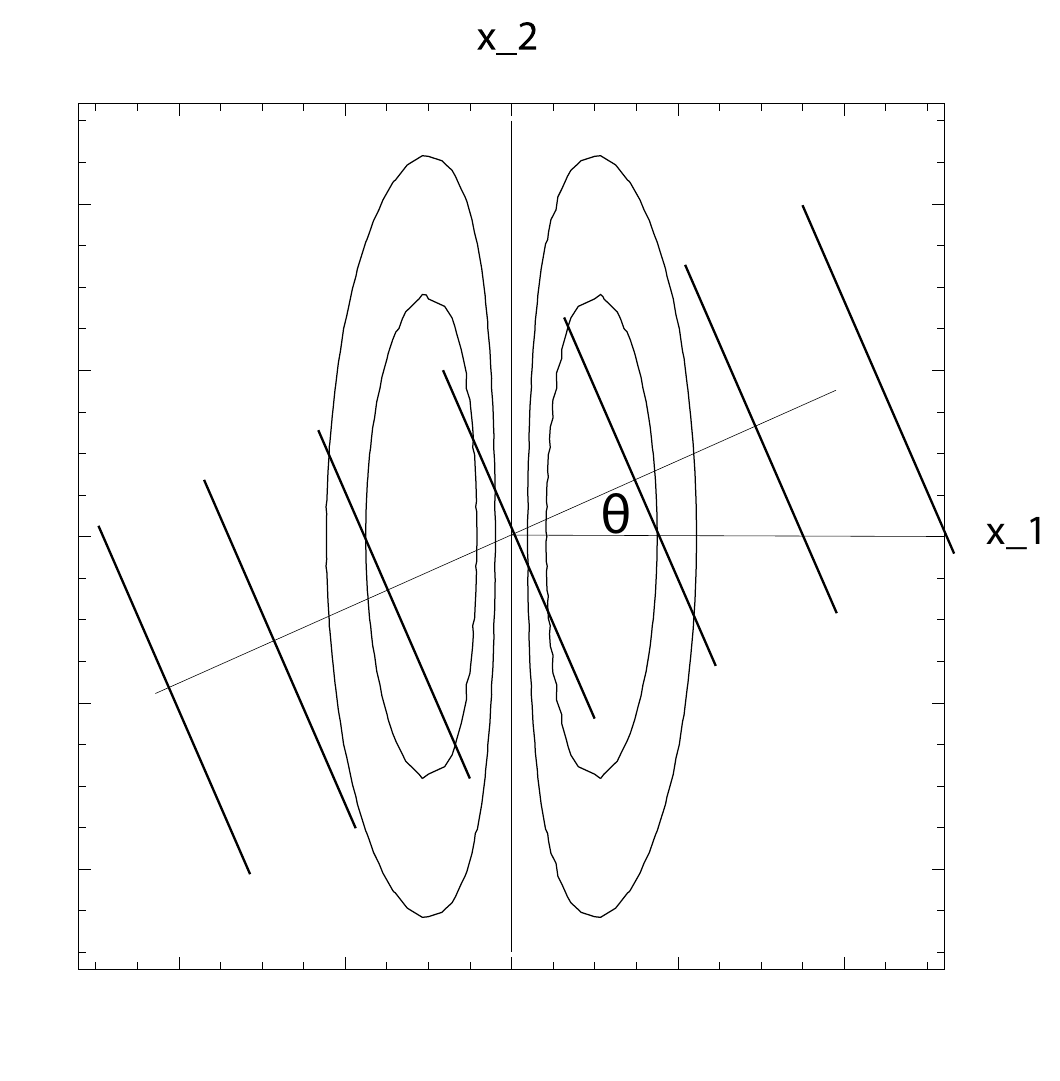}
  \end{center}
  \caption{Schematic illustration of the modelling situation studied
    in the theoretical analysis, where
    the coordinate system is aligned to the preferred orientation
    $\varphi = 0$ of the receptive field, and the receptive field is
    then exposed to a sine wave
    pattern with inclination angle $\theta$. In this figure, the sine
    wave pattern is schematically illustrated by a set of level lines,
  overlayed onto a few level curves of a first-order affine Gaussian
  derivative kernel. (Horizontal axis: spatial coordinate $x_1$.
  Vertical axis: spatial coordinate $x_2$.)}
  \label{fig-schem-ill-model-rf-sine}
\end{figure}

\begin{figure*}[btp]
\begin{center}
    \begin{tabular}{cccccc}
      $\kappa= 1$
      & $\kappa = 2 \sqrt{2}$
      & $\kappa = 2$
      & $\kappa = 2\sqrt{2}$
      & $\kappa = 4$        
      & $\kappa = 4\sqrt{2}$ \\
     \includegraphics[width=0.14\textwidth]{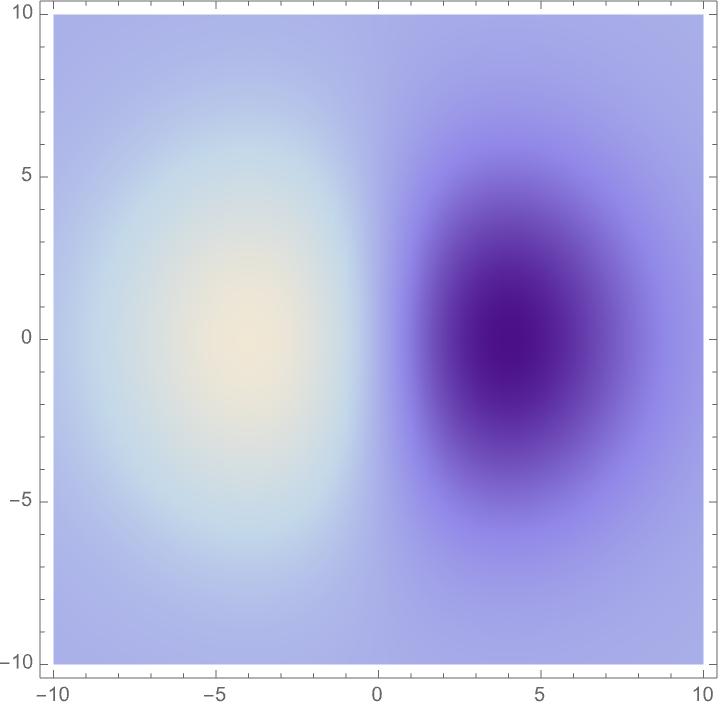}
      & \includegraphics[width=0.14\textwidth]{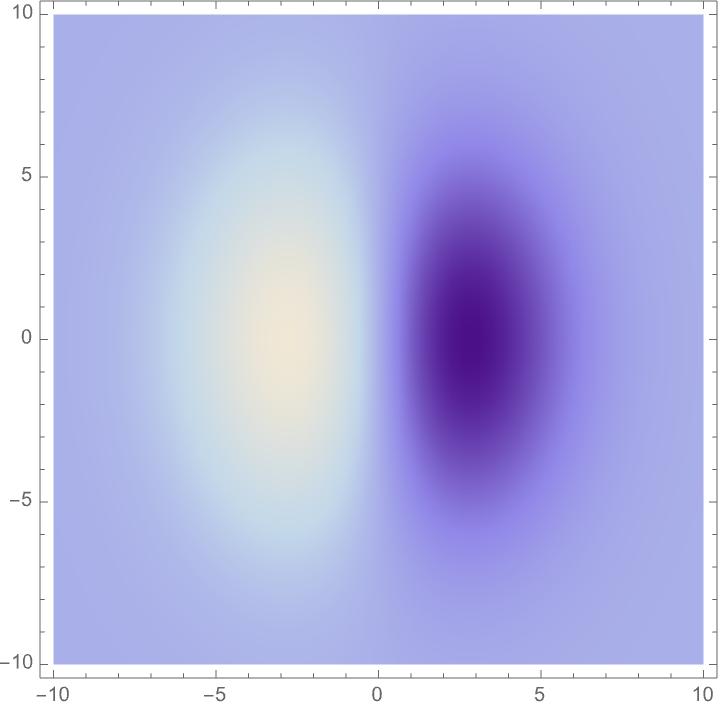}
      & \includegraphics[width=0.14\textwidth]{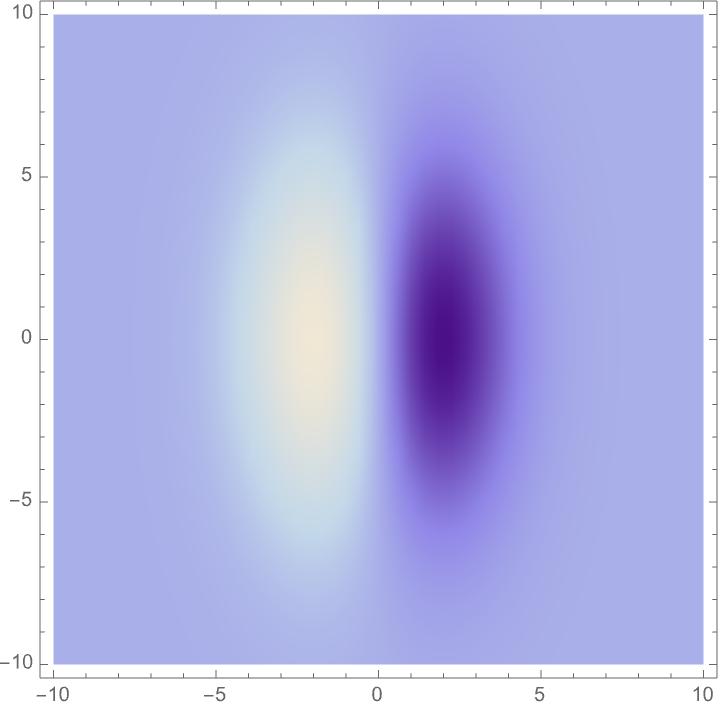}
      & \includegraphics[width=0.14\textwidth]{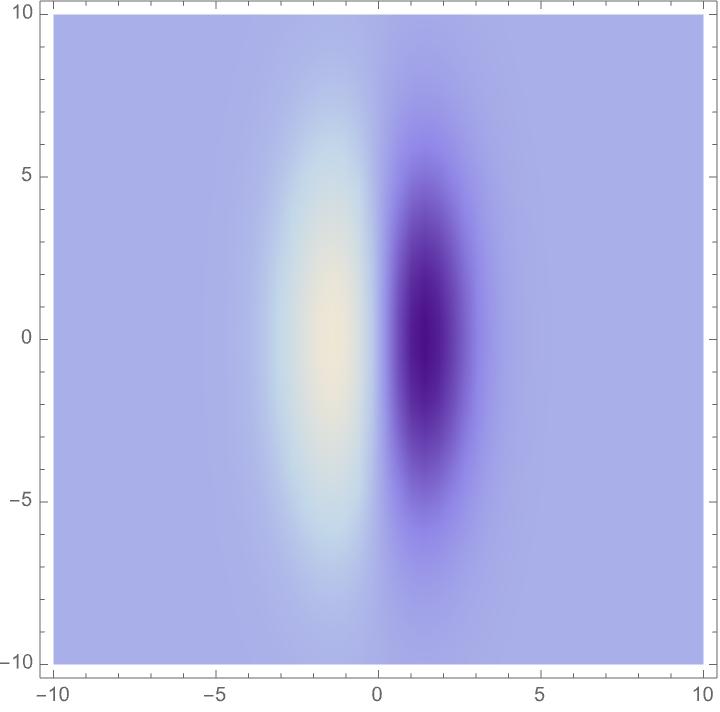}
      & \includegraphics[width=0.14\textwidth]{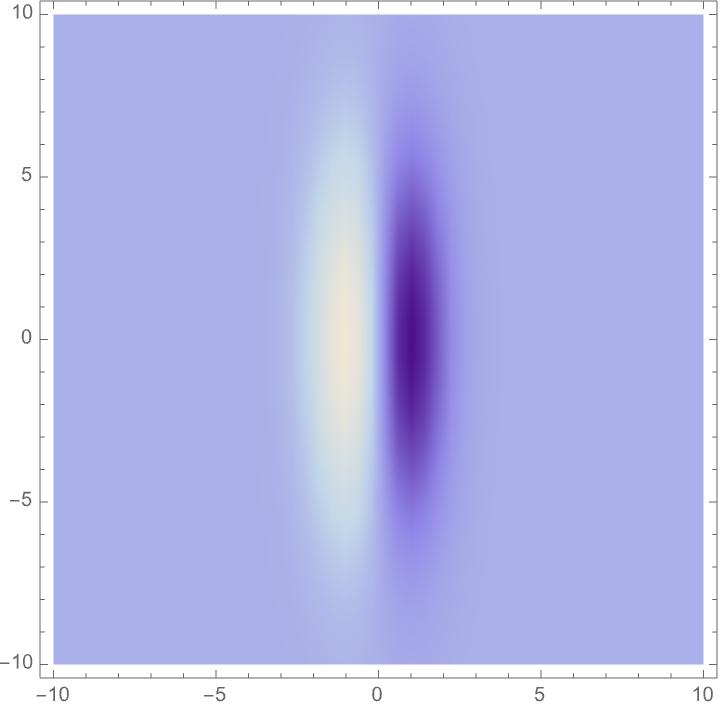}
      & \includegraphics[width=0.14\textwidth]{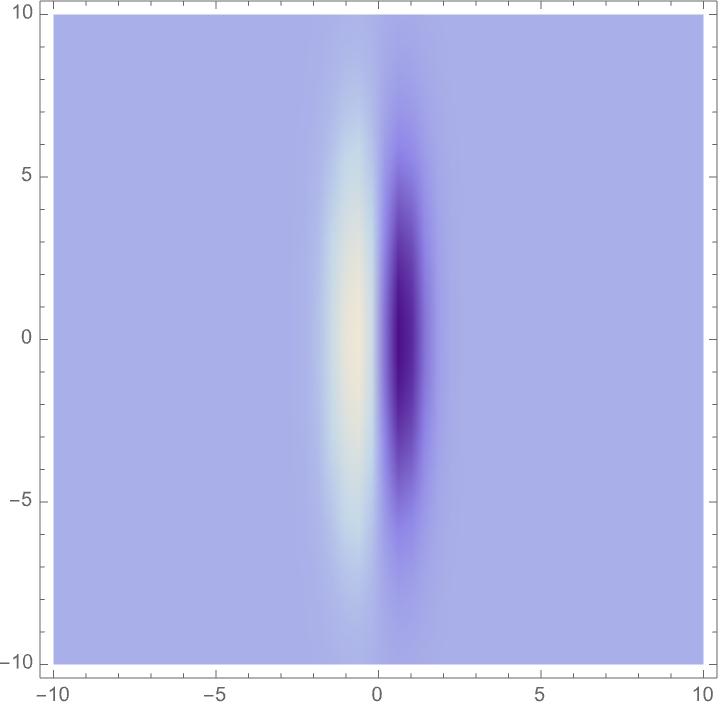} \\
     \includegraphics[width=0.14\textwidth]{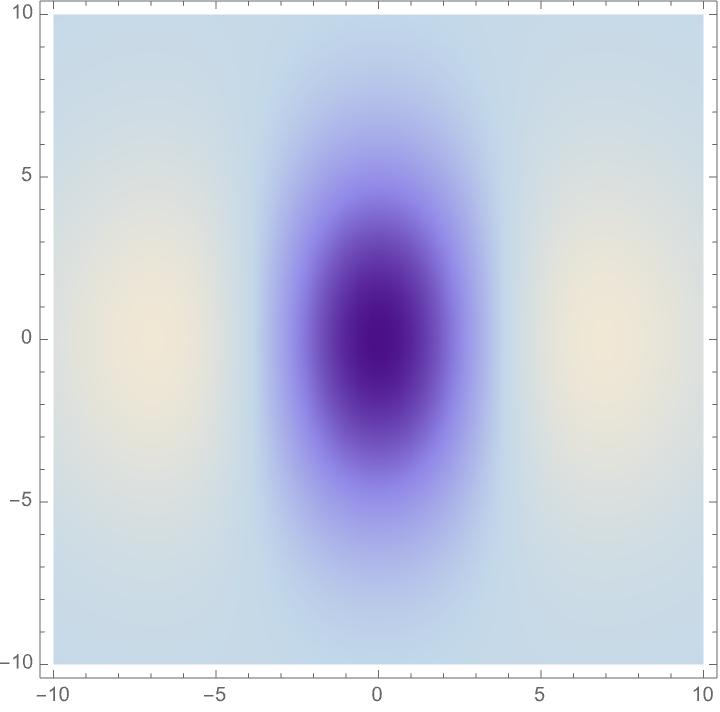}
      & \includegraphics[width=0.14\textwidth]{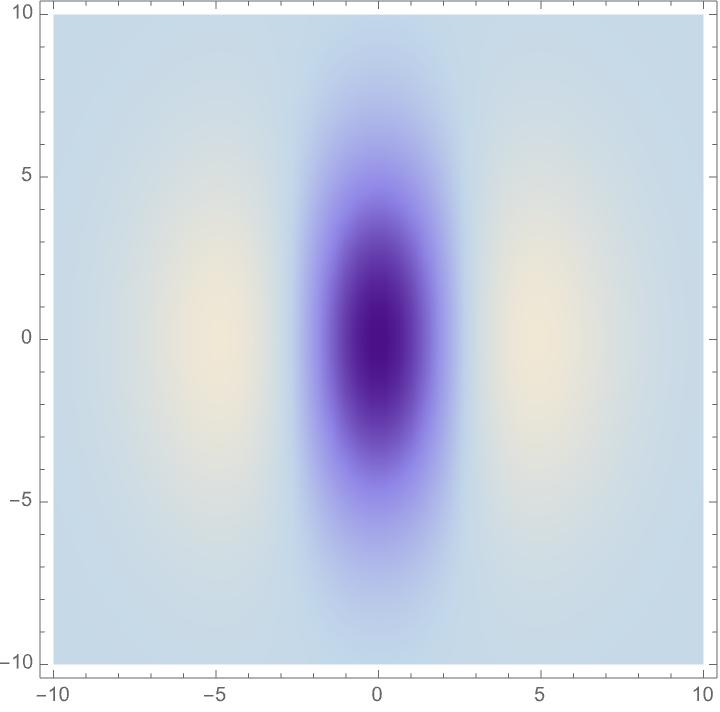}
      & \includegraphics[width=0.14\textwidth]{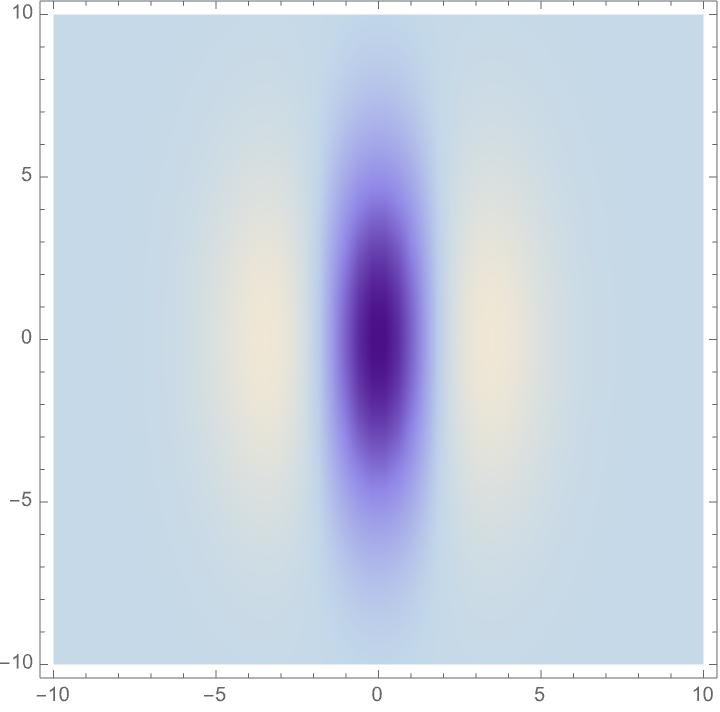}
      & \includegraphics[width=0.14\textwidth]{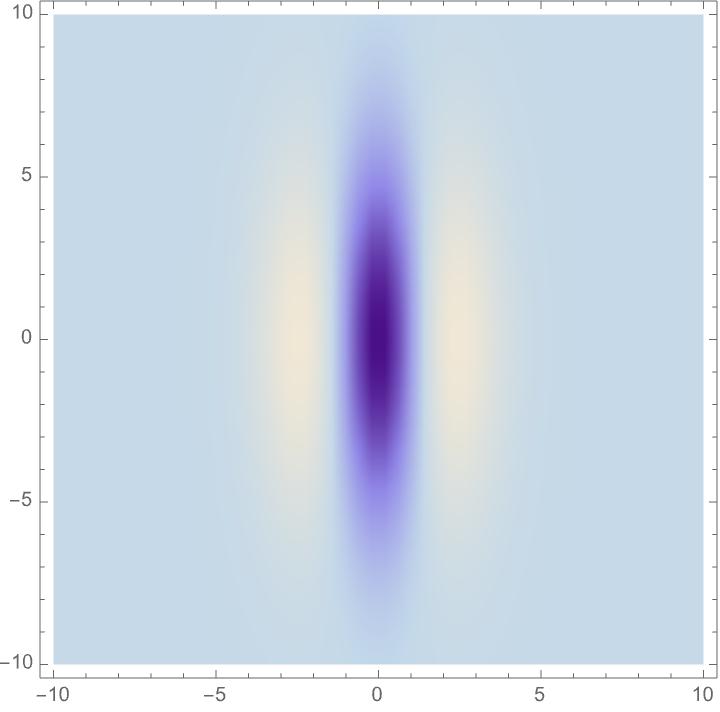}
      & \includegraphics[width=0.14\textwidth]{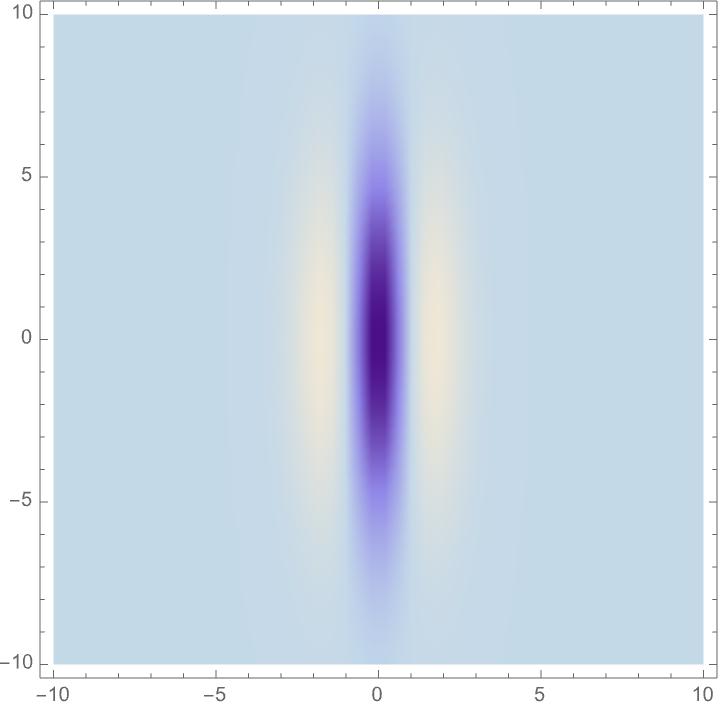}
      &  \includegraphics[width=0.14\textwidth]{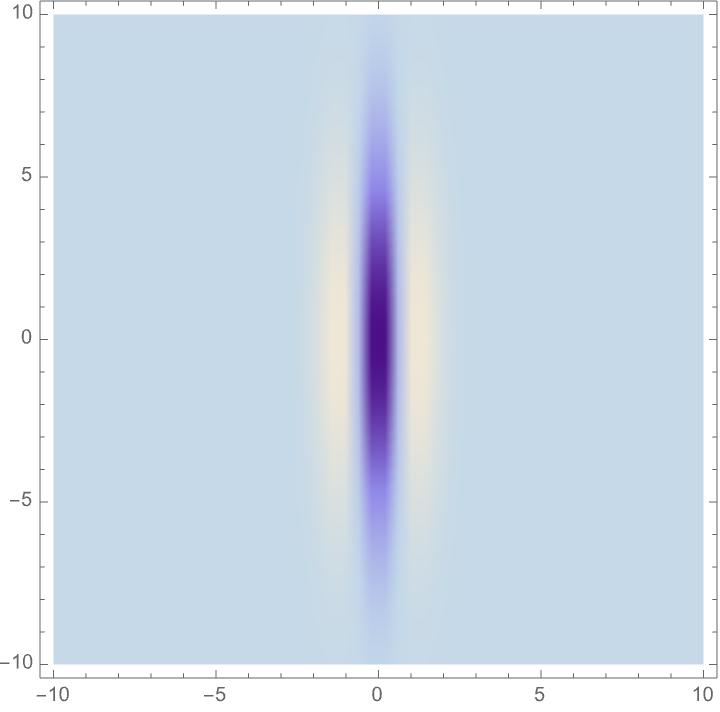} 
    \end{tabular}
  \end{center}
  \caption{Affine Gaussian derivative
    receptive fields of orders 1 and 2 for the image orientation $\varphi = 0$,
    with the scale parameter ratio 
    $\kappa = \sigma_2/\sigma_1$ increasing from $1$ to $4\sqrt{2}$
    according to a logarithmic distribution, from left to right,
    with the larger of these scale parameters $\sigma_2$ kept constant.
    (top row) First-order directional derivatives of affine Gaussian
    kernels according to (\ref{eq-spat-RF-model}) for $m = 1$.
    (bottom row) Second-order directional derivatives of affine
    Gaussian kernels according to (\ref{eq-spat-RF-model}) for $m = 2$.
    (Horizontal axes: image coordinate $x_1$.
     Vertical axes: image coordinate $x_2$.)}
  \label{fig-ecc-variability}
\end{figure*}

For simplicity of analysis, and without loss of generality,
we will henceforth align the coordinate system with the preferred
orientation of the receptive field, in other words choosing the
coordinate system such that the orientation angle $\varphi = 0$.
Then, we will expose this receptive field to static or moving
sine wave patterns, that are oriented with respect to an inclination angle
$\theta$ relative to the resulting horizontal $x_1$-direction, as
illustrated in Figure~\ref{fig-schem-ill-model-rf-sine}.
Figure~\ref{fig-ecc-variability} shows maps of the underlying first-order
and second-order Gaussian derivatives for different degrees of
elongation of the receptive fields.

By necessity, the presentation that will follow will be somewhat
technical, since we will analyze the properties of our mathematical
models for the receptive fields of simple and complex cells for three
different main cases of either (i)~purely spatial receptive fields,
(ii)~space-time separable spatio-temporal receptive fields and
(iii)~velocity-adapted spatio-temporal receptive fields.

Readers, who may be more interested in the final results
only and their biological implications, while not in the details of the
mathematical modelling with its associated theoretical analysis,
can without major loss of continuity proceed directly to
Section~\ref{sec-summ-theor-ori-anal}, where a condensed overview is
given of the derived orientation selectivity results.

Readers, who additionally is interested in getting just a brief overview of how the
theoretical analysis is carried out, and the assumptions regarding the
probing of the receptive fields that it rests upon, would be
recommended to additionally read at least
one of the theoretical modelling cases, where the purely spatial analysis in the
following Section~\ref{sec-pure-spat-anal} then constitutes the simplest case.

\begin{figure*}[hbtp]
  \begin{center}
    \begin{tabular}{cccc}
      & {\em\footnotesize First-order simple cell\/}
      &       {\em\footnotesize Second-order simple cell\/}
      &       {\em\footnotesize Complex cell\/} \\
      {\footnotesize $\kappa = 1$}
      & \includegraphics[width=0.29\textwidth]{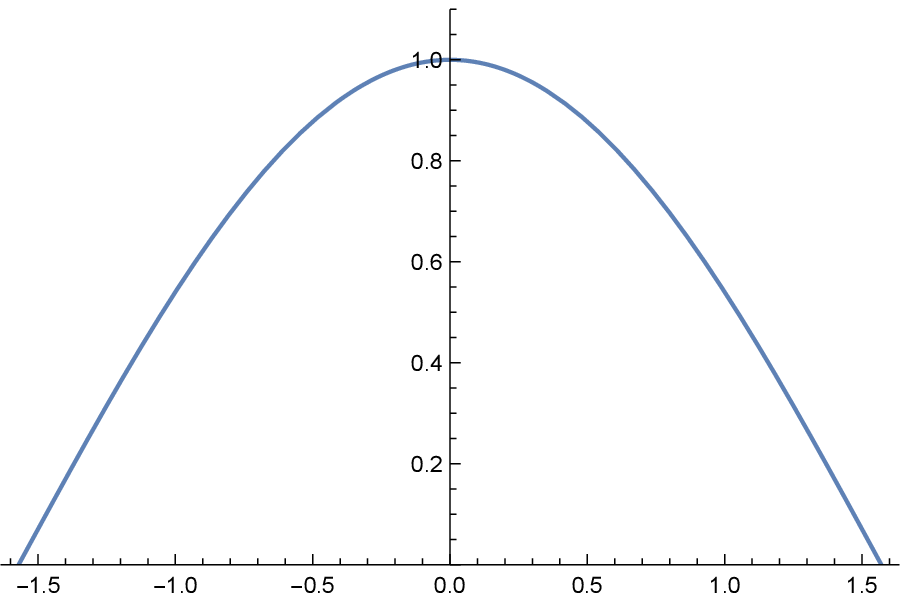}
      & \includegraphics[width=0.29\textwidth]{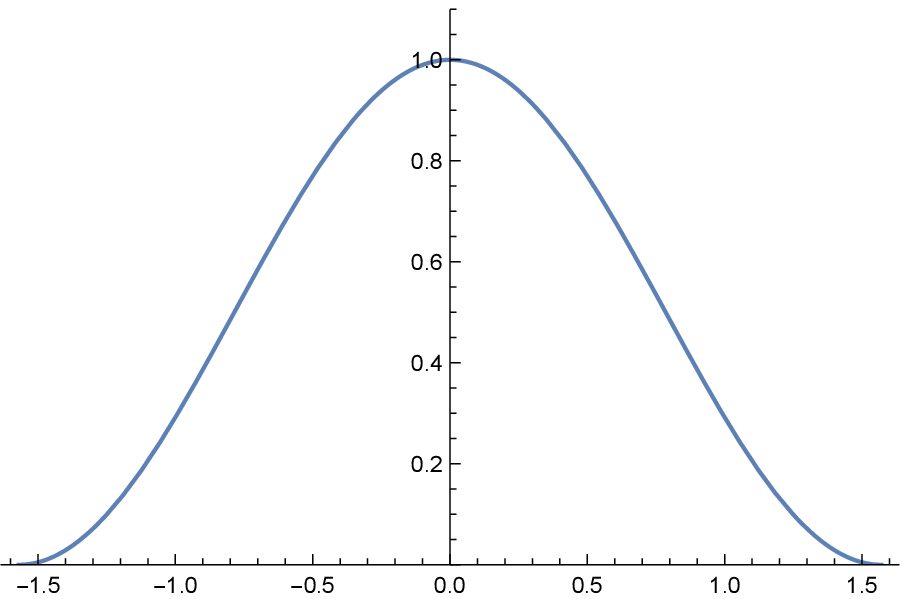}
      & \includegraphics[width=0.29\textwidth]{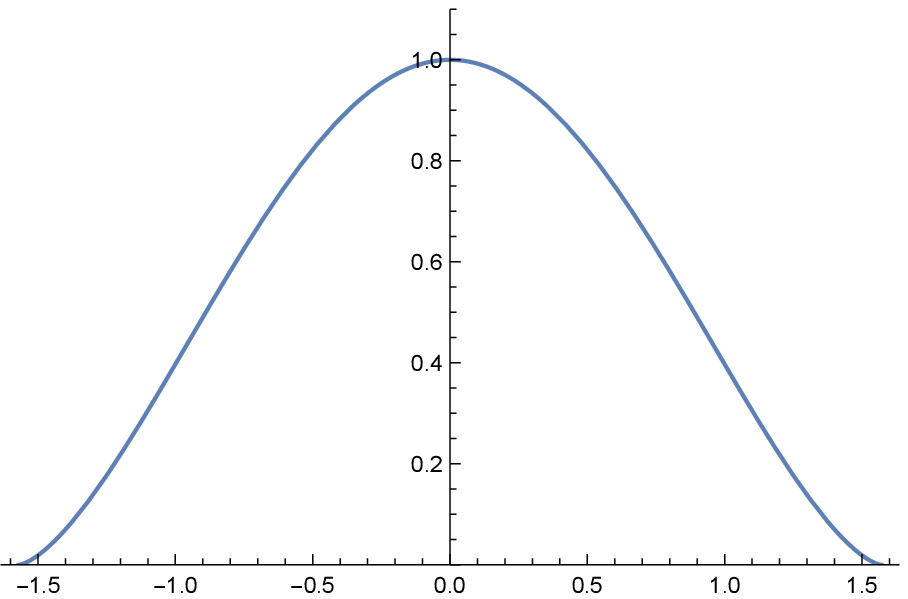} \\    
      {\footnotesize $\kappa = 2$}
      & \includegraphics[width=0.29\textwidth]{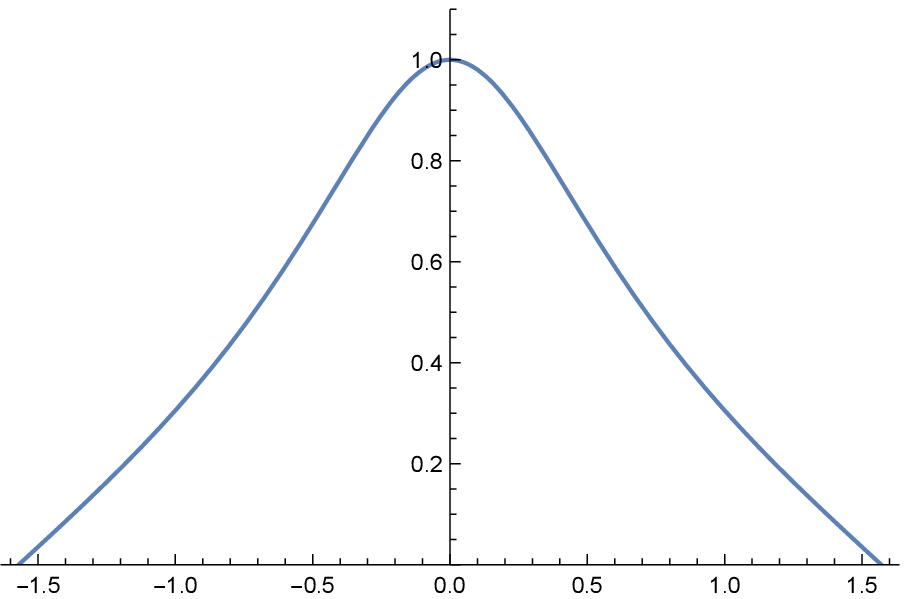}
      & \includegraphics[width=0.29\textwidth]{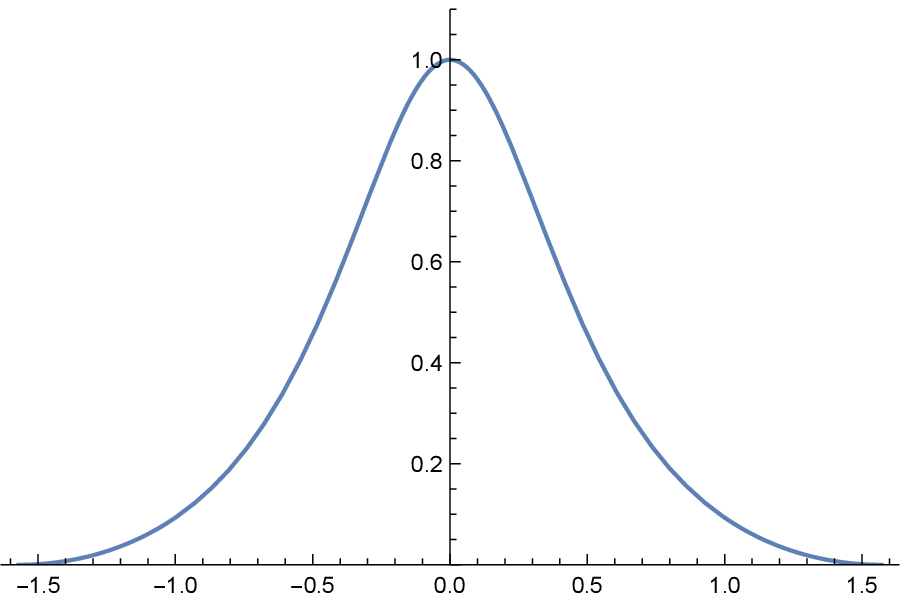}
      & \includegraphics[width=0.29\textwidth]{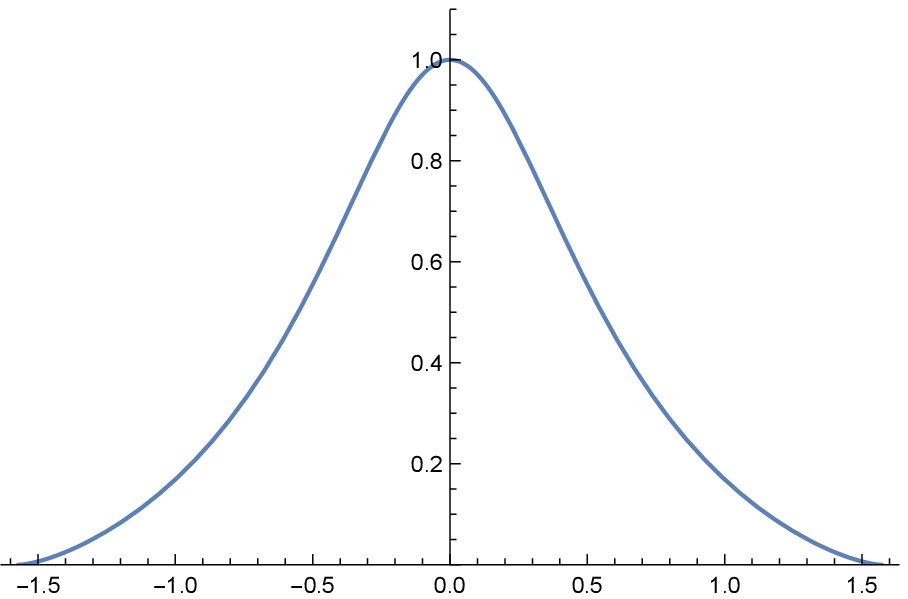} \\    
      {\footnotesize $\kappa = 4$}
      & \includegraphics[width=0.29\textwidth]{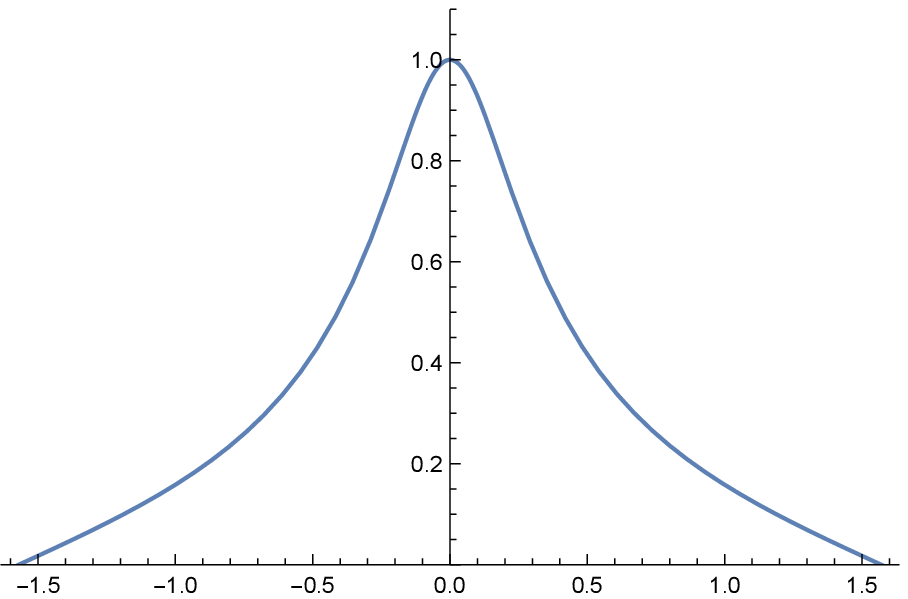}
      & \includegraphics[width=0.29\textwidth]{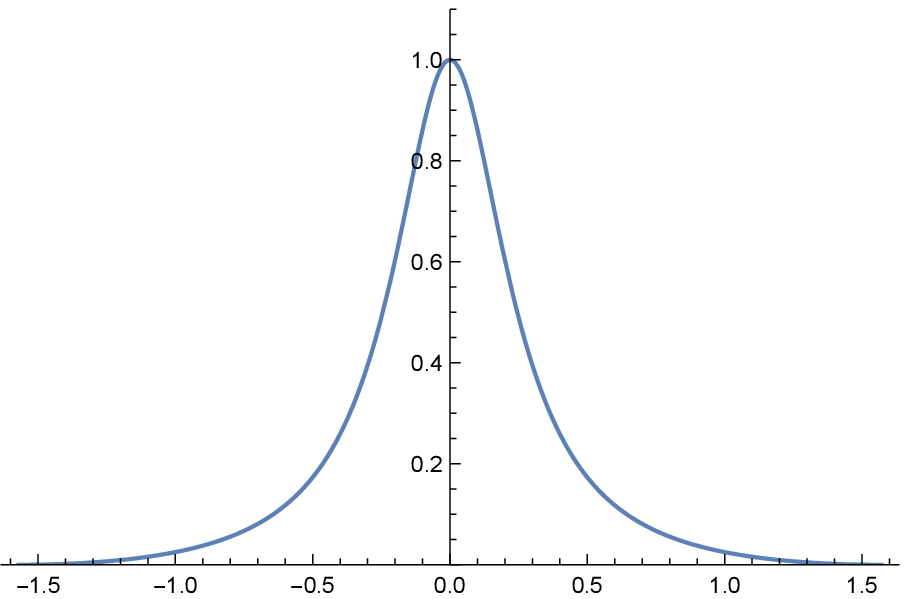}
      & \includegraphics[width=0.29\textwidth]{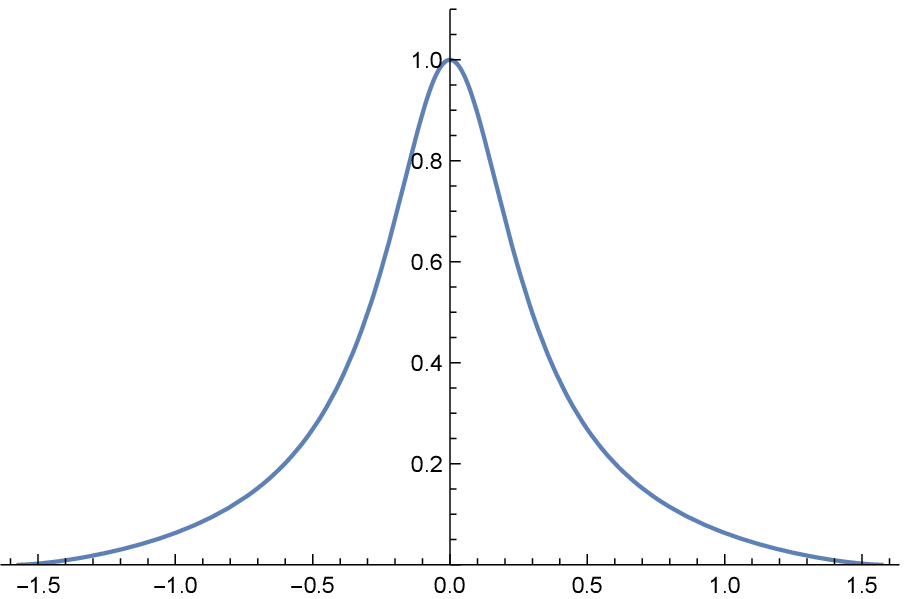} \\    
      {\footnotesize $\kappa = 8$}
      & \includegraphics[width=0.29\textwidth]{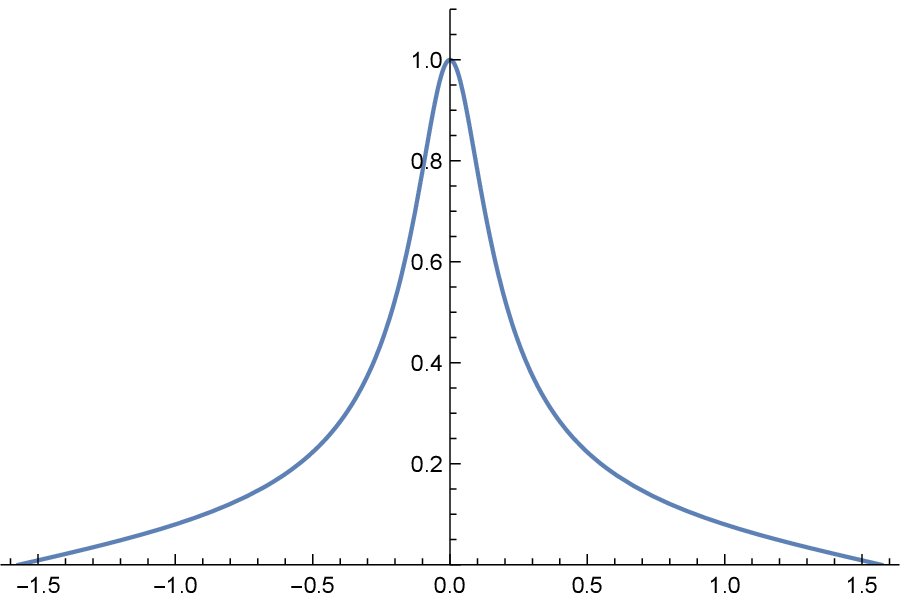}
      & \includegraphics[width=0.29\textwidth]{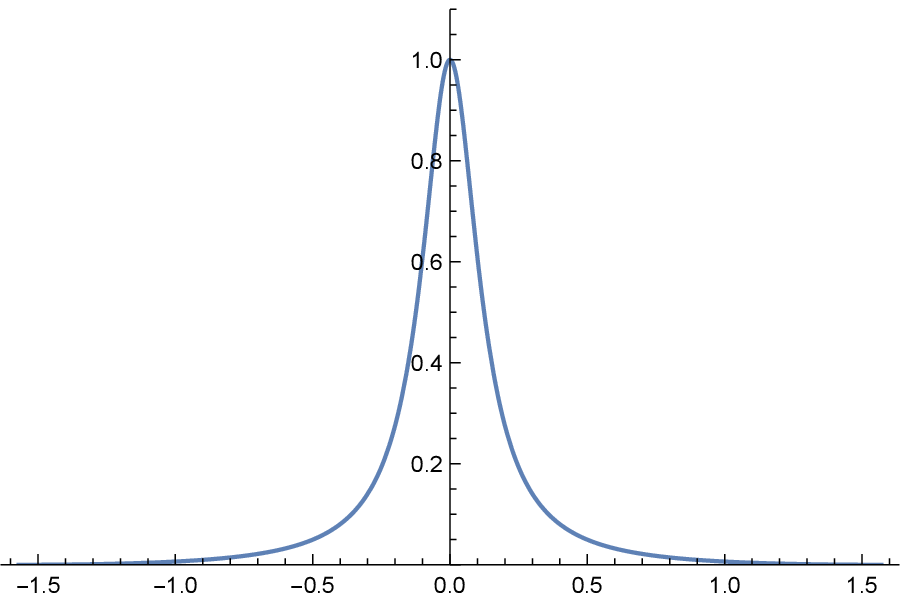}
      & \includegraphics[width=0.29\textwidth]{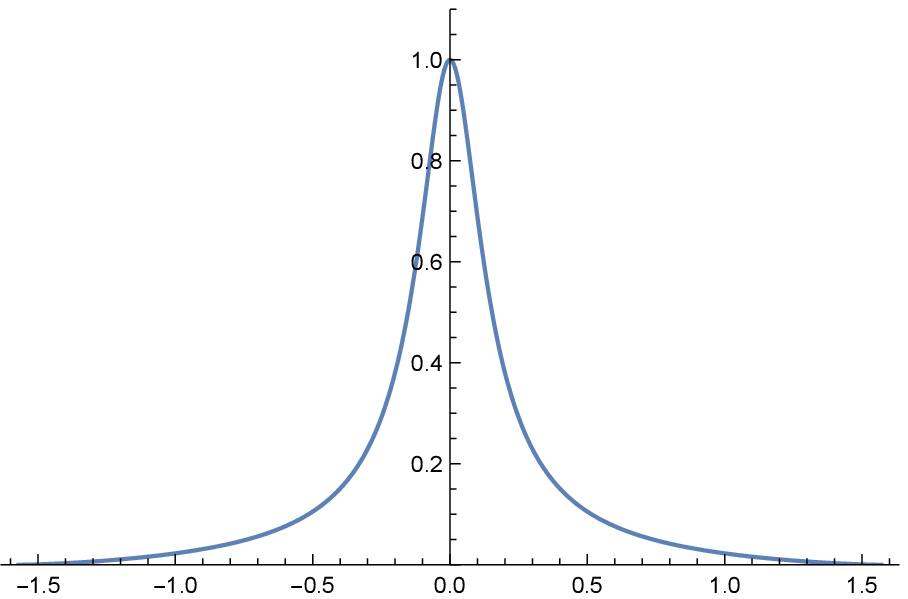} \\    
     \end{tabular}
  \end{center}
  \caption{Graphs of the orientation selectivity for {\em purely spatial
    models\/} of (left column) simple cells in terms of first-order
    directional derivatives of affine Gaussian kernels, (middle column) simple
    cells in terms of second-order directional derivatives of affine Gaussian
    kernels and (right column) complex cells in terms of directional
    quasi-quadrature measures that combine the first- and second-order
    simple cell responses in a Euclidean way for $C = 1/\sqrt{2}$,
    and shown for different values of the ratio
    $\kappa$ between the spatial scale parameters in the vertical
    {\em vs.\/}\ the horizontal directions. Observe how the degree of orientation
    selectivity varies strongly depending on the eccentricity
    $\epsilon = 1/\kappa$ of the receptive fields.
    (top row) Results for $\kappa = 1$.
    (second row) Results for $\kappa = 2$.
    (third row) Results for $\kappa = 4$.
    (bottom row) Results for $\kappa = 8$.
    (Horizontal axes: orientation $\theta \in [-\pi/2, \pi/2]$.
     Vertical axes: Amplitude of the receptive field response relative
     to the maximum response obtained for $\theta = 0$.)}
  \label{fig-ori-sel-spat-anal}
\end{figure*}

\begin{figure*}
 \begin{align}
    \begin{split}
      \label{eq-spat-compl-cell}
      {\cal Q}_{0,\spat,\norm} L
      = \sqrt[4]{2} e^{-\frac{1}{\sqrt{2}}}
          \frac{\left| \cos \theta \right|}{\cos ^2\theta + \kappa ^2 \sin ^2\theta}
          \sqrt{\cos ^2\theta + \kappa ^2 \sin ^2\theta
            \cos ^2
            \left(
              \frac{\sqrt[4]{2}}{\sqrt{\cos ^2\theta+\kappa ^2 \sin ^2\theta}}
                  \left(
                    \frac{x_1 \cos \theta}{\sigma_1}
                    + \frac{x_2 \sin \theta}{\sigma_1}
                  \right)
                  + \beta
             \right)}
     \end{split}
  \end{align}
  \caption{The expression for the oriented spatial quasi-quadrature measure
                 ${\cal Q}_{0,\spat,\norm} L$ in the purely spatial
                 model (\ref{eq-quasi-quad-dir-pure-spat-anal})
                 of a complex cell, when applied to a sine wave pattern
                 of the form (\ref{eq-sine-wave-model-spat-anal}),
                 for $\omega = \omega_{\cal Q}$ according to (\ref{eq-omegaQ-spat}).}
   \label{fig-eq-spat-compl-cell}
\end{figure*}

\subsection{Analysis for purely spatial models of receptive fields}
\label{sec-pure-spat-anal}

For the forthcoming purely spatial analysis, we will analyze the
response properties of our spatial models of simple and complex cells
to sine wave patterns with angular frequency $\omega$
and orientation $\theta$ of the form
\begin{equation}
  \label{eq-sine-wave-model-spat-anal}
  f(x_1, x_2) =
  \sin
  \left(
    \omega \cos (\theta) \, x_1 + \omega \sin (\theta) \, x_2+ \beta
  \right).
\end{equation}

\subsubsection{First-order simple cell}

Consider a simple cell that is modelled as a first-order
scale-normalized derivative of an affine Gaussian kernel
(according to (\ref{eq-spat-RF-model}) for $m = 1$), and oriented in the
horizontal $x_1$-direction (for $\varphi = 0$) with spatial scale parameter $\sigma_1$ in
the horizontal $x_1$-direction and spatial scale parameter $\sigma_2$
in the vertical $x_2$-direction, and thus a spatial covariance matrix
of the form $\Sigma_0 = \diag(\sigma_1^2, \sigma_2^2)$:
\begin{align}
  \begin{split}
    & T_{0,\norm}(x_1, x_2;\; \sigma_1, \sigma_2) =
  \end{split}\nonumber\\
  \begin{split}
     & = \frac{\sigma_1}{2 \pi \sigma_1 \sigma_2} \,
            \partial_{x_1}
              \left( e^{-x_1^2/2\sigma_1^2 - x_2^2/2\sigma_2^2} \right)
  \end{split}\nonumber\\
  \begin{split}
    & = - \frac{x_1}{2 \pi \sigma_1^2 \sigma_2} \,
               e^{-x_1^2/2\sigma_1^2 - x_2^2/2\sigma_2^2}.
  \end{split}
\end{align}
The corresponding receptive field response is then, after solving the
convolution integral%
\footnote{Note, however, that, since the input signals in this analysis
  are throughout sine waves, we could also get the receptive field
  responses to the sine waves by multiplying the amplitudes of the
  sine waves by the absolute valued of the Fourier transforms of the
  receptive field models, while simultaneously shifting the phases of the sine waves
  with the phases of the Fourier transforms of the receptive field models.
  For convenience, we do, however, throughout this treatment solve
  the convolution integrals as explicit integrals in Mathematica,
  since symbolic computations are really helpful in this mathematical
  analysis, when optimizing the
  scale selectivity properties of the receptive fields, by
  differentiating the receptive field responses with respect to the
  angular frequencies of the sine waves, and also to
  handle the more complex non-linear models of complex cells, 
  considered later in this paper.}
in Mathematica,
\begin{align}
   \begin{split}
     L_{0,\norm}(x_1, x_2;\; \sigma_1, \sigma_2) =
  \end{split}\nonumber\\
  \begin{split}
    & = \int_{\xi_1 = -\infty}^{\infty}  \int_{\xi_2 = -\infty}^{\infty}
             T_{0,\norm}(\xi_1, \xi_2;\; \sigma_1, \sigma_2)
  \end{split}\nonumber\\
  \begin{split}
    & \phantom{= = \int_{\xi_1 = -\infty}^{\infty}  \int_{\xi_2 = -\infty}^{\infty}}
             \times f(x_1 - \xi_1, x_2 - \xi_2) \, d \xi_1 \xi_2
  \end{split}\nonumber\\
  \begin{split}
    & = \omega \, \sigma_1 \cos (\theta) \,
           e^{-\frac{1}{2} \omega^2 (\sigma_1^2 \cos^2 \theta + \sigma_2^2 \sin^2 \theta)}
  \end{split}\nonumber\\
  \begin{split}
    \label{eq-L0-pure-spat-anal}
    & \phantom{= =}
           \times \cos
             (
                \omega \cos (\theta) \, x_1 + \omega \sin (\theta) \, x_2+ \beta
             ),
   \end{split}         
\end{align}
{\em i.e.\/},\ a cosine wave with amplitude
\begin{equation}
  \label{eq-A-varphi}
  A_{\varphi}(\theta, \omega;\; \sigma_1, \sigma_2)
  = \omega \, \sigma_1 \left| \cos \theta \right| \,
             e^{-\frac{1}{2} \omega^2 (\sigma_1^2 \cos^2 \theta + \sigma_2^2 \sin^2 \theta)}.
\end{equation}
If we assume that the receptive field is fixed, then the amplitude of the
response will depend strongly on the angular frequency $\omega$ of the
sine wave. The value first increases because of the
factor $\omega$ and then decreases because of the exponential
decrease with $\omega^2$.

If we assume that a biological experiment regarding orientation selectivity is
carried out in such a way that the angular frequency is varied for
each inclination angle $\theta$, and then that the result is for each
value of $\theta$ reported
for the angular frequency $\hat{\omega}$ that leads to the maximum
response, then we can determine this value of $\hat{\omega}$ by
differentiating $A_{\varphi}(\theta, \omega;\; \sigma_1, \sigma_2)$ with
respect to $\omega$ and setting the derivative to zero, which gives:
\begin{equation}
  \label{eq-omega1-spat-prel}
    \hat{\omega}_{\varphi}  = \frac{1}{\sqrt{\sigma_1^2 \cos^2 \theta + \sigma_2^2 \sin^2 \theta}}.
\end{equation}
Inserting this value into $A_{\varphi}(\theta, \omega;\; \sigma_1, \sigma_2)$,
and introducing a scale parameter ratio $\kappa$ such that
\begin{equation}
   \sigma_2 = \kappa \, \sigma_1,
 \end{equation}
 which implies
\begin{equation}
  \label{eq-omega1-spat}
    \hat{\omega}_{\varphi}  = \frac{1}{\sigma_1 \sqrt{\cos^2 \theta + \kappa^2 \sin^2 \theta}},
\end{equation}
then gives the following orientation selectivity measure
\begin{equation}
  \label{eq-ori-sel-simple-1der}
  A_{\varphi,\max}(\theta, \; \kappa)
  = \frac{\left| \cos \theta \right|}{\sqrt{e} \sqrt{\cos ^2 \theta + \kappa ^2 \sin ^2\theta}}.
\end{equation}
Note, specifically, that this amplitude measure is independent of the
spatial scale parameter $\sigma_1$ of the receptive field, which,
in turn, is a consequence of the
scale-invariant nature of differential expressions in terms of
scale-normalized derivatives for scale normalization
parameter $\gamma = 1$.

The left column in Figure~\ref{fig-ori-sel-spat-anal} shows the result of plotting the
measure $A_{\varphi,\max}(\theta;\; \kappa)$  of the orientation selectivity
as function of the inclination angle $\theta$ for a few values of the
scale parameter ratio $\kappa$, with the values rescaled such that the
peak value for each graph is equal to 1. As can be seen from the graphs, the
degree of orientation selectivity increases strongly with the value of
the spatial scale ratio parameter $\kappa$.

\subsubsection{Second-order simple cell}

Consider next a simple cell that can be modelled as a second-order
scale-normalized derivative of an affine Gaussian kernel
(according to (\ref{eq-spat-RF-model}) for $m = 2$), and oriented in the
horizontal $x_1$-direction (for $\varphi = 0$) with spatial scale parameter $\sigma_1$ in
the horizontal $x_1$-direction and spatial scale parameter $\sigma_2$
in the vertical $x_2$-direction, and thus again with a spatial covariance matrix
of the form $\Sigma_0 = \diag(\sigma_1^2, \sigma_2^2)$:
\begin{align}
  \begin{split}
    & T_{00,\norm}(x_1, x_2;\; \sigma_1, \sigma_2) =
  \end{split}\nonumber\\
  \begin{split}
     & = \frac{\sigma_1^2}{2 \pi \sigma_1 \sigma_2} \,
            \partial_{x_1 x_1}
              \left( e^{-x_1^2/2\sigma_1^2 - x_2^2/2 \sigma_2^2} \right)
  \end{split}\nonumber\\
  \begin{split}
    & = \frac{(x_1^2 - \sigma_1^2)}{2 \pi \sigma_1^3 \sigma_2} \,
               e^{-x_1^2/2\sigma_1^2 - x_2^2/2 \sigma_2^2}.
  \end{split}
\end{align}
The corresponding receptive field response is then, again after solving the
convolution integral in Mathematica,
\begin{align}
   \begin{split}
     L_{00,\norm}(x_1, x_2;\; \sigma_1, \sigma_2) =
  \end{split}\nonumber\\
  \begin{split}
    & = \int_{\xi_1 = -\infty}^{\infty}  \int_{\xi_2 = -\infty}^{\infty}
             T_{00,\norm}(\xi_1, \xi_2;\; \sigma_1, \sigma_2)
  \end{split}\nonumber\\
  \begin{split}
    & \phantom{= = \int_{\xi_1 = -\infty}^{\infty}  \int_{\xi_2 = -\infty}^{\infty}}
             \times f(x_1 - \xi_1, x_2 - \xi_2) \, d \xi_1 \xi_2
  \end{split}\nonumber\\
  \begin{split}
    & = - \omega^2 \, \sigma_1^2 \cos^2 (\theta) \,
           e^{-\frac{1}{2} \omega^2 (\sigma_1^2 \cos^2 \theta + \sigma_2^2 \sin^2 \theta)}
  \end{split}\nonumber\\
  \begin{split}
    \label{eq-L00-pure-spat-anal}
    & \phantom{= =}
           \times \sin
             (
                \omega \cos (\theta) \, x_1 + \omega \sin (\theta) \, x_2 + \beta
             ),
   \end{split}         
\end{align}
{\em i.e.\/},\ a sine wave with amplitude
\begin{equation}
  \label{eq-A-varphivarphi}
  A_{\varphi\varphi}(\theta, \omega;\; \sigma_1, \sigma_2)
  = \omega^2 \, \sigma_1^2 \cos^2 (\theta) \,
      e^{-\frac{1}{2} \omega^2 (\sigma_1^2 \cos^2 \theta + \sigma_2^2 \sin^2 \theta)}.
\end{equation}
Again, also this expression first increases and then increases with
the angular frequency $\omega$.
Selecting again the value of $\hat{\omega}$ at which the amplitude of
the receptive
field response assumes its maximum over $\omega$ gives
\begin{equation}
  \label{eq-omega2-spat}  
  \hat{\omega}_{\varphi\varphi}
  = \frac{\sqrt{2}}{\sigma_1 \sqrt{\cos^2 \theta + \kappa^2 \sin^2 \theta}},
\end{equation}
and implies that the maximum amplitude over spatial scales as
function of the inclination angle $\theta$ and the scale parameter
ratio $\kappa$ can be written
\begin{equation}
  \label{eq-ori-sel-simple-2der}
  A_{\varphi\varphi,\max}(\theta;\; \kappa)
  = \frac{2 \cos^2 \theta}
             {e \left( \cos ^2 \theta + \kappa ^2 \sin ^2\theta \right)}. 
\end{equation}
Again, this amplitude measure is also independent of the spatial scale
parameter $\sigma_1$ of the receptive field, because of the
scale-invariant property of scale-normalized
derivatives, when the scale normalization parameter $\gamma$ is chosen
as $\gamma = 1$.

The middle column in Figure~\ref{fig-ori-sel-spat-anal} shows the result of plotting the
measure $A_{\varphi\varphi,\max}(\theta;\; \kappa)$ of the orientation selectivity
as function of the inclination angle $\theta$ for a few values of the
scale parameter ratio $\kappa$, with the values rescaled such that the
peak value for each graph is equal to 1. Again, the
degree of orientation selectivity increases strongly with the value of
$\kappa$, as for the first-order model of a simple cell.

\subsubsection{Complex cell}
\label{sec-pure-spat-compl-cell-def}
  
To model the spatial response of a complex cell
according to the spatial quasi-quadrature measure
(\ref{eq-quasi-quad-dir}), we combine the responses of the first- and
second-order simple cells for $\Gamma = 0$:
\begin{equation}
  \label{eq-quasi-quad-dir-pure-spat-anal}
  {\cal Q}_{0,\spat,\norm} L
  = \sqrt{L_{0,\norm}^2 + C_{\varphi} \, L_{00,\norm}^2},
\end{equation}
with $L_{0,\norm}$ according to (\ref{eq-L0-pure-spat-anal}) and
$L_{00,\norm}$ according to (\ref{eq-L00-pure-spat-anal}).
Choosing the angular frequency $\hat{\omega}$ as the geometric average of
the angular frequencies for which the first- and second-order components of this entity
assume their maxima over angular frequencies, respectively,
\begin{equation}
  \label{eq-omegaQ-spat}
  \hat{\omega}_{\cal Q}
  = \sqrt{\hat{\omega}_{\varphi} \, \hat{\omega}_{\varphi\varphi}}
  = \frac{\sqrt[4]{2}}{\sigma_1 \sqrt{\cos^2 \theta + \kappa^2 \sin^2 \theta}},
\end{equation}
with $\hat{\omega}_{\varphi}$ according to (\ref{eq-omega1-spat}) and
$\hat{\omega}_{\varphi\varphi}$ according to (\ref{eq-omega2-spat}).
Again letting $\sigma_1 = \kappa \, \sigma_1$,
and setting%
\footnote{Concerning the choice of the weighting factor $C_{\varphi}$
  between first- and second-order information, it holds that
  $C_{\varphi} = 1/\sqrt{2}$ implies that the spatial quasi-quadrature
  measure will assume a constant value (be phase independent) for a
  sine wave at the scale level that is the geometric average of the
  scale levels at which the scale-normalized amplitudes of the
  first- and the second-order components in the quasi-quadrature measure
  assume their maxima over scale, for the specific choice of
  $\gamma = 1$ and $\Gamma = 0$. We will later see manifestations of
  this property, in that the responses of the different
  quasi-quadrature measures, that we use for modelling
  complex cells, will be phase independent for inclination angle
  $\theta = 0$, for an angular frequency that is the geometric average
  of the angular frequencies for which the first- and second-order
  components in the quasi-quadrature measures will assume their maximum
  amplitude over scales (see Equations~(\ref{eq-spat-compl-cell}),
  (\ref{eq-sep-compl-cell}) and (\ref{eq-vel-adapt-compl-cell})).}
the relative weight between first- and second-order information to
$C_{\varphi} = 1/\sqrt{2}$ according to (Lindeberg \citeyear{Lin18-SIIMS}),
gives the expression according to Equation~(\ref{eq-spat-compl-cell})
in Figure~\ref{fig-eq-spat-compl-cell}.

For inclination angle $\theta = 0$, that measure is spatially
constant, in agreement with previous work on closely related isotropic
purely spatial isotropic quasi-quadrature measures
(Lindeberg \citeyear{Lin18-SIIMS}). Then, the
spatial phase dependency increases with increasing values of the
inclination angle $\theta$.
To select a single representative of those differing representations,
let us choose the geometric average of the extreme values, which then
assumes the form
\begin{equation}
  \label{eq-ori-sel-simple-quasi}
  A_{{\cal Q},\spat}(\theta;\; \kappa)
  = \frac{\sqrt[4]{2} \, \left| \cos \theta \right|^3}
             {\sqrt{e} \left( \cos ^2 \theta + \kappa ^2 \sin ^2\theta \right)^{3/4}}. 
\end{equation}
The right column in Figure~\ref{fig-ori-sel-spat-anal} shows the
result of plotting the measure $A_{{\cal Q},\spat}(\theta;\; \kappa)$ of the orientation
selectivity as function of the inclination angle $\theta$ for a few
values of the scale parameter ratio $\kappa$, with the values rescaled
such that the peak value for each graph is equal to 1. As can be seen from the
graphs, the degree of orientation selectivity increases strongly with
the value of $\kappa$ also for this model of a complex cell, and in a
qualitatively similar way as for the simple cell models.

\subsection{Analysis for space-time separable models of
  spatio-temporal receptive fields}

To simplify the main flow through the paper, the detailed analysis of
the orientation selectivity properties for the space-time separable
models for the receptive fields of simple and complex cells is given in
Appendix~\ref{app-anal-space-time-sep-rf}, with the main results
summarized in Section~\ref{sec-summ-theor-ori-anal}.
Readers, who are mainly interested in understanding the principles
by which the theoretical analysis is carried out, can proceed directly
to Section~\ref{sec-vel-adapt-spat-temp-anal},
without major loss of continuity.

The main conceptual difference with the analysis of the space-time
separable spatio-temporal receptive field models in
Appendix~\ref{app-anal-space-time-sep-rf}, compared to the purely spatial
models in Section~\ref{sec-pure-spat-anal} or the velocity-adapted
spatio-temporal receptive field models in Section~\ref{sec-vel-adapt-spat-temp-anal},
is that the space-time separable spatio-temporal receptive field
models in Appendix~\ref{app-anal-space-time-sep-rf}
do additionally involve derivatives with respect to time.

\begin{figure*}[hbtp]
  \begin{center}
    \begin{tabular}{cccc}
      & {\em\footnotesize First-order first-order simple cell\/}
      &       {\em\footnotesize Second-order second-order simple cell\/}
      &       {\em\footnotesize Complex cell\/} \\
      {\footnotesize $\kappa = 1$}
      & \includegraphics[width=0.29\textwidth]{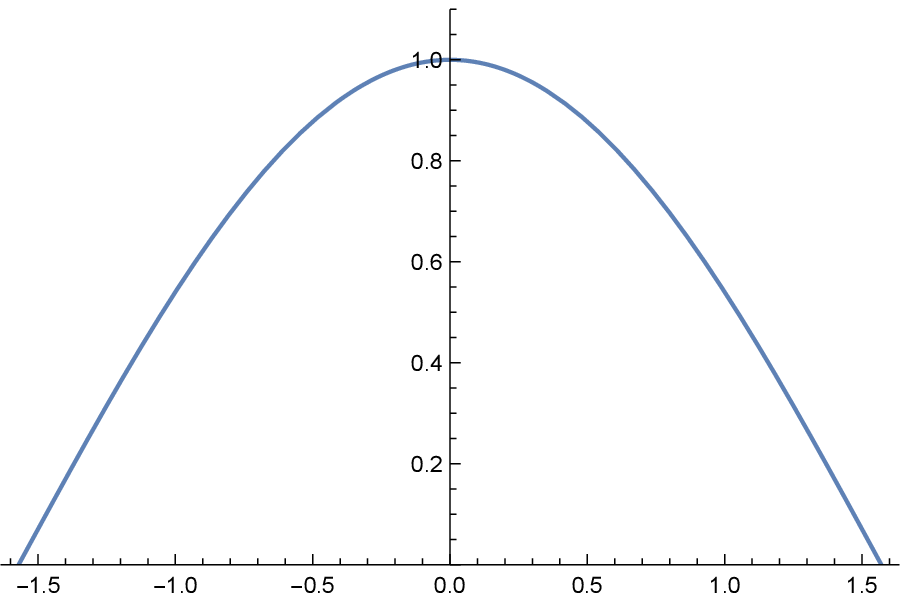}
      & \includegraphics[width=0.29\textwidth]{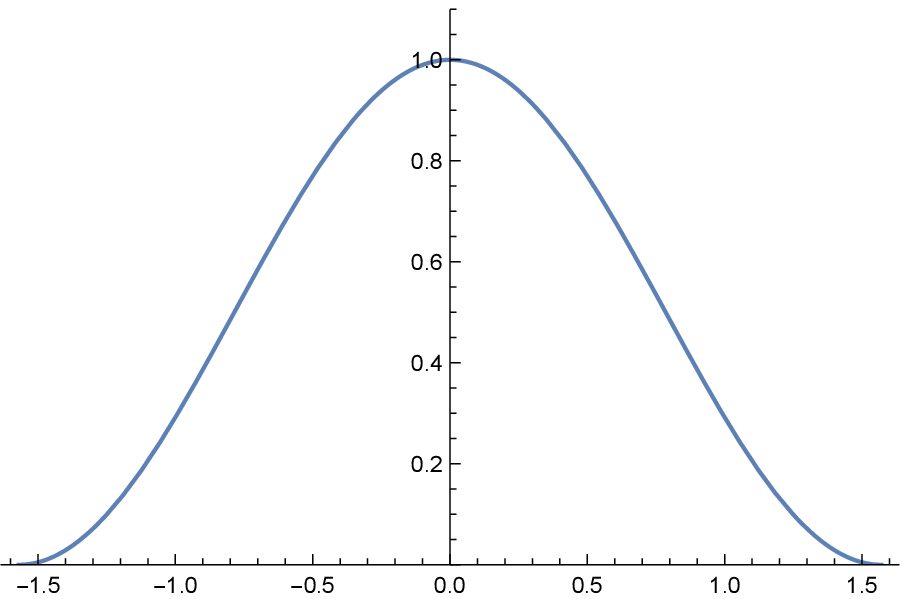}
      & \includegraphics[width=0.29\textwidth]{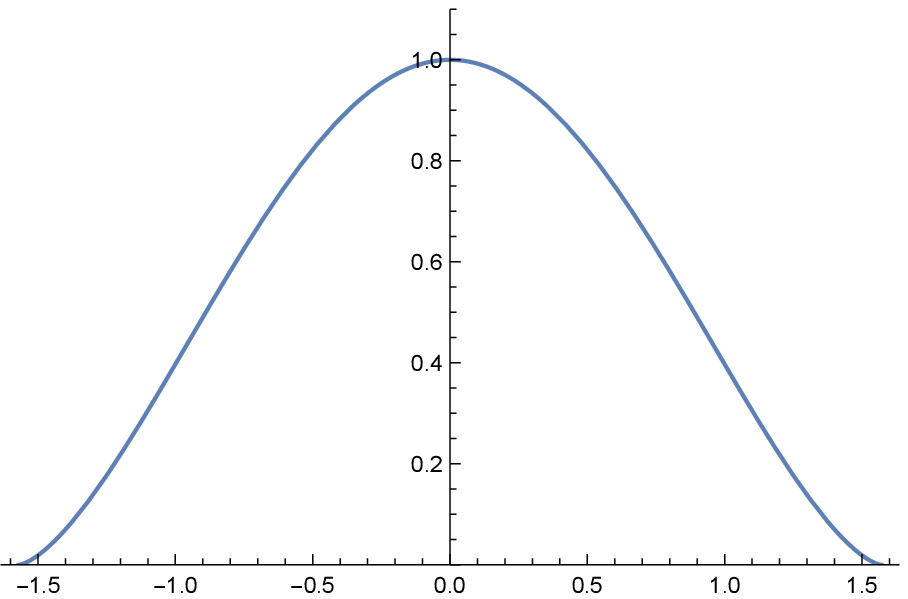} \\    
      {\footnotesize $\kappa = 2$}
      & \includegraphics[width=0.29\textwidth]{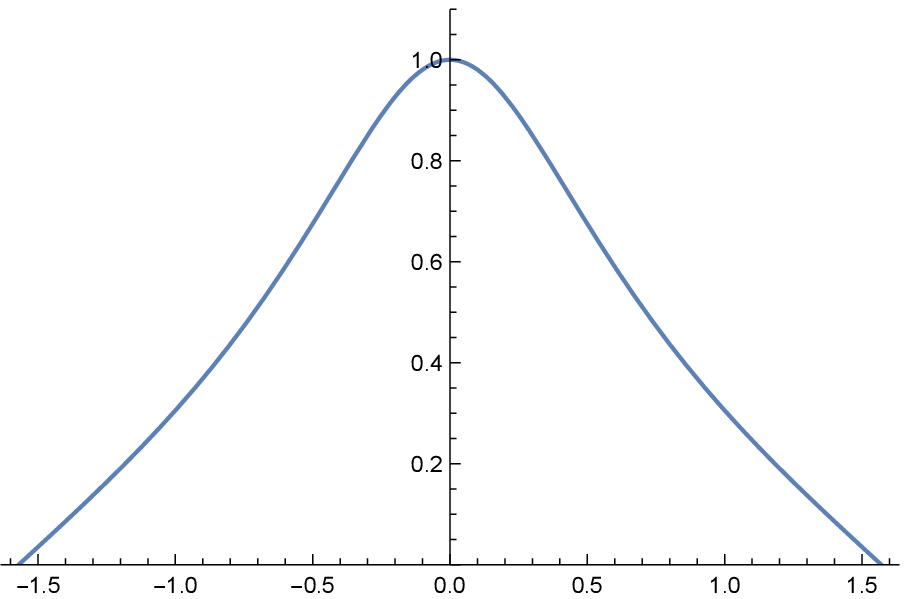}
      & \includegraphics[width=0.29\textwidth]{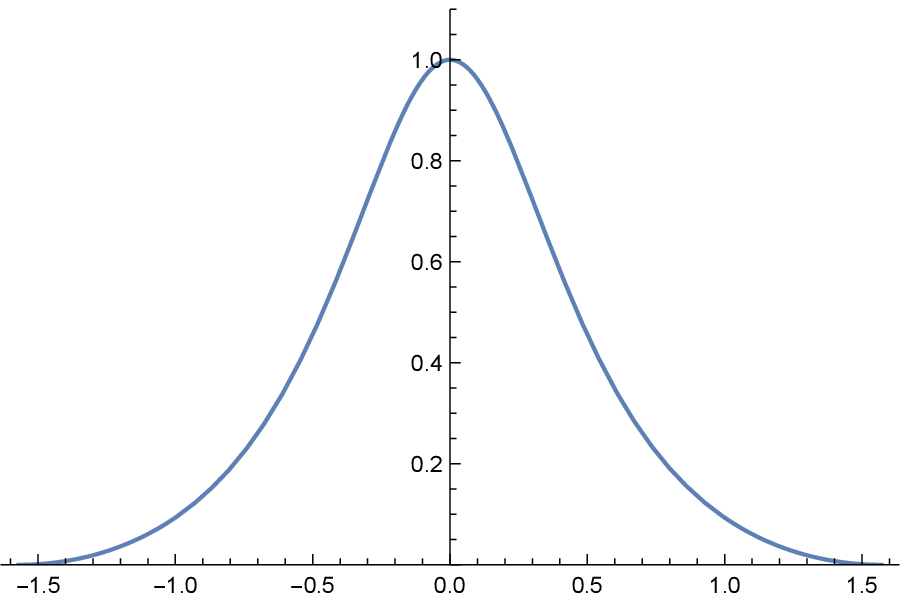}
      & \includegraphics[width=0.29\textwidth]{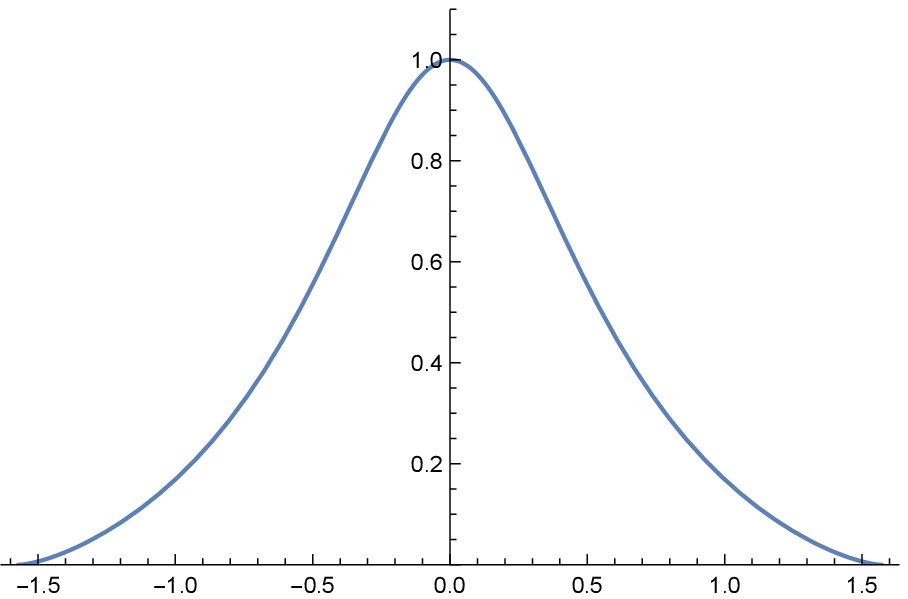} \\    
      {\footnotesize $\kappa = 4$}
      & \includegraphics[width=0.29\textwidth]{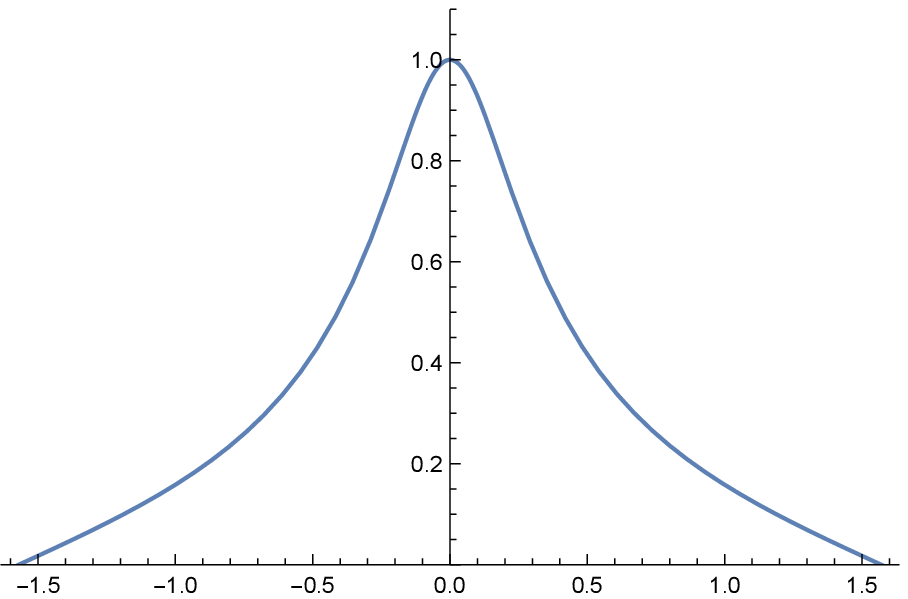}
      & \includegraphics[width=0.29\textwidth]{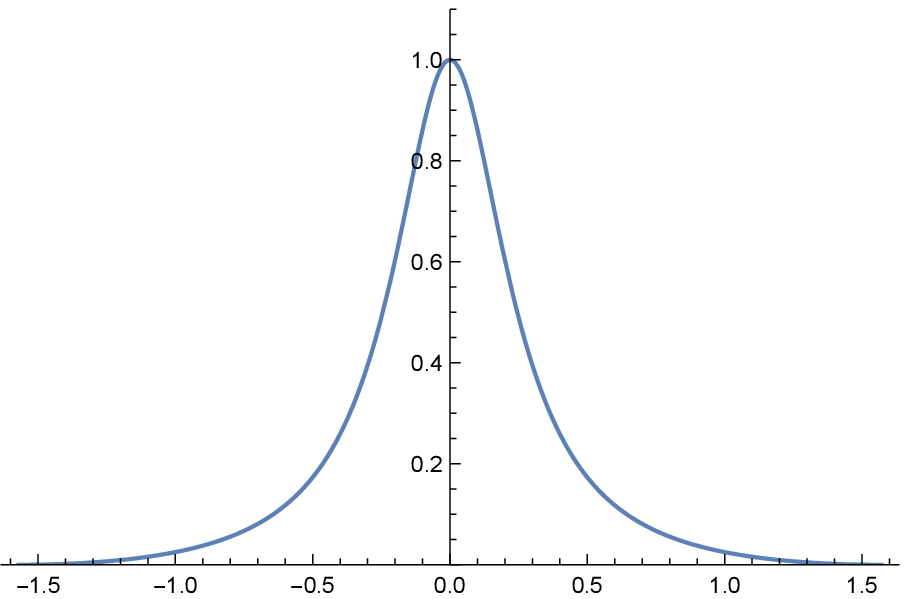}
      & \includegraphics[width=0.29\textwidth]{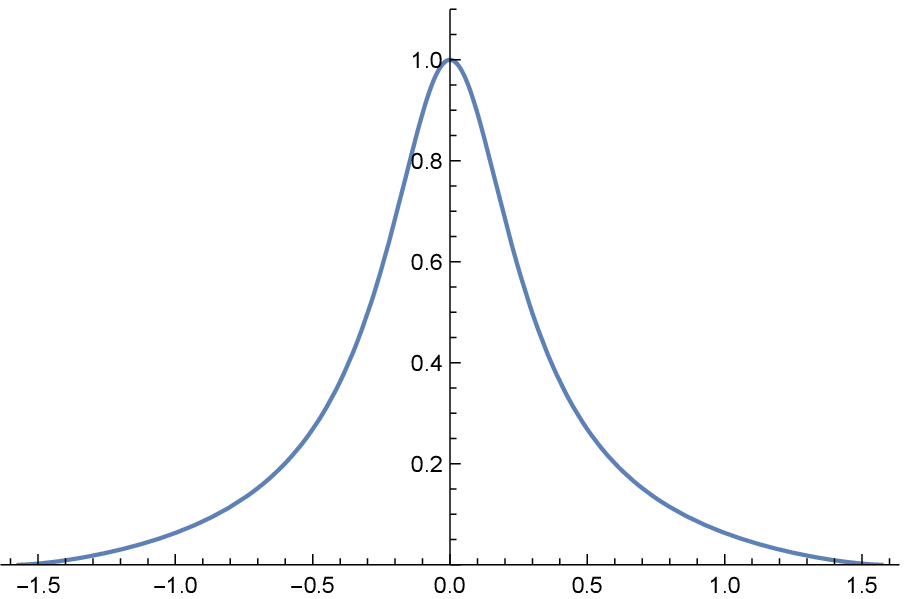} \\    
      {\footnotesize $\kappa = 8$}
      & \includegraphics[width=0.29\textwidth]{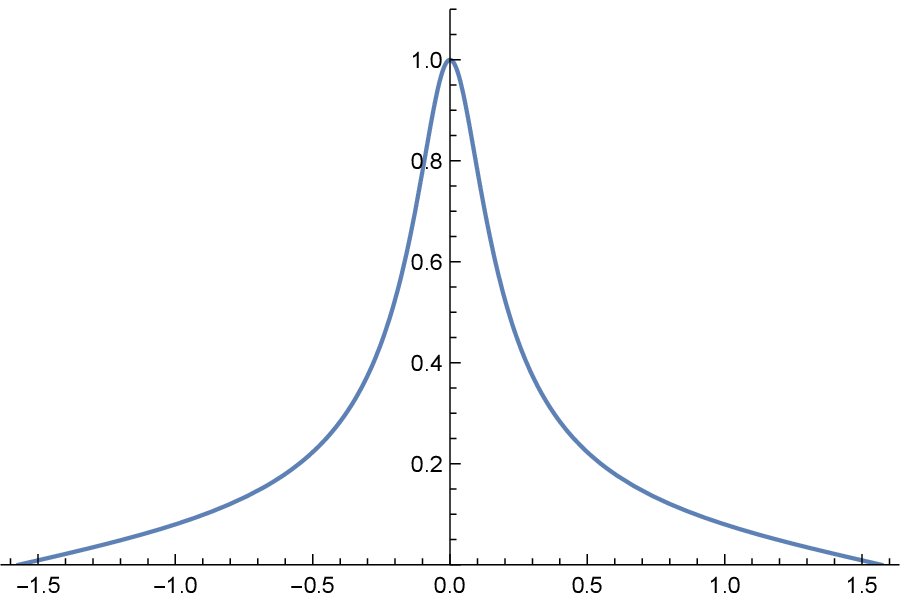}
      &
        \includegraphics[width=0.29\textwidth]{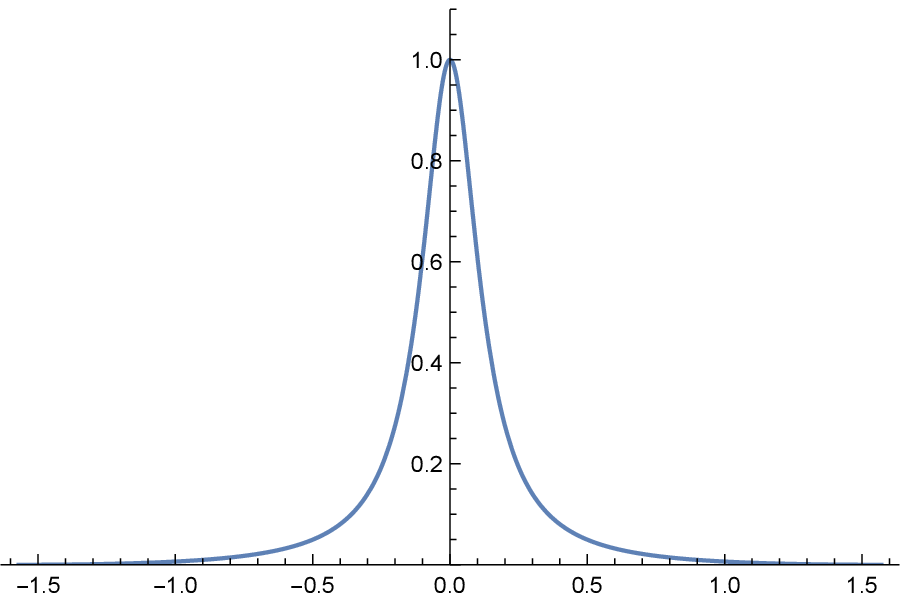}
      & \includegraphics[width=0.29\textwidth]{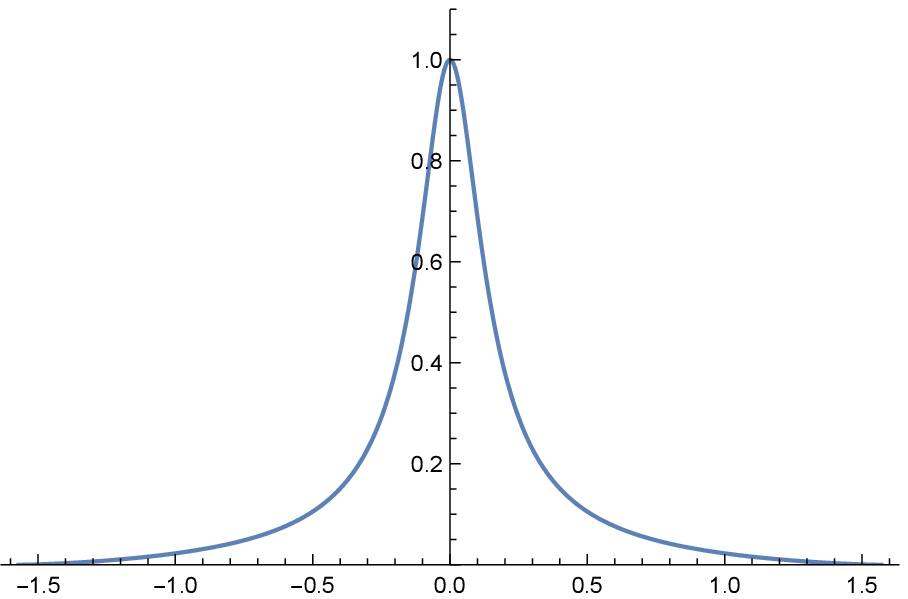} \\    
     \end{tabular}
  \end{center}
  \caption{Graphs of the orientation selectivity for {\em velocity-adapted
    spatio-temporal models\/} of (left column) simple cells in terms of first-order
    directional derivatives of affine Gaussian kernels combined with
    zero-order temporal Gaussian kernels, (middle column) simple
    cells in terms of second-order directional derivatives of affine Gaussian
    kernels combined with zero-order
    temporal Gaussian kernels
    and (right column) complex cells in terms of directional
    quasi-quadrature measures that combine the first- and second-order
    simple cell responses in a Euclidean way for $C_{\varphi} = 1/\sqrt{2}$
    shown for different values of the ratio
    $\kappa$ between the spatial scale parameters in the vertical
    {\em vs.\/}\ the horizontal directions. Observe how the degree of orientation
    selectivity varies strongly depending on the eccentricity
    $\epsilon = 1/\kappa$ of the receptive fields.
    (top row) Results for $\kappa = 1$.
    (second row) Results for $\kappa = 2$.
    (third row) Results for $\kappa = 4$.
    (bottom row) Results for $\kappa = 8$.
    (Horizontal axes: orientation $\theta \in [-\pi/2, \pi/2]$.
     Vertical axes: Amplitude of the receptive field response relative
     to the maximum response obtained for $\theta = 0$.)}
  \label{fig-ori-sel-veladapt-anal}
\end{figure*}

\begin{figure*}
 \begin{align}
    \begin{split}
      \label{eq-vel-adapt-compl-cell}
      {\cal Q}_{0,\vel,\norm} L
      = \sqrt[4]{2} \, e^{-\frac{1}{\sqrt{2}}}
          \frac{\left| \cos \theta \right|}
                  {\cos^2\theta + \kappa^2 \sin^2\theta}
          \sqrt{\cos^2 \theta+\kappa ^2 \sin^2\theta \,
                     \cos ^2
                     \left(
                        \frac{2^{3/4} (\cos (\theta) \, (x_1 - v t) + \sin (\theta) \, x_2)}
                                {\sigma_1 \sqrt{1 + \kappa^2-\left(\kappa ^2-1\right) \cos 2 \theta}}
                   + \beta
                 \right)}
     \end{split}
  \end{align}
  \caption{The expression for the oriented spatio-temporal quasi-quadrature measure
                 ${\cal Q}_{0,\vel,\norm} L$ in the velocity-adapted spatio-temporal
                 model (\ref{eq-quasi-quad-dir-vel-adapt-spat-temp})
                 of a complex cell, when applied to a sine wave pattern
                 of the form (\ref{eq-sine-wave-model-vel-adapt-anal}),
                 for $\omega = \omega_{\cal Q}$ according to
                 (\ref{eq-omegaQ-vel-adapt}) and $u = u_{\cal Q}$
                 according to (\ref{eq-uQ-vel-adapt}).}
   \label{fig-eq-vel-adapt-compl-cell}
\end{figure*}

\subsection{Analysis for velocity-adapted spatio-temporal models of
  receptive fields}
\label{sec-vel-adapt-spat-temp-anal}

Let us next analyze the response
properties of non-separable velocity-adapted spatio-temporal receptive fields to a moving
sine wave of the form 
\begin{multline}
   \label{eq-sine-wave-model-vel-adapt-anal}
  f(x_1, x_2, t) = \\ 
  = \sin
     \left(
       \omega \cos (\theta) \, x_1  + \omega \sin (\theta) \, x_2 -  u \, t + \beta
     \right).
\end{multline}
Based on the observation that the response properties of temporal
derivatives will be zero, if the (scalar) velocity $v$ of the spatio-temporal
receptive field model is adapted to the (scalar) velocity $u$ of the moving sine
wave, we will study the case when the temporal order of
differentiation $n$ is zero.

\subsubsection{First-order simple cell}

Consider a velocity-adapted receptive field
corresponding to a {\em first-order\/} scale-normalized
Gaussian derivative with scale parameter $\sigma_1$ and velocity $v$ in the horizontal
$x_1$-direction, a zero-order Gaussian kernel with scale parameter
$\sigma_2$ in the vertical $x_2$-direction, and a zero-order
Gaussian derivative with scale parameter $\sigma_t$
in the temporal direction, corresponding to $\varphi = 0$, $v = 0$,
$\Sigma_0 = \diag(\sigma_1^2, \sigma_2^2)$,
$m = 1$ and $n = 0$ in (\ref{eq-spat-temp-RF-model-der-norm-caus}):
\begin{align}
  \begin{split}
    & T_{0,\norm}(x_1, x_2, t;\; \sigma_1, \sigma_2, \sigma_t) =
  \end{split}\nonumber\\
  \begin{split}
     & = \frac{\sigma_1}{(2 \pi)^{3/2} \, \sigma_1 \sigma_2 \sigma_t} \,
            \partial_{x_1} 
            \left. \left(
                e^{-x_1^2/2\sigma_1^2 - x_2^2/2 \sigma_2^2 -  t^2/2\sigma_t^2}
            \right) \right|_{x_1 \rightarrow x_1-v t}
  \end{split}\nonumber\\
  \begin{split}
    & = \frac{(x_1 - v t)}{(2 \pi)^{3/2} \, \sigma_1^2 \sigma_2 \sigma_t} \,
               e^{-(x_1- vt)^2/2\sigma_1^2 - x_2^2/2 \sigma_2^2 - t^2/2\sigma_t^2}.
  \end{split}
\end{align}
The corresponding receptive field response is then, after solving the
convolution integral in Mathematica,
\begin{align}
   \begin{split}
     L_{0,\norm}(x_1, x_2, t;\; \sigma_1, \sigma_2, \sigma_t) =
  \end{split}\nonumber\\
  \begin{split}
    & = \int_{\xi_1 = -\infty}^{\infty}  \int_{\xi_2 = -\infty}^{\infty} \int_{\zeta = -\infty}^{\infty}
             T_{0,\norm}(\xi_1, \xi_2, \zeta;\; \sigma_1, \sigma_2, \sigma_t)
  \end{split}\nonumber\\
  \begin{split}
    & \phantom{= = \int_{\xi_1 = -\infty}^{\infty}  \int_{\xi_2 = -\infty}^{\infty}}
             \times f(x_1 - \xi_1, x_2 - \xi_2, t - \zeta) \, d \xi_1 \xi_2 d\zeta
  \end{split}\nonumber\\
  \begin{split}
    & = \omega  \, \sigma_1 \cos \theta \, 
  \end{split}\nonumber\\
  \begin{split}
   & \phantom{= =}
       \times e^{-\frac{\omega^2}{2} 
           \left( \left(\sigma_1^2+\sigma_t^2 v^2\right) \cos^2 (\theta)
             +\sigma_2^2 \sin ^2 \theta  -2 \sigma_t^2 u v \cos \theta +\sigma_t^2 u^2\right)}
  \end{split}\nonumber\\
  \begin{split}
    \label{eq-L0-vel-adapt-anal}
    & \phantom{= =}
           \times \cos
             \left(
               \cos (\theta) \, x_1 + \sin (\theta) \, x_2 -\omega  \, u \, t  + \beta
             \right),
   \end{split}         
\end{align}
{\em i.e.\/},\ a cosine wave with amplitude
\begin{align}
   \begin{split}
      & A_{\varphi}(\theta, u, \omega;\; \sigma_1, \sigma_2, \sigma_t) = 
   \end{split}\nonumber\\
   \begin{split}
     & = \omega \, \sigma_1 \left| \cos \theta \right| \, 
   \end{split}\nonumber\\
  \begin{split}
    \label{eq-A-varphi-spat-temp-anal}
      & \phantom{= =}         
          \times
          e^{-\frac{\omega^2}{2} 
               \left(\cos^2 \theta  \left(\sigma_1^2+\sigma_t^2 v^2\right)
               +\sigma_2^2 \sin ^2 \theta  -2 \sigma_t^2 u v \cos \theta +\sigma_t^2 u^2\right)}.
  \end{split}
\end{align}
Assume that a biological experiment regarding the response properties of the
receptive field is performed by varying both the angular frequency
$\omega$ and the image velocity $u$ to get the maximum value of the
response over these parameters. Differentiating the amplitude $A_{\varphi}$
with respect to $\omega$ and $u$ and setting these derivatives to
zero then gives
\begin{equation}
  \label{eq-omega1-vel-adapt}  
   \hat{\omega}_{\varphi} = \frac{1}{\sigma_1 \sqrt{\cos^2 \theta + \kappa^2 \sin^2 \theta}},
\end{equation}
\begin{equation}
  \label{eq-u1-vel-adapt}    
    \hat{u}_{\varphi} =  v \cos \theta.
\end{equation}
Inserting these values into $A_{\varphi}(\theta, u, \omega;\; \sigma_1, \sigma_2, \sigma_t)$
then gives the following orientation selectivity measure
\begin{equation}
  A_{\varphi,\max}(\theta, \; \kappa)
  = \frac{\left| \cos \theta \right|}{\sqrt{e} \, \sqrt{\cos ^2 \theta + \kappa ^2 \sin ^2\theta}}.
\end{equation}
The left column in Figure~\ref{fig-ori-sel-veladapt-anal} shows
the result of plotting the measure $A_{\varphi,\max}(\theta;\; \kappa)$ of
the orientation selectivity as function of the inclination angle $\theta$
for a few values of the scale parameter ratio $\kappa$, with the values rescaled
such that the peak value for each graph is equal to 1. As we can see
from the graphs, as for the previous purely spatial models of the
receptive fields, as well as for the previous space-time separable
model of the receptive fields, the degree of orientation selectivity increases
strongly with the value of $\kappa$.

\subsubsection{Second-order simple cell}

Consider next a velocity-adapted receptive field
corresponding to a {\em second-order\/} scale-normalized
Gaussian derivative with scale parameter $\sigma_1$ and velocity $v$ in the horizontal
$x_1$-direction, a zero-order Gaussian kernel with scale parameter
$\sigma_2$ in the vertical $x_2$-direction, and a zero-order
Gaussian derivative with scale parameter $\sigma_t$
in the temporal direction, corresponding to $\varphi = 0$, $v = 0$,
$\Sigma_0 = \diag(\sigma_1^2, \sigma_2^2)$,
$m = 2$ and $n = 0$ in (\ref{eq-spat-temp-RF-model-der-norm-caus}):
\begin{align}
  \begin{split}
    & T_{00,\norm}(x_1, x_2, t;\; \sigma_1, \sigma_2, \sigma_t) =
  \end{split}\nonumber\\
  \begin{split}
     & = \frac{\sigma_1^2}{(2 \pi)^{3/2} \, \sigma_1 \sigma_2 \sigma_t} \,
            \partial_{x_1 x_1} 
            \left. \left(
                e^{-x_1^2/2\sigma_1^2 - x_2^2/2 \sigma_2^2 - t^2/2\sigma_t^2}
            \right) \right|_{x_1 \rightarrow x_1-v t}
  \end{split}\nonumber\\
  \begin{split}
    & = \frac{((x_1 - v t)^2 - \sigma_1^2)}{(2 \pi)^{3/2} \, \sigma_1^3 \sigma_2 \sigma_t} \,
               e^{-(x_1 - v t)^2/2\sigma_1^2 - x_2^2/2 \sigma_2^2 - t^2/2\sigma_t^2}.
  \end{split}
\end{align}
The corresponding receptive field response is then, after solving the
convolution integral in Mathematica,
\begin{align}
   \begin{split}
     L_{00,\norm}(x_1, x_2, t;\; \sigma_1, \sigma_2, \sigma_t) =
  \end{split}\nonumber\\
  \begin{split}
    & = \int_{\xi_1 = -\infty}^{\infty}  \int_{\xi_2 = -\infty}^{\infty} \int_{\zeta = -\infty}^{\infty}
             T_{00,\norm}(\xi_1, \xi_2, \zeta;\; \sigma_1, \sigma_2, \sigma_t)
  \end{split}\nonumber\\
  \begin{split}
    & \phantom{= = \int_{\xi_1 = -\infty}^{\infty}  \int_{\xi_2 = -\infty}^{\infty}}
             \times f(x_1 - \xi_1, x_2 - \xi_2, t - \zeta) \, d \xi_1 \xi_2 d\zeta
  \end{split}\nonumber\\
  \begin{split}
    & = -\omega^2  \sigma_1^2 \cos^2 \theta \, 
 \end{split}\nonumber\\
  \begin{split}
    & \phantom{= =}
    \times
       e^{-\frac{\omega^2}{2} 
           \left( \left(\sigma_1^2+\sigma_t^2 v^2\right) \cos^2 \theta
             +\sigma_2^2 \sin ^2 \theta  -2 \sigma_t^2 u v \cos \theta +\sigma_t^2 u^2\right)}
  \end{split}\nonumber\\
  \begin{split}
    \label{eq-L00-vel-adapt-anal}
    & \phantom{= =}
           \times \cos
             \left(
               \sin (\theta) \, x_1 + \sin (\theta) \, x_2 - \omega \, u \, t  + \beta
             \right),
   \end{split}         
\end{align}
{\em i.e.\/},\ a sine wave with amplitude
\begin{align}
   \begin{split}
      & A_{\varphi\varphi}(\theta, u, \omega;\; \sigma_1, \sigma_2, \sigma_t) = \\
  \end{split}\nonumber\\
  \begin{split}
      & = \omega^2  \sigma_1^2 \cos^2 \theta \, 
   \end{split}\nonumber\\
  \begin{split}
    \label{eq-A-varphivarphi-spat-temp-anal}
    & \phantom{= =}    
        \times
        e^{-\frac{\omega^2}{2} 
             \left(\cos^2 \theta  \left(\sigma_1^2+\sigma_t^2 v^2\right)
              +\sigma_2^2 \sin ^2 \theta  -2 \sigma_t^2 u v \cos \theta +\sigma_t^2 u^2\right)}.
  \end{split}         
\end{align}
Assume that a biological experiment regarding the response properties of the
receptive field is performed by varying both the angular frequency
$\omega$ and the image velocity $u$ to get the maximum value of the
response over these parameters. Differentiating the amplitude $A_{\varphi\varphi}$
with respect to $\omega$ and $u$ and setting these derivatives to
zero then gives
\begin{equation}
  \label{eq-omega2-vel-adapt}
   \hat{\omega}_{\varphi\varphi}
   = \frac{\sqrt{2}}{\sigma_1 \sqrt{\cos^2 \theta + \kappa^2 \sin^2 \theta}},
\end{equation}
\begin{equation}
  \label{eq-u2-vel-adapt}      
    \hat{u}_{\varphi\varphi} =  v \cos \theta.
\end{equation}
Inserting these values into
$A_{\varphi\varphi}(\theta, u, \omega;\; \sigma_1, \sigma_2, \sigma_t)$
then gives the following orientation selectivity measure
\begin{equation}
  A_{\varphi\varphi,\max}(\theta, \; \kappa)
  = \frac{2 \cos^2 \theta}{e \, (\cos ^2 \theta + \kappa ^2 \sin ^2\theta)}.
\end{equation}
The middle column in Figure~\ref{fig-ori-sel-veladapt-anal} shows
the result of plotting the measure $A_{\varphi\varphi,\max}(\theta;\; \kappa)$ of
the orientation selectivity as function of the inclination angle $\theta$
for a few values of the scale parameter ratio $\kappa$, with the values rescaled
such that the peak value for each graph is equal to 1.
Again, the degree of orientation selectivity increases
strongly with the value of $\kappa$.

\subsubsection{Complex cell}

To model the spatial response of a complex cell
according to the spatio-temporal quasi-quadrature measure
(\ref{eq-quasi-quad-dir-vel-adapt-spat-temp}) based on velocity-adapted spatio-temporal
receptive fields, we combine the responses of the first- and
second-order simple cells (for $\Gamma = 0$)
\begin{multline}
  \label{eq-quasi-quad-dir-vel-adapt-spat-temp-anal}
  ({\cal Q}_{0,\vel,\norm} L)
  = \sqrt{\frac{L_{0,\norm}^2 
              + \, C_{\varphi} \, L_{00,\norm}^2}{\sigma_{\varphi}^{2\Gamma}}},
\end{multline}
with $L_{0,\norm}$ according to (\ref{eq-L0-vel-adapt-anal}) and
$L_{00,\norm}$ according to (\ref{eq-L00-vel-adapt-anal}).

Selecting the angular frequency as the geometric average of the angular
frequency values at which the above spatio-temporal simple cell models
assume their maxima over angular frequencies,
as well as using the same value of $u$,
\begin{equation}
   \label{eq-omegaQ-vel-adapt}
   \hat{\omega}_{\cal Q}
   = \sqrt{\hat{\omega}_{\varphi} \, \hat{\omega}_{\varphi\varphi}}
   = \frac{\sqrt[4]{2}}{\sigma_1 \sqrt{\cos^2 \theta + \kappa^2 \sin^2 \theta}},
 \end{equation}
with $\hat{\omega}_{\varphi}$ according to (\ref{eq-omega1-vel-adapt}) and
$\hat{\omega}_{\varphi\varphi}$ according to (\ref{eq-omega2-vel-adapt}),
as well as choosing the image velocity $\hat{u}$ as the same value as for
which the above spatio-temporal simple cell models assume their maxima
over the image velocity ((\ref{eq-u1-vel-adapt}) and (\ref{eq-u2-vel-adapt}))
\begin{equation}
   \label{eq-uQ-vel-adapt}
    \hat{u}_{\cal Q} =  v \cos \theta,
\end{equation}
as well as letting $\sigma_1 = \kappa \, \sigma_1$,
and setting the relative weights between first- and second-order
information to $C_{\varphi} = 1/\sqrt{2}$ and $C_t = 1/\sqrt{2}$
according to (Lindeberg \citeyear{Lin18-SIIMS}), then gives
the expression according to Equation~(\ref{eq-vel-adapt-compl-cell})
in Figure~\ref{fig-eq-vel-adapt-compl-cell}.

For inclination angle $\theta = 0$, that measure is spatially
constant, in agreement with our previous purely spatial analysis, as
well as in agreement with previous work on closely related isotropic spatio-temporal
quasi-quadrature measures (Lindeberg \citeyear{Lin18-SIIMS}).
When the inclination angle increases, the phase dependency of the
quasi-quadrature measure will, however, increase.
To select a single representative of those differing representations,
let us choose the geometric average of the extreme values, which then
assumes the form
\begin{equation}
  A_{{\cal Q},\vel,\max}(\theta;\; \kappa)
  = \frac{\sqrt[4]{2} \, \left|\cos \theta\right|^{3/2} }
             {e^{1/\sqrt{2}} \, (\cos ^2 \theta + \kappa ^2 \sin ^2\theta)^{3/2}}.
\end{equation}
The right column in Figure~\ref{fig-ori-sel-veladapt-anal} shows
the result of plotting the measure $A_{{\cal Q},\vel,\max}(\theta;\; \kappa)$ of
the orientation selectivity as function of the inclination angle $\theta$
for a few values of the scale parameter ratio $\kappa$, with the values rescaled
such that the peak value for each graph is equal to 1. Again, the degree of
orientation selectivity increases strongly with the value of $\kappa$.

\begin{table*}[hbtp]
  \begin{tabular}{cccc}
    \hline
    & Purely spatial model
    & Space-time separable spatio-temporal model
    & Velocity-adapted spatio-temporal model \\
    \hline
    First-order simple cell
       & $\frac{\left| \cos \theta \right|}
                    {\sqrt{\cos ^2 \theta + \kappa ^2 \sin ^2\theta}}$ 
       & $\frac{\left| \cos \theta \right|}
                    {\sqrt{\cos ^2 \theta + \kappa ^2 \sin ^2\theta}}$
       & $\frac{\left| \cos \theta \right|}
                    {\sqrt{\cos ^2 \theta + \kappa ^2 \sin ^2\theta}}$ \\
       Second-order simple cell
       & $\frac{\cos^2 \theta}
                    {\cos ^2 \theta + \kappa ^2 \sin ^2\theta}$ 
       & $\frac{\cos^2 \theta}
                    {\cos ^2 \theta + \kappa ^2 \sin ^2\theta}$ 
       & $\frac{\cos^2 \theta}
                    {\cos ^2 \theta + \kappa ^2 \sin ^2\theta}$ \\
    Complex cell
       & $\frac{\left| \cos \theta \right|^{3/2}}
                      {\left( \cos ^2 \theta + \kappa ^2 \sin ^2\theta \right)^{3/4}}$
        & $\frac{\left| \cos \theta \right| \,
                      \sqrt{2 + \kappa^2 + (2 - \kappa^2) \cos 2 \theta}}
              {\cos ^2\theta + \kappa ^2 \sin^2\theta}$
       & $\frac{\left| \cos \theta \right|^{3/2}}
                    {\left( \cos ^2 \theta + \kappa ^2 \sin ^2\theta \right)^{3/4}}$ \\
  \end{tabular}
  \caption{Summary of the forms of the orientation selectivity functions derived
    from the theoretical models of simple cells and complex cells
    based on the generalized Gaussian derivative model for visual
    receptive fields, in the cases
    of either (i)~purely spatial models, (ii)~space-time separable spatio-temporal
    models and (iii)~velocity-adapted spatio-temporal
    models. Concerning the notation, the term ``first-order simple
    cell'' means a model of a simple cell that corresponds to a
    first-order directional derivative of an affine Gaussian kernel
    over the spatial domain, whereas the term ``second-order simple
    cell'' means a model of a simple cell that corresponds to a
    second-order directional derivative of an affine Gaussian
    kernel over the spatial domain.
    As we can see from the table, the form of the orientation selectivity
    function is similar for all the models of first-order simple cells. The
    form of the orientation selectivity function is also similar for all the models of
    second-order simple cells. For complex cells, the orientation
    selectivity function of the space-time separable model is, however,
    different from the orientation selectivity function of the purely spatial
    model and the velocity-adapted spatio-temporal model, which both
    have similar orientation selectivity functions. Note, in particular, that
    common for all these models is the fact that the degree of
    orientation selectivity increases with the scale parameter ratio
    $\kappa = \sigma_2/\sigma_1$, which is the ratio between the scale
    parameter $\sigma_2$ in the direction $\orth \varphi$ perpendicular
    to the preferred orientation $\varphi$ of the receptive field and the
    scale parameter $\sigma_1$ in the preferred orientation $\varphi$
    of the receptive field. The shapes of the theoretically derived
    orientation selectivity curves do, however, notably differ between the
    classes of (i)~first-order simple cells, (ii)~second-order simple
    cells and (iii)~complex cells.}
    \label{tab-summ-ori-sel-diff-models}
\end{table*}

\subsection{Resulting models for orientation selectivity}
\label{sec-summ-theor-ori-anal}

Table~\ref{tab-summ-ori-sel-diff-models} summarizes the results from
the above theoretical analysis of the orientation selectivity for
our idealized models of simple
cells and complex cells, based on the generalized Gaussian derivative
model for visual receptive fields, in the cases
of either (i)~purely spatial models, (ii)~space-time separable spatio-temporal
models and (iii)~velocity-adapted spatio-temporal
models. The overall methodology that we have used for deriving these results is by
exposing each theoretical receptive field model to either purely spatial
or joint spatio-temporal sine wave patterns, and measuring the
response properties for different inclination angles $\theta$, at the
angular frequency of the sine wave, as well as the image velocity of
the spatio-temporal sine wave, at which these models assume their
maximum response over variations of these probing parameters.

As can be seen from the table, the form of the orientation selectivity curve
is similar for all the models of first-order simple cells, which
correspond to first-order derivatives of affine Gaussian kernels over
the spatial domain.
The form of the orientation selectivity curve is also similar for all the models of
second-order simple cells, which correspond to second-order
derivatives of affine Gaussian kernels over the spatial domain.
For complex cells, the form of the orientation selectivity curve for the space-time
separable model is, however,
different from the form of the orientation selectivity curve for the purely spatial
model and the velocity-adapted spatio-temporal model, which both
have a similar form for their orientation selectivity curves.

Note, in particular, that common for all these models is
the fact that the degree of
orientation selectivity increases with the scale parameter ratio
$\kappa = \sigma_2/\sigma_1$, which is the ratio between the scale
parameter $\sigma_2$ in the direction $\orth \varphi$ perpendicular
to the preferred orientation $\varphi$ of the receptive field and the
scale parameter $\sigma_1$ in the preferred orientation $\varphi$
of the receptive field. In other words, for higher values of $\kappa$,
the form of the orientation selectivity curve is more narrow than the
form of the orientation selectivity curve for a lower value of $\kappa$.
The form of the orientation selectivity curve is also more narrow for
a simple cell that can be modelled as a second-order directional
derivative of an affine Gaussian kernel, than for a simple cell that
can be modelled as a first order derivative of an affine Gaussian
kernel.

In this respect, the theoretical analysis supports the conclusion that the
degree of orientation selectivity of the receptive fields increases with the degree of
anisotropy or elongation of the receptive fields, specifically the fact that highly
anisotropic or elongated affine Gaussian derivative based receptive fields have
higher degree of orientation selectivity than more isotropic affine
Gaussian derivative based receptive fields.

The shapes of the resulting orientation selectivity curves do,
however, notably differ between the classes of (i)~first-order simple cells,
(ii)~second-order simple cells and (iii)~complex cells.
This property is important to take into account, if one aims at
fitting parameterized models of orientation selectivity curves to
neurophysiological measurements of corresponding data.

\begin{figure*}[hbtp]
  \begin{center}
    \begin{tabular}{cccc}
      & {\em\footnotesize $\sigma_1 \nu = 1/2$\/}
      &       {\em\footnotesize $\sigma_1 \nu = 1$\/}
      &       {\em\footnotesize $\sigma_1 \nu = 2$\/} \\
      {\footnotesize $\kappa = 1$}
      & \includegraphics[width=0.29\textwidth]{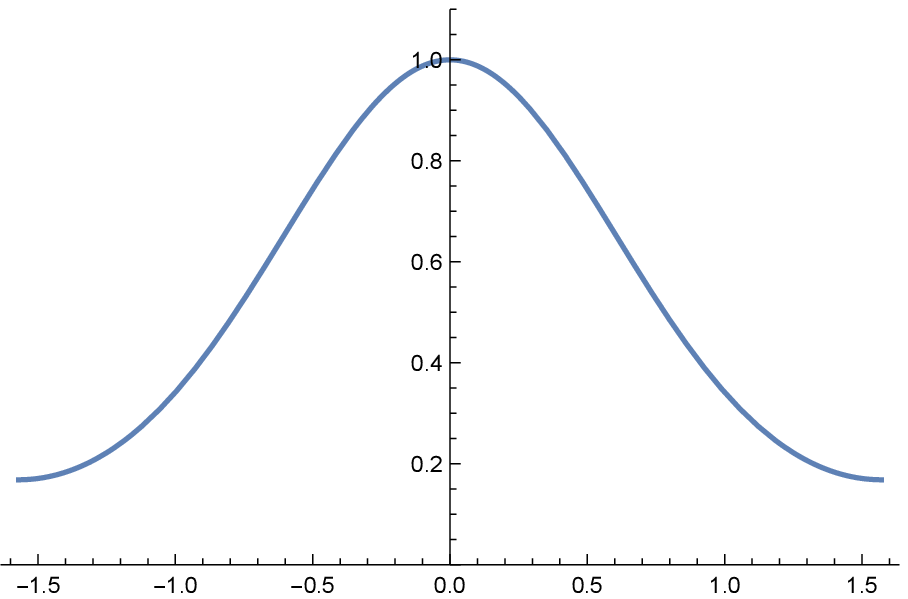}
      & \includegraphics[width=0.29\textwidth]{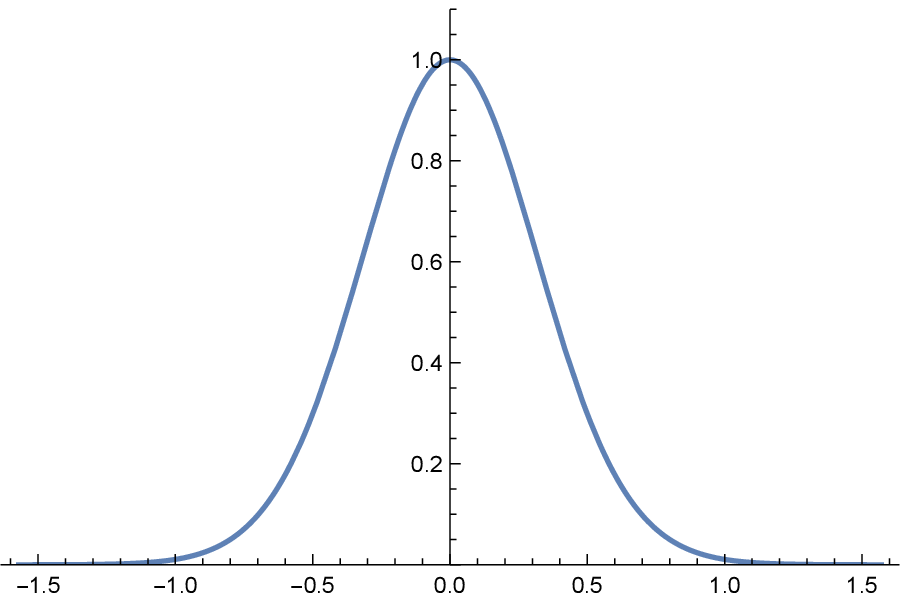}
      & \includegraphics[width=0.29\textwidth]{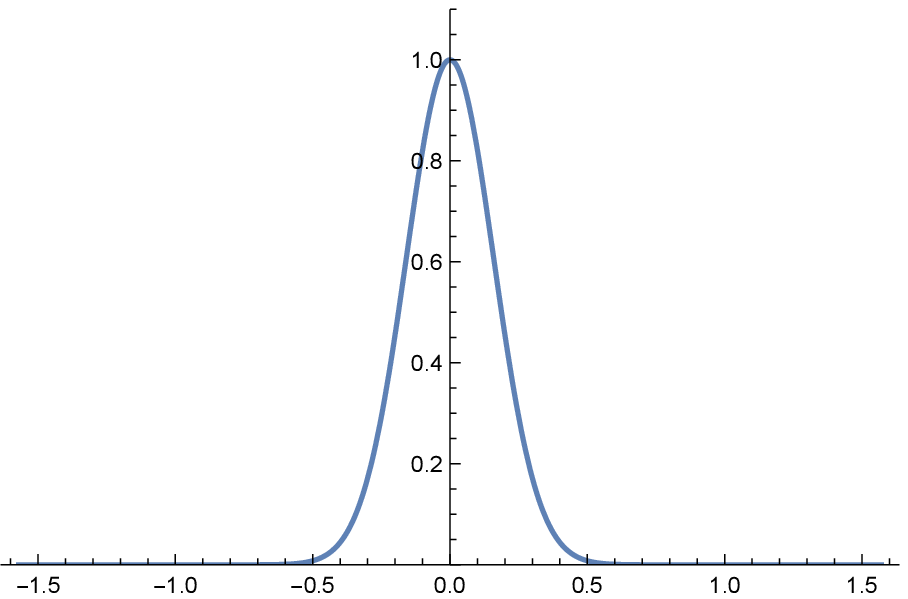} \\    
      {\footnotesize $\kappa = 2$}
      & \includegraphics[width=0.29\textwidth]{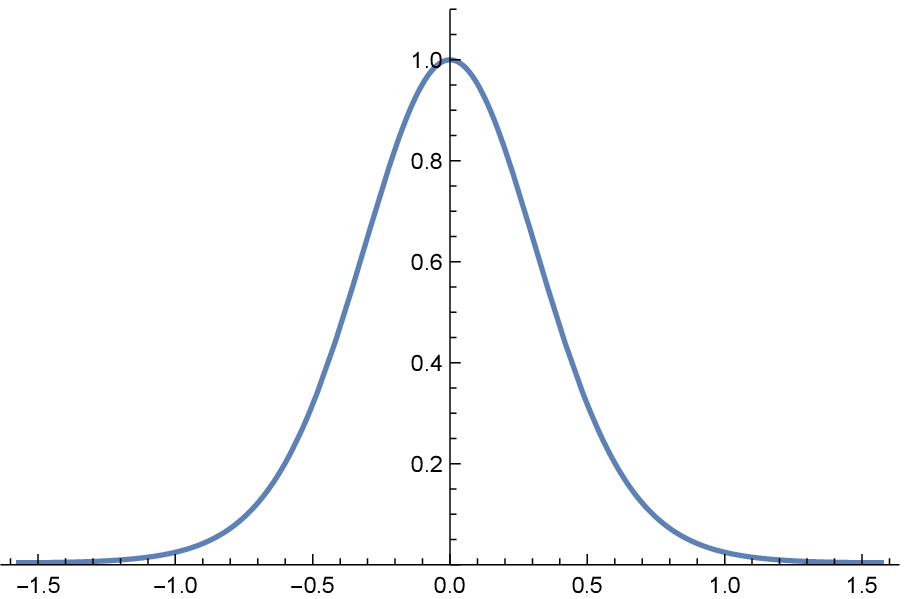}
      & \includegraphics[width=0.29\textwidth]{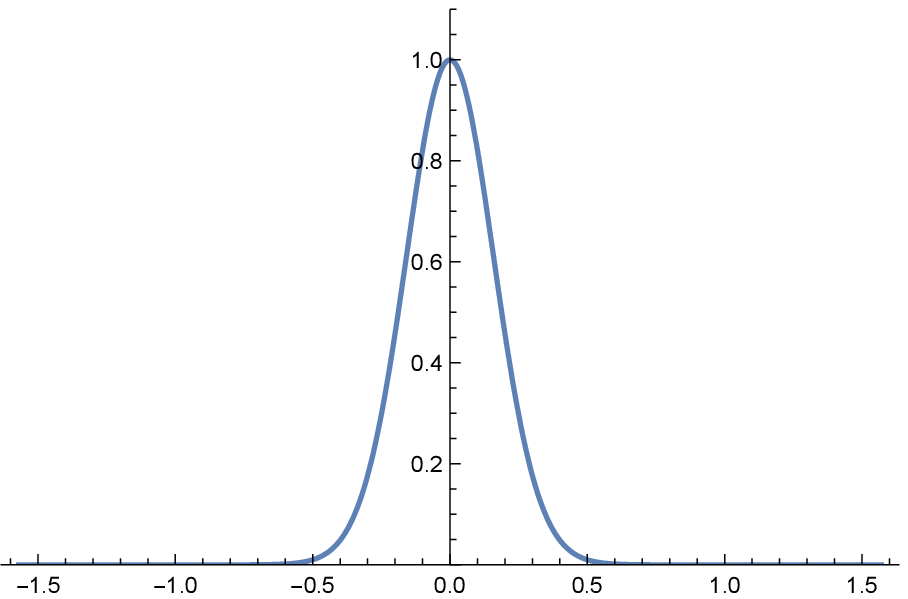}
      & \includegraphics[width=0.29\textwidth]{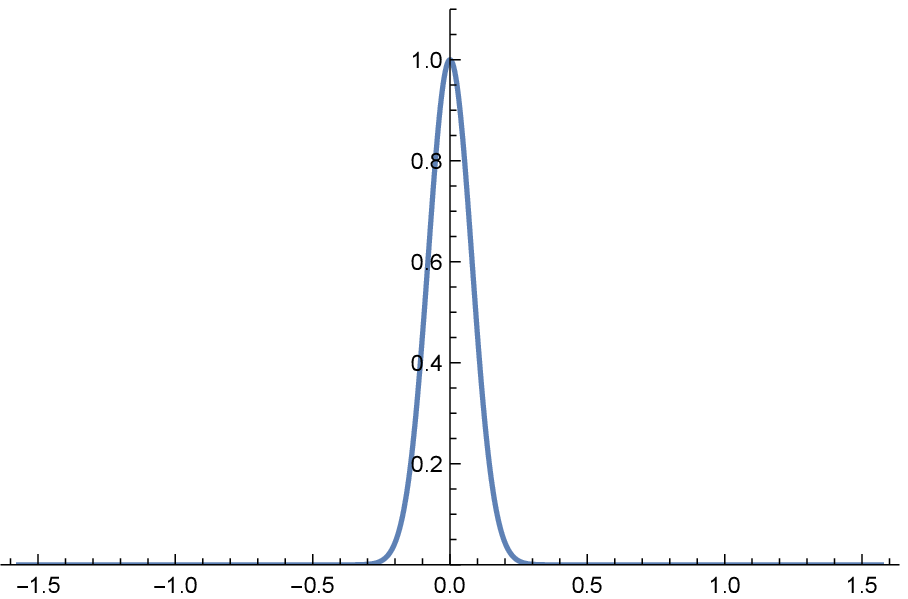} \\    
      {\footnotesize $\kappa = 4$}
      & \includegraphics[width=0.29\textwidth]{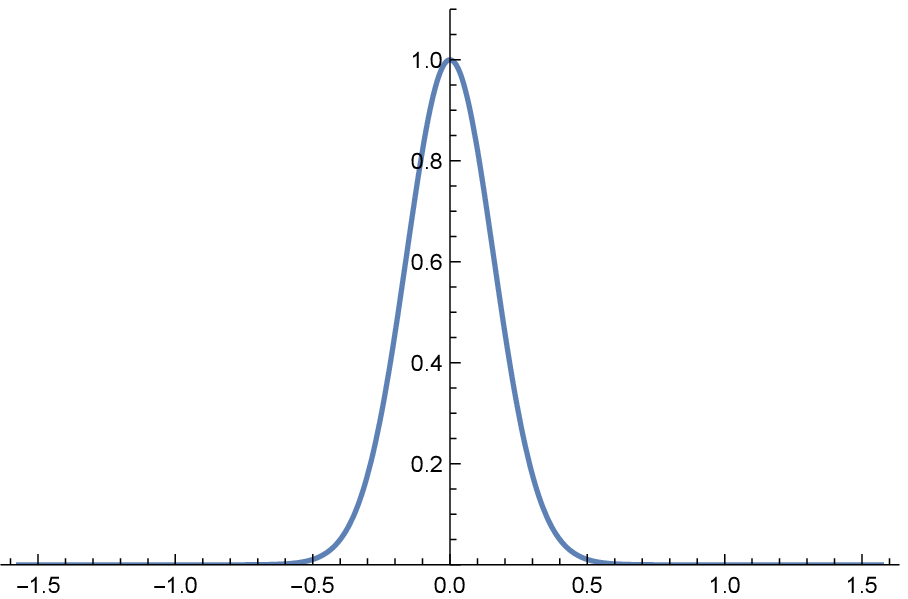}
      & \includegraphics[width=0.29\textwidth]{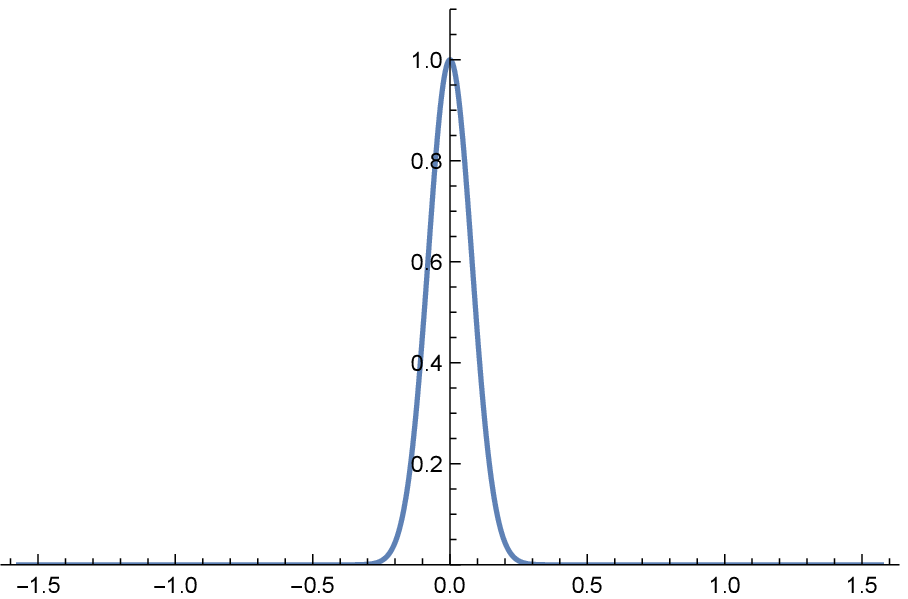}
      & \includegraphics[width=0.29\textwidth]{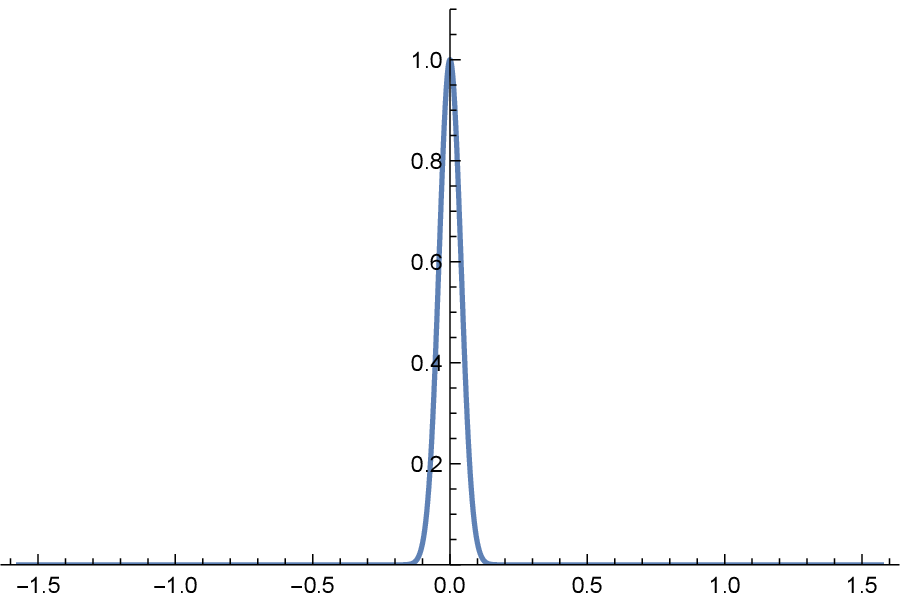} \\    
      {\footnotesize $\kappa = 8$}
      & \includegraphics[width=0.29\textwidth]{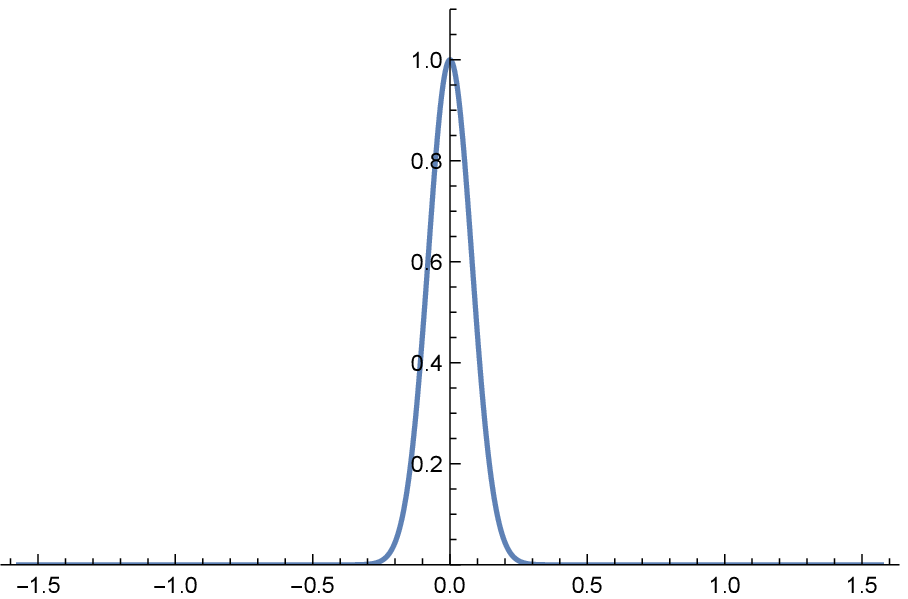}
      & \includegraphics[width=0.29\textwidth]{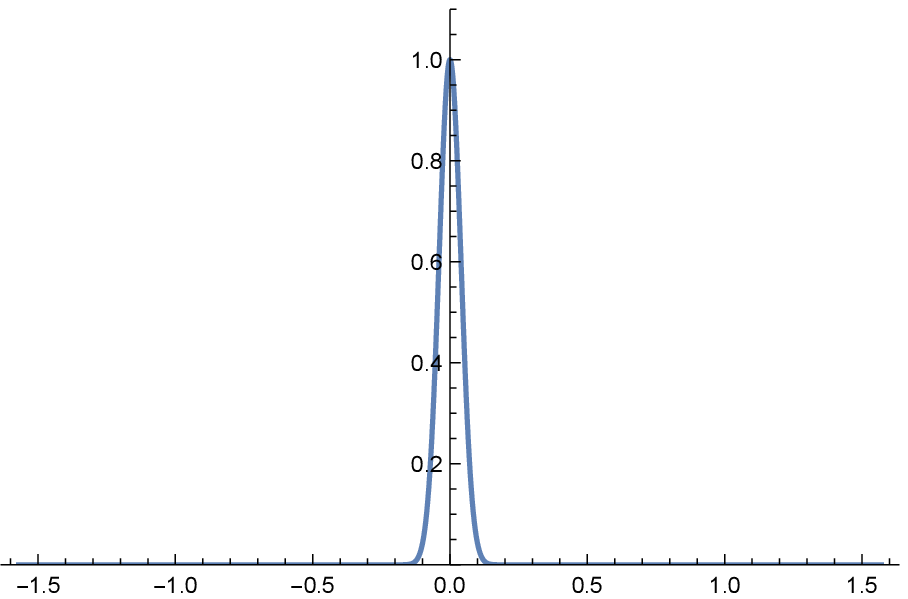}
      & \includegraphics[width=0.29\textwidth]{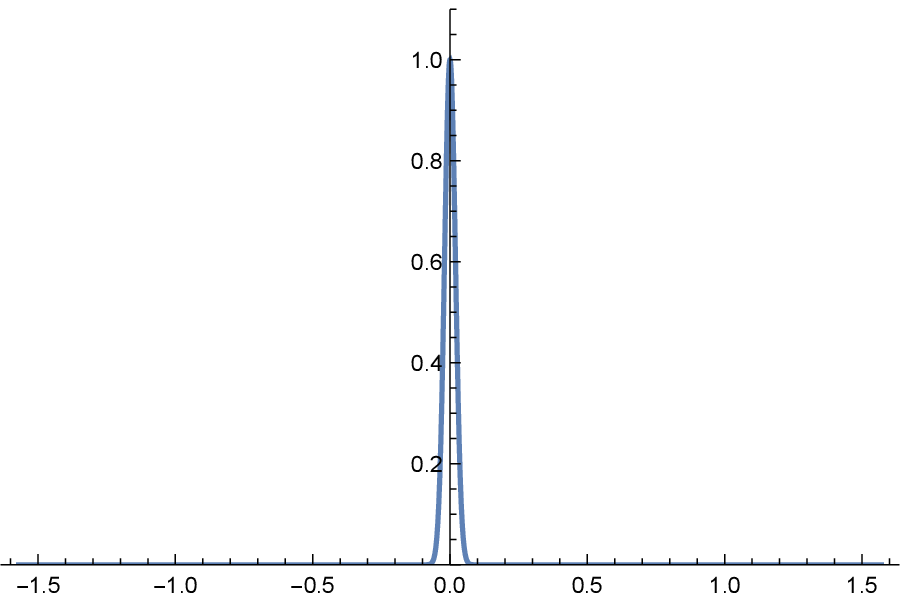} \\    
     \end{tabular}
  \end{center}
  \caption{Graphs of the orientation selectivity of {\em purely
      spatial models\/} of a simple cell, using the {\em even component of
    the affine Gabor model\/}
    according to (\ref{eq-aff-gabor-mod-even}) and
    (\ref{eq-aff-gabor-mod-odd}).
    Observe how the degree of orientation selectivity depends on the
    scale parameter ratio $\kappa$, in a qualitatively similar manner
    as for the affine Gaussian derivative model for purely spatial
    receptive fields. For the affine Gabor model, the degree of
    orientation selectivity does, however, also depend strongly on the
    product of the two remaining parameters $\sigma_1 \, \nu$ in the model.
     (top row) Results for $\kappa = 1$.
    (second row) Results for $\kappa = 2$.
    (third row) Results for $\kappa = 4$.
    (bottom row) Results for $\kappa = 8$.
    (left column) Results for $\sigma_1 \nu = 1/2$.
    (middle column) Results for $\sigma_1 \nu = 1$.
    (right column) Results for $\sigma_1 \nu = 2$.
    (Horizontal axes: orientation $\theta \in [-\pi/2, \pi/2]$.
     Vertical axes: Amplitude of the receptive field response relative
     to the maximum response obtained for $\theta = 0$.)}
  \label{fig-ori-sel-spat-anal-gabor}
\end{figure*}

\section{Orientation selectivity analysis for an affine Gabor model of
  simple cells}
\label{app-ori-sel-anal-gabor}

In this section, we will perform an orientation
selectivity analysis, as analytically similar as possible, for idealized receptive
fields according to an affine Gabor model for purely spatial receptive fields.

Given two spatial scale parameters $\sigma_1$ and $\sigma_2$ and an
angular frequency $\nu$, consider an affine Gabor model with a pair of
purely spatial receptive fields of the form
\begin{align}
  \begin{split}
    \label{eq-aff-gabor-mod-even}
    T_{\even}(x_1, x_2;\; \sigma_1, \sigma_2, \nu)
    & = \frac{1}{2 \pi \sigma_1 \sigma_2} \,
            e^{- x_1^2/2 \sigma_1^2 - x_2^2/2 \sigma_2^2}
            \cos(\nu \, x_1),
    \end{split}\\
  \begin{split}
    \label{eq-aff-gabor-mod-odd}    
    T_{\odd}(x_1, x_2;\; \sigma_1, \sigma_2, \nu)
    & = \frac{1}{2 \pi \sigma_1 \sigma_2} \,
            e^{- x_1^2/2 \sigma_1^2 - x_2^2/2 \sigma_2^2}
            \sin(\nu \, x_1).
    \end{split}
\end{align}
Let us, in analogy with the treatment in
Section~\ref{sec-pure-spat-anal}, subject these receptive fields to
test probes of the form (\ref{eq-sine-wave-model-spat-anal})
\begin{equation}
  \label{eq-sine-wave-model-spat-anal-app}
  f(x_1, x_2) =
  \sin
  \left(
    \omega \cos (\theta) \, x_1 + \omega \sin (\theta) \, x_2+ \beta
  \right)
\end{equation}
for different inclination angles $\theta$. The responses of the even
and odd components of the Gabor pair to such test functions are 
given by
\begin{align}
   \begin{split}
     L_{\even}(x_1, x_2;\; \sigma_1, \sigma_2, \nu) =
   \end{split}\nonumber\\
  \begin{split}
    & = \int_{\xi_1 = -\infty}^{\infty}  \int_{\xi_2 = -\infty}^{\infty}
             T_{\even}(\xi_1, \xi_2;\; \sigma_1, \sigma_2, \nu)
  \end{split}\nonumber\\
  \begin{split}
    & \phantom{= = \int_{\xi_1 = -\infty}^{\infty}  \int_{\xi_2 = -\infty}^{\infty}}
             \times f(x_1 - \xi_1, x_2 - \xi_2) \, d \xi_1 \xi_2,
  \end{split}\nonumber\\
   \begin{split}
     L_{\odd}(x_1, x_2;\; \sigma_1, \sigma_2, \nu) =
   \end{split}\nonumber\\
  \begin{split}
    & = \int_{\xi_1 = -\infty}^{\infty}  \int_{\xi_2 = -\infty}^{\infty}
             T_{\odd}(\xi_1, \xi_2;\; \sigma_1, \sigma_2, \nu)
  \end{split}\nonumber\\
  \begin{split}
    & \phantom{= = \int_{\xi_1 = -\infty}^{\infty}  \int_{\xi_2 = -\infty}^{\infty}}
             \times f(x_1 - \xi_1, x_2 - \xi_2) \, d \xi_1 \xi_2.
     \end{split}
\end{align}
Solving these convolution integrals in Mathematica then gives
\begin{align}
  \begin{split}
     L_{\even}(x_1, x_2;\; \sigma_1, \sigma_2, \nu) =
   \end{split}\nonumber\\
  \begin{split}
    & = \frac{1}{2} \left(e^{2 \nu  \omega  \sigma_1^2 \cos \theta}+1\right) \sin (\beta
    +\omega  \, x_1  \cos \theta+\omega  \, x_2 \sin \theta) \times
  \end{split}\nonumber\\
  \begin{split}
    & \phantom{=} \quad
    e^{\frac{1}{2} \left(-\nu
   ^2 \sigma_1^2-2 \nu  \omega  \sigma_1^2 \cos \theta-\omega ^2
   \sigma_1^2 \cos ^2\theta-\omega ^2 \sigma_2^2 \sin ^2 \theta\right)},
 \end{split}\\
  \begin{split}
     L_{\odd}(x_1, x_2;\; \sigma_1, \sigma_2, \nu) =
   \end{split}\nonumber\\
  \begin{split}
    & = -\frac{1}{2} \left(e^{2 \nu  \omega  \sigma_1^2 \cos \theta}-1\right) \cos (\beta
    +\omega  \, x_1 \cos \theta+\omega \,  x_2 \sin \theta) \times
   \end{split}\nonumber\\
  \begin{split}
    & \phantom{=} \quad
    e^{\frac{1}{2} \left(-\nu
   ^2 \sigma_1^2-2 \nu  \omega  \sigma_1^2 \cos \theta-\omega ^2
   \sigma_1^2 \cos ^2\theta-\omega ^2 \sigma_2^2 \sin ^2 \theta \right)}.
   \end{split}
\end{align}
Unfortunately, it is hard to analytically determine the angular
frequency $\omega$ of the test function that gives the strongest
amplitude for the Gabor response obtained with a given frequency $\nu$.
Therefore, we will in the following simply set $\omega = \nu$,
implying that this analysis for the Gabor model will not be
methodologically identical to the previous analysis for the affine
Gaussian derivative model, in the sense that we will not optimize the
angular frequency $\omega$ of the test probe for each inclination
angle $\theta$.

Letting additionally $\sigma_2 = \kappa \, \sigma_1$ for $\kappa > 1$, we obtain that
the amplitudes of the two Gabor responses can be written:
\begin{align}
  \begin{split}
    A_{\even}(\theta)
    = \frac{1}{2} \left(e^{2 \nu ^2 \sigma_1^2 \cos \theta}+1 \right) 
        e^{- \nu ^2 \sigma_1^2 \cos ^2 \left(\frac{\theta }{2}\right)
        (\kappa^2-1)(1 -  \cos \theta)},
  \end{split}\\
  \begin{split}
    A_{\odd}(\theta)
    = \frac{1}{2} \left( e^{2 \nu ^2 \sigma_1^2 \cos \theta}-1 \right)
        e^{- \nu ^2 \sigma_1^2 \cos ^2 \left(\frac{\theta }{2}\right)
        (\kappa^2-1)(1 -  \cos \theta)}.
  \end{split}
\end{align}
In contrast to the previous analysis for affine Gaussian derivative
models of simple cells, where we determined the angular frequency of
the probing sine wave, to maximize the response given a value of the
spatial scale parameter, and in this way eliminated the dependency of
the orientation selectivity on the size of the spatial scale parameter in the
idealized receptive field model, and in this way reduced the explicit
orientation dependency to only depend on the ratio $\kappa$ between the scale
parameters, we cannot, however, here eliminate the remaining
parameters of the idealized Gabor model from the orientation
selectivity analysis.

The affine Gabor model comprises
three internal parameters for the receptive fields (the spatial scale
parameters $\sigma_1$ and $\sigma_2$, the
angular frequency $\nu$ as well as the here not explicitly modelled
orientation angle $\varphi$),
while the purely spatial affine Gaussian
derivative model comprises only three internal parameters
(the spatial scale parameters $\sigma_1$ and $\sigma_2$
as well as the in Section~\ref{sec-anal-ori-sel} neither explicitly modelled
orientation angle $\varphi$).

The affine Gabor model, thus, comprises one
more parameter than the affine Gaussian derivative model.
The dependency on the remaining degrees of freedom
(beyond the orientation $\varphi$), due to variations
in $\sigma_1$ and $\nu$, is, however, of a special form, in the respect that the
dependency only depends on the product between $\sigma_1$ and $\nu$.
Therefore, we will have to, beyond a variability with respect to the
scale parameter ratio $\kappa$, also investigate the behaviour with
respect to variations of the product of the angular frequency $\nu$ and the
remaining spatial scale parameter $\sigma_1$.

Figure~\ref{fig-ori-sel-spat-anal-gabor} shows orientation selectivity
curves for the even component of the affine Gabor pair for different combinations of the scale
parameter ratio $\kappa = 1, 2, 4$ or $8$, and the product
$\sigma_1 \nu = 1/2, 1$ or $2$. As can be seen from these graphs, the
orientation selectivity becomes more narrow with increasing values of the
scale parameter ratio $\kappa$, as for the affine Gaussian derivative
model. The orientation selectivity does, however,
also become more narrow when the product $\sigma_1 \, \nu$ increases.

Hence, from the conceptual background of whether one could possibly
infer that the receptive fields ought to be more elongated,
when the orientation selectivity becomes more narrow,
one would not be able to make such a logical inference from a population of
neurons according to the affine Gabor model, if the only
{\em a priori\/} would be that the population would be generated
from an expansion over all the four
internal parameters $\sigma_1$, $\kappa$, $\nu$ and the orientation
$\varphi$. If it, on the
other hand, would be {\em a priori\/} known that the relationship
between the angular frequency $\nu$ and the remaining spatial scale
parameter $\sigma_1$ would be fixed to their product being
held constant,
then one would, in principle, be able to make such a logical inference.

To save space, we do not show the graphs for the corresponding
orientation selectivity curves for the odd component of the Gabor
pair, nor for the idealized model of a complex cell,
obtained
by squaring the responses from the odd and even components
and then adding them, and finally taking the
square root of the result:
\begin{align}
  \begin{split}
    Q^2 = L_{\even}^2 + L_{\odd}^2
  \end{split}\nonumber\\
  \begin{split}
    = \frac{1}{4} e^{-\nu ^2 \sigma_1^2 \left(\kappa ^2 \sin ^2\theta+\cos
        ^2\theta+2 \cos \theta+1\right)} \times
  \end{split}\nonumber\\
  \begin{split}
    \phantom{=} 
    \left(
      1 +
      e^{4 \nu ^2  \sigma_1^2 \cos \theta}
      \right.
 \end{split}\nonumber\\
  \begin{split}
     \phantom{=} 
     \left.
      \quad -2 \, e^{2 \nu ^2 \sigma_1^2 \cos
   \theta} \cos (2 (\beta +\nu  \, x_1 \cos \theta+\nu  \, x_2 \sin
 \theta))
 \right).
 \end{split}
\end{align}
The qualitative results, that receptive
fields become more narrow as the scale parameter ratio $\kappa$ increases, or
the product $\sigma_1 \nu$ of the two remaining parameters increases,
are, however, similar. So are the results that the orientation
selectivity of the receptive fields, however, also depends strongly
upon the product $\sigma_1 \nu$.

We refrain from extending the analysis to possible
spatio-temporal models, since it may not be fully clear how 
the spatio-temporal Gabor models should then be defined.

\section{Relations to compact quantitative measures of the degree of orientation selectivity}
\label{sec-rel-comp-quant-measures}

In (Ringach {\em et al.\/} \citeyear{RinShaHaw03-JNeurSci}), two compact measures of the orientation selectivity of biological receptive fields are used, in terms of either the circular variance (Mardia \citeyear{Mar72})
\begin{equation}
  V = 1 - |R|,
\end{equation}
where the complex-valued resultant $R$ is given by
\begin{equation}
  \label{eq-def-resultant}
  R = \frac{\int_{\theta = - \pi}^{\pi} r(\theta) \, e^{2 i \theta} d\theta}
                {\int_{\theta = - \pi}^{\pi} r(\theta) \, d\theta},
\end{equation}
or the orientation bandwidth $B$, defined as the value of $B$ for which
\begin{equation}
  \frac{r(B)}{r(0)} = \frac{1}{\sqrt{2}}.
\end{equation}
In the following, we will compute these compact measures for the orientation selectivity curves derived from the affine-Gaussian-derivative-based models of visual receptive fields, as summarized in Table~\ref{tab-summ-ori-sel-diff-models}.

\begin{figure*}[tbp]
  \begin{equation}
    R_{\complex}
    = \frac{\kappa ^2 \left(48 \left(\kappa ^2-1\right)^{3/4} \Gamma \left(\frac{5}{4}\right)
   \, _2F_1\left(\frac{1}{2},1;\frac{3}{4};\frac{1}{\kappa ^2}\right)-16 \left(\kappa
   ^2-1\right)^{3/4} \Gamma \left(\frac{5}{4}\right)+3 \sqrt{2 \pi } \, \kappa  \, \Gamma
   \left(-\frac{1}{4}\right)\right)}{2 \left(\kappa ^2-1\right) \left(16 \left(\kappa
   ^2-1\right)^{3/4} \Gamma \left(\frac{5}{4}\right) \,
   _2F_1\left(\frac{1}{2},1;\frac{3}{4};\frac{1}{\kappa ^2}\right)+\sqrt{2 \pi } \, \kappa \,
   \Gamma \left(-\frac{1}{4}\right)\right)}
  \end{equation}
  \caption{Explicit expression for the resultant $R$ for (i)~the purely spatial model of complex cells based on affine Gaussian derivatives, as well as for (ii)~the velocity-adapted spatio-temporal model of complex cells based on affine Gaussian derivatives. The function $_2F_1(a, b; c; z)$ is the hypergeometric function $\operatorname{Hypergeometric2F1}[a, b, c, z]$ in Mathematica, while $\Gamma(z)$ denotes Euler's Gamma function.}
  \label{fig-expl-expr-R-complex-2-cases}
\end{figure*}

\begin{figure*}[tbp]
  \begin{equation}
     B_{\complex} = \cos ^{-1}\left(\sqrt{\frac{A}{\kappa
   ^6-3 \kappa ^4+3 \kappa ^2+3}}\right)
  \end{equation}
  where
  \begin{equation}
    A = \kappa^2 - 2 \kappa^4+ \kappa^6 
    + \sqrt[3]{2} \, \sqrt[3]{\sqrt{\kappa ^{12} \left(\kappa ^6-3 \kappa ^4+3 \kappa
          ^2+3\right)^2}-\kappa ^6 \left(\kappa ^6-3 \kappa ^4+3 \kappa ^2-5\right)}
    + C
  \end{equation}
  and
  \begin{equation}
     C = -\frac{2 \times 2^{2/3} \, \kappa ^4 \left(\kappa ^2-1\right)}{\sqrt[3]{-\kappa ^{12}+3 \kappa
   ^{10}-3 \kappa ^8+5 \kappa ^6+\sqrt{\kappa ^{12} \left(\kappa ^6-3 \kappa ^4+3 \kappa
   ^2+3\right)^2}}}
  \end{equation}
  \caption{Explicit expression for the orientation selectivity bandwidth $B$ for (i)~the purely spatial model of complex cells based on affine Gaussian derivatives, as well as for (ii)~the velocity-adapted spatio-temporal model of complex cells based on affine Gaussian derivatives. }
  \label{fig-expl-expr-B-complex-2-cases}
\end{figure*}

\begin{figure*}[hbtp]
  \begin{center}
    \begin{tabular}{ccc}
      {\em\footnotesize First-order simple cell\/}
      &       {\em\footnotesize Second-order simple cell\/}
      &       {\em\footnotesize Complex cell\/} \\
      \includegraphics[width=0.30\textwidth]{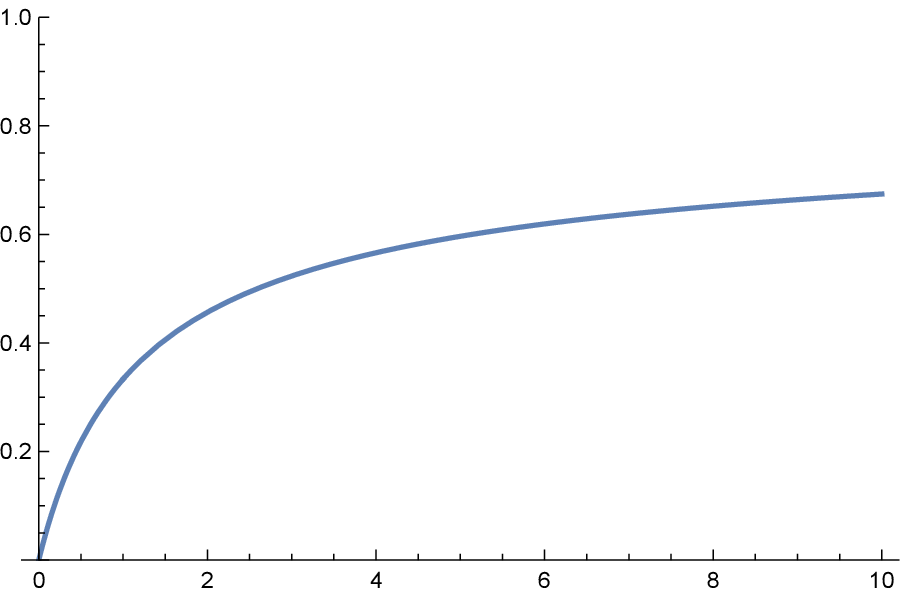}
      & \includegraphics[width=0.30\textwidth]{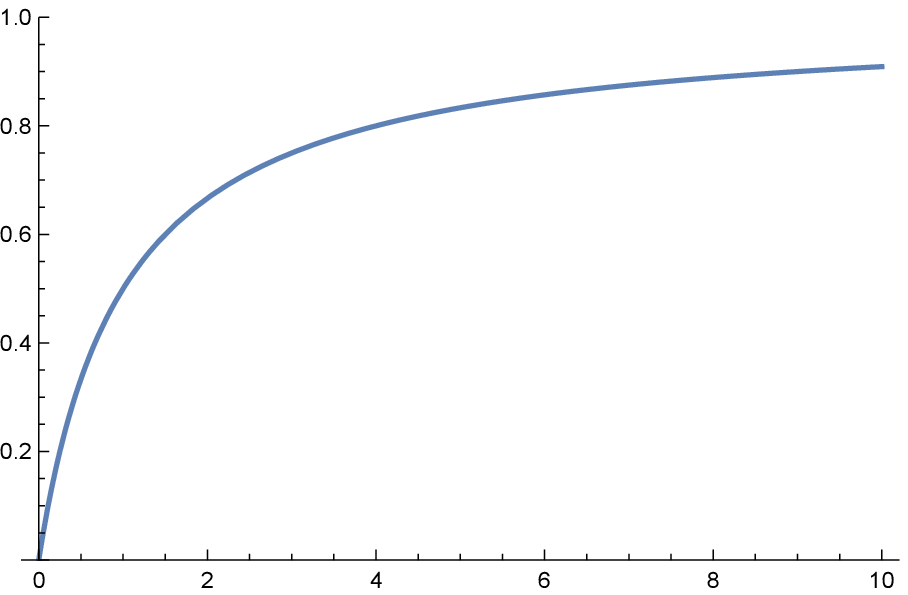}
      & \includegraphics[width=0.30\textwidth]{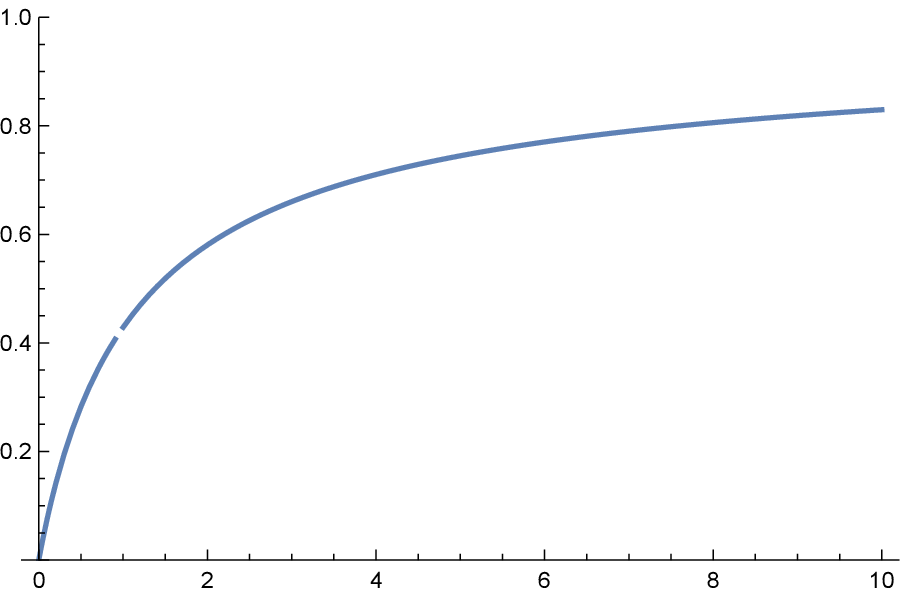}\\
    \end{tabular}
  \end{center}
  \caption{Graphs of the {\em resultant\/} $R$ of the orientation selectivity curves as function of the scale parameter ratio $\kappa$ for the different models of visual receptive fields based on affine Gaussian derivatives. (left) First-order model of simple cell. (middle) Second-order model of simple cell. (right) Purely spatial model of complex cell or velocity-adapted model of complex cell. (Horizontal axes: scale parameter ratio $\kappa$. Vertical axes: resultant $R$.)}
  \label{fig-R-graphs}
\end{figure*}

\begin{figure*}[hbtp]
  \begin{center}
    \begin{tabular}{ccc}
      {\em\footnotesize First-order simple cell\/}
      &       {\em\footnotesize Second-order simple cell\/}
      &       {\em\footnotesize Complex cell\/} \\
      \includegraphics[width=0.30\textwidth]{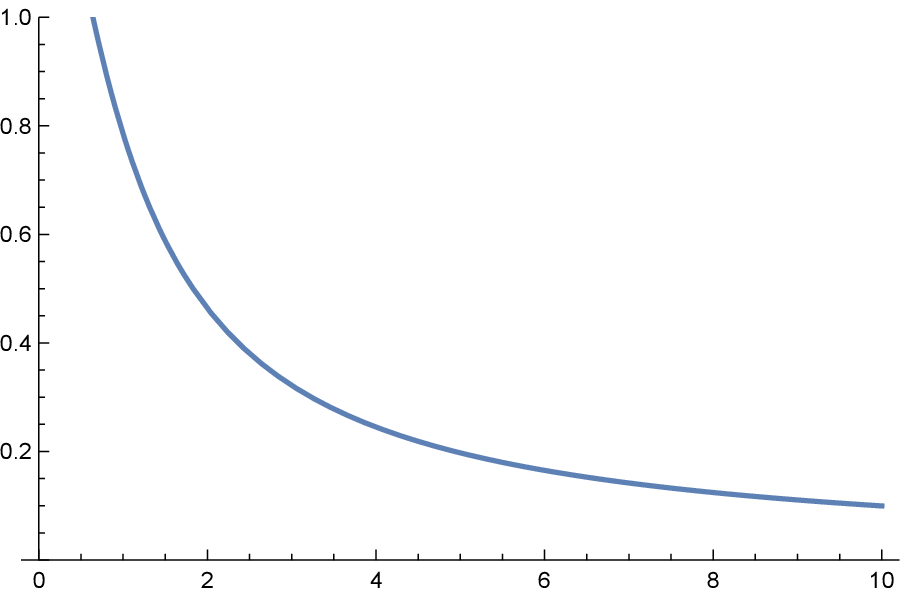}
      & \includegraphics[width=0.30\textwidth]{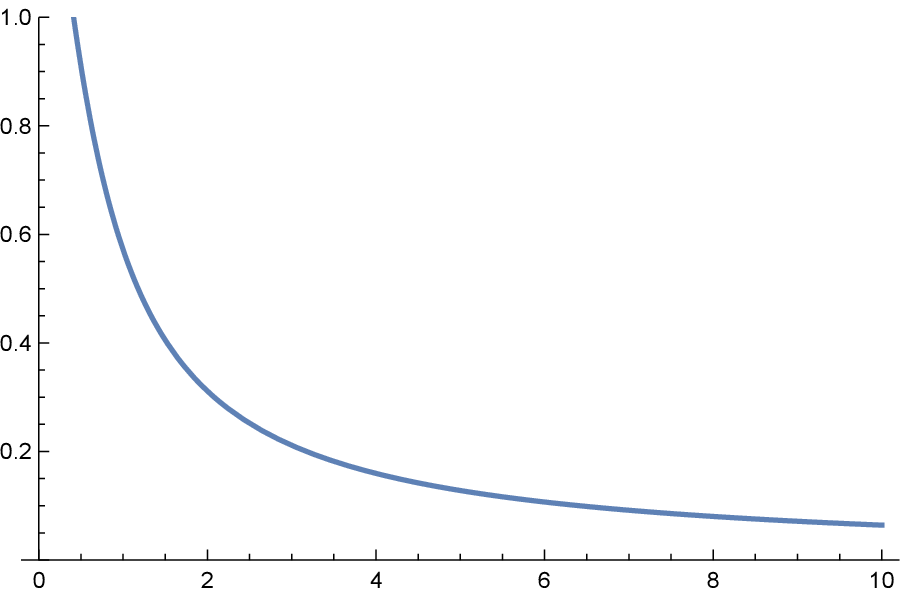}
      & \includegraphics[width=0.30\textwidth]{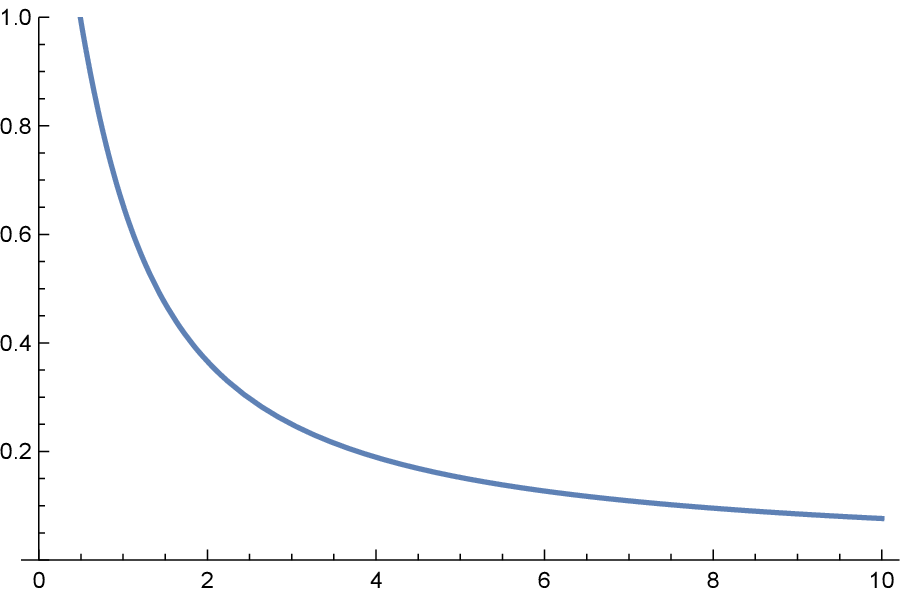}\\
    \end{tabular}
  \end{center}
  \caption{Graphs of the {\em orientation bandwidth\/} $B$ of the orientation selectivity curves as function of the scale parameter ratio $\kappa$ for the different models of visual receptive fields based on affine Gaussian derivatives. (left) First-order model of simple cell. (middle) Second-order model of complex cell. (right) Purely spatial model of complex cell or velocity-adapted model of complex cell. (Horizontal axes: scale parameter ratio $\kappa$. Vertical axes: orientation bandwidth $B$.)}
  \label{fig-B-graphs}
\end{figure*}

\subsection{The resultants $R$ of the orientation selectivity curves computed for the affine-Gaussian-derivative-based models of simple and complex cells}

Due to the symmetry of the orientation selectivity curves in Table~\ref{tab-summ-ori-sel-diff-models}, the purely imaginary component of the resultant $R$ will be zero for all the considered models of simple and complex cells in terms of affine-Gaussian-derivative-based receptive fields. In the following, we therefore focus solely on the purely real component of the resultants.

For the first-order simple cells, we obtain, after solving the resulting integrals in Mathematica:
\begin{align}
  \begin{split}
    R_{\simple,1}
    & = \frac{\int_{\theta = -\pi/2}^{\pi/2}
                    \frac{ \cos \theta }
                                {\sqrt{\cos ^2 \theta + \kappa ^2 \sin ^2\theta}}
                        \, \cos 2 \theta \, d\theta}
                     {\int_{\theta = -\pi/2}^{\pi/2}
                         \frac{ \cos \theta}
                                 {\sqrt{\cos ^2 \theta + \kappa ^2 \sin ^2\theta}}
                         \, d\theta}
  \end{split}\nonumber\\
  \begin{split}
    \label{eq-R-simple-cell-1}
    & = \frac{\kappa  \left(\kappa  \cosh ^{-1} \kappa -\sqrt{\kappa^2-1}\right)}
                   {\left(\kappa ^2-1\right) \cosh ^{-1} \kappa},
     \end{split}
\end{align}
and for the second-order simple cells:
\begin{align}
  \begin{split}
    R_{\simple,2}
    & = \frac{\int_{\theta = -\pi/2}^{\pi/2}
                         \frac{\cos^2 \theta}
                                 {\cos ^2 \theta + \kappa ^2 \sin ^2\theta}
                         \, \cos 2 \theta \, d\theta}
                    {\int_{\theta = -\pi/2}^{\pi/2}
                         \frac{\cos^2 \theta}
                                 {\cos ^2 \theta + \kappa ^2 \sin ^2\theta}
                         \, d\theta}
  \end{split}\nonumber\\
  \begin{split}
    \label{eq-R-simple-cell-2}    
    & = \frac{\kappa }{\kappa +1}.
  \end{split}
\end{align}
For both the purely spatial model of complex cells based on affine Gaussian derivatives, as well as for the velocity-adapted model of complex cells based on affine Gaussian derivatives, we do in a similar manner get
\begin{align}
   \begin{split}
     R_{\complex}
     & = \frac{\int_{\theta = -\pi/2}^{\pi/2}
                        \frac{\cos ^{3/2} \theta }
                                {\left( \cos ^2 \theta + \kappa ^2 \sin ^2\theta \right)^{3/4}}
                                \, \cos 2 \theta \, d\theta}
                    {\int_{\theta = -\pi/2}^{\pi/2}
                        \frac{\cos ^{3/2} \theta}
                                {\left( \cos ^2 \theta + \kappa ^2 \sin ^2\theta \right)^{3/4}}
                                \, d\theta}                       
    \end{split}
\end{align}
with the explicit expression for that result in Figure~\ref{fig-expl-expr-R-complex-2-cases}.

For the space-time separable model of complex cells, the corresponding formulation
\begin{align}
   \begin{split}
     R_{\complex-\sep}
     & = \frac{\int_{\theta = -\pi/2}^{\pi/2}
                       \frac{\cos \theta \,
                                  \sqrt{2 + \kappa^2 + (2 - \kappa^2) \cos 2 \theta}}
                                {\cos ^2\theta + \kappa ^2 \sin^2\theta}
                                \, \cos 2 \theta \, d\theta}
                    {\int_{\theta = -\pi/2}^{\pi/2}
                       \frac{\cos \theta \,
                                  \sqrt{2 + \kappa^2 + (2 - \kappa^2) \cos 2 \theta}}
                                {\cos ^2\theta + \kappa ^2 \sin^2\theta}
                                \, d\theta}                        
    \end{split}
\end{align}
can be handled in closed form in Mathematica, but does unfortunately lead to an explicit expression for the result that is too complex to be reported here.

Figure~\ref{fig-R-graphs} shows the resulting resultants of the orientation selectivity curves obtained in this way, for the first-order models of simple cells, the second-order models of simple cells as well as for the purely spatial model of complex cells and the velocity-adapted models of complex cells.

\subsection{The orientation bandwidths $B$ of the orientation selectivity curves computed for the affine-Gaussian-derivative-based models of simple and complex cells}

To measure the orientation selectivity bandwidth for the first-order simple cells, we solve the equation
\begin{equation}
   \frac{\cos \theta}
           {\sqrt{\cos ^2 \theta + \kappa ^2 \sin ^2\theta}}
   = \frac{1}{\sqrt{2}},
\end{equation}
which gives the bandwidth measure
\begin{equation}
   \label{eq-B-simple-cell-1}
   B_{\simple,1} = \cos ^{-1}\left(\frac{\kappa }{\sqrt{\kappa ^2+1}}\right).
\end{equation}
Correspondingly, to determine the orientation selectivity bandwidth for the second-order simple cells, we solve the equation
\begin{equation}
  \frac{\cos^2 \theta}
         {\cos ^2 \theta + \kappa ^2 \sin ^2\theta}
  = \frac{1}{\sqrt{2}},
\end{equation}
which gives the bandwidth measure
{\small \begin{multline}
  \label{eq-B-simple-cell-2}   
  B_{\simple,2} = \\ =
  2 \tan ^{-1}\left(\sqrt{\frac{-2 \sqrt{2} \, \kappa ^2+2 \sqrt{2} \sqrt{\kappa
   ^2+\sqrt{2}-1} \, \kappa +\sqrt{2}-2}{\sqrt{2}-2}}\right).
\end{multline}}
To determine the corresponding orientation selectivity measure for the
purely spatial model of complex cells based on affine Gaussian derivatives, as well as for the velocity-adapted model of complex cells based on affine Gaussian derivatives, we solve the equation
\begin{equation}
  \frac{\left| \cos \theta \right|^{3/2}}
                      {\left( \cos ^2 \theta + \kappa ^2 \sin ^2\theta \right)^{3/4}}
  = \frac{1}{\sqrt{2}},
\end{equation}
which gives the explicit result for the bandwidth measure $B_{\complex}$ shown in Figure~\ref{fig-expl-expr-B-complex-2-cases}.

Figure~\ref{fig-B-graphs} shows the resulting orientation bandwidths of the orientation selectivity curves obtained in this way, for the first-order models of simple cells, the second-order models of simple cells as well as for the purely spatial model of complex cells and the velocity-adapted models of complex cells.

\section{Perturbation analysis regarding the choices of the angular
  frequencies and the velocities for the probing sine wave stimuli}
\label{sec-perturb-anal}

In the above theoretical analysis, we have for simplicity of
theoretical analysis throughout assumed that the spatial frequency of
the probing sine wave has been adapted to the spatial
scale parameter of the receptive field.
For the velocity-adapted receptive field models, we have additionally
assumed that the velocity of the stimulus has been optimally adapted
to the velocity parameter, of the receptive field.

If one in a biological experiment would, however, not adapt these
parameters of the stimuli to their optimal values with regard to each
receptive field and each image orientation,
one may therefore ask to what extent the above stated
theoretical results would be affected by a not optimal match between
the parameters of the stimuli to the parameters of the receptive
field. To address that
question, we will in this section perform different types of perturbation analyses,
where the parameters of the stimuli are, thus, not optimally adapted to the
parameters of the receptive fields.

\begin{figure*}[hbtp]
  \begin{center}
    \begin{tabular}{cccc}
      & {\em\footnotesize First-order simple cell\/}
      &       {\em\footnotesize Second-order simple cell\/}
      &       {\em\footnotesize Complex cell\/} \\
      {\footnotesize $\kappa = 1$}
      & \includegraphics[width=0.29\textwidth]{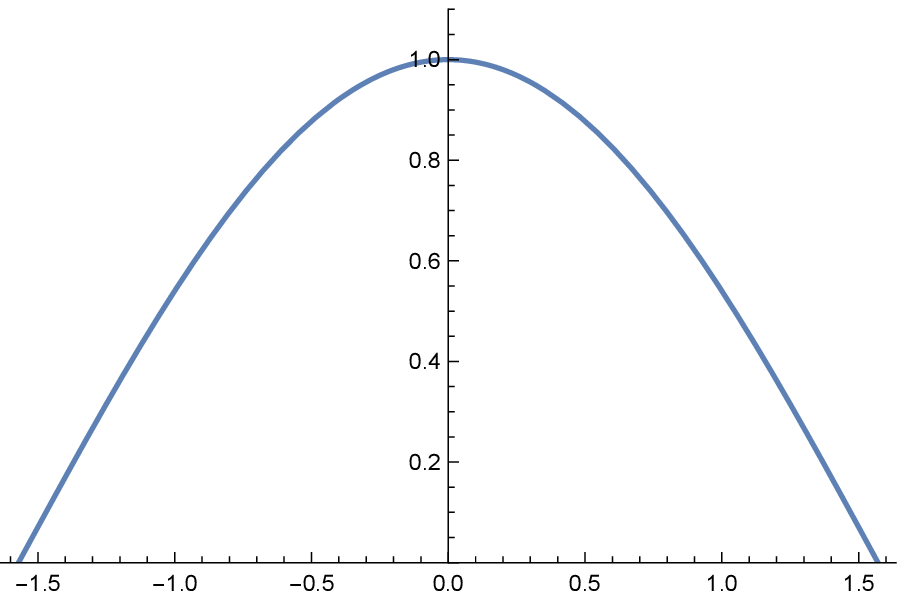}
      & \includegraphics[width=0.29\textwidth]{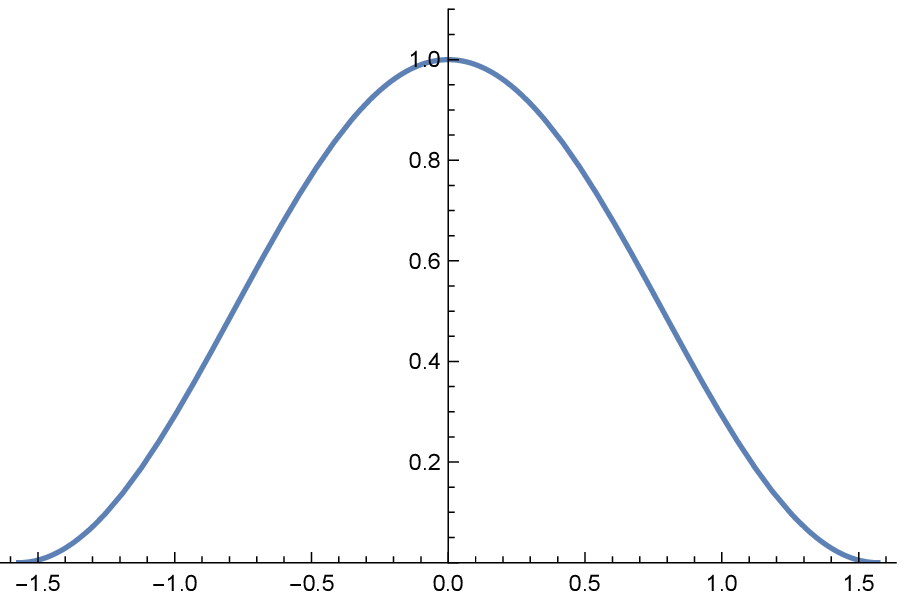}
      & \includegraphics[width=0.29\textwidth]{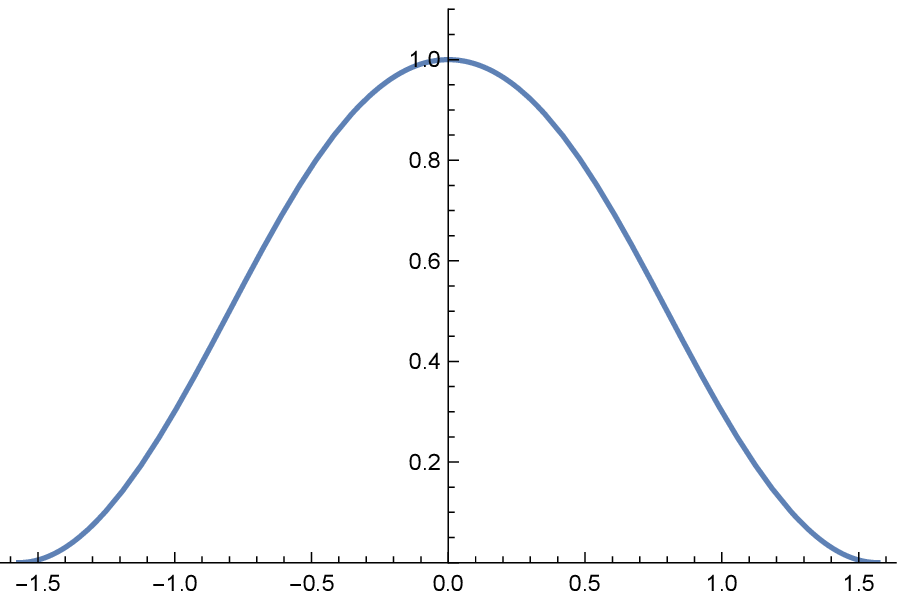} \\    
      {\footnotesize $\kappa = 2$}
      & \includegraphics[width=0.29\textwidth]{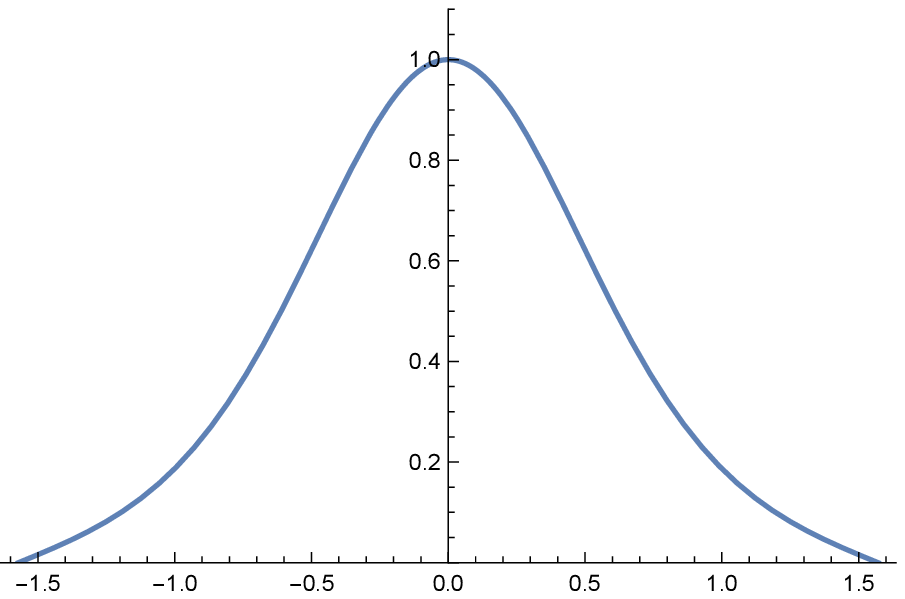}
      & \includegraphics[width=0.29\textwidth]{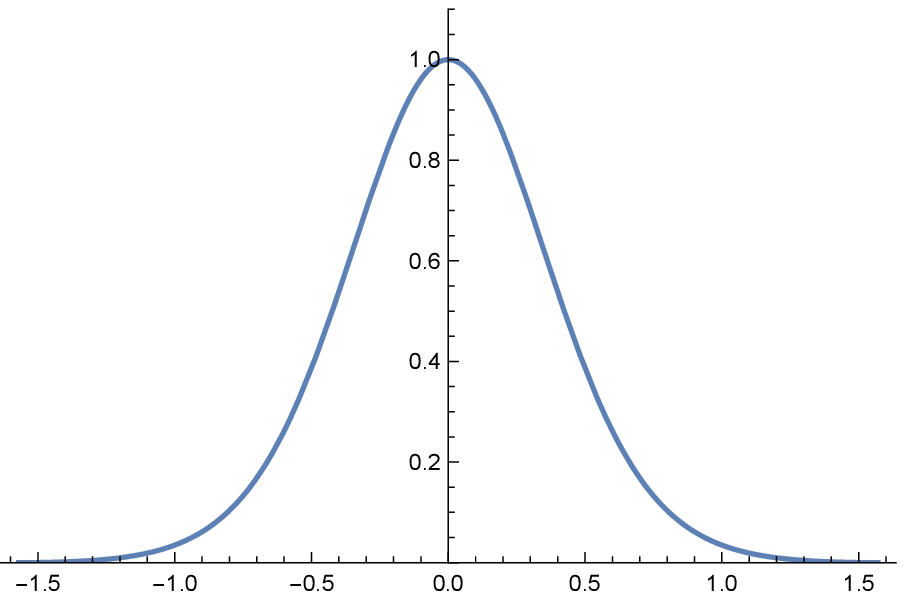}
      & \includegraphics[width=0.29\textwidth]{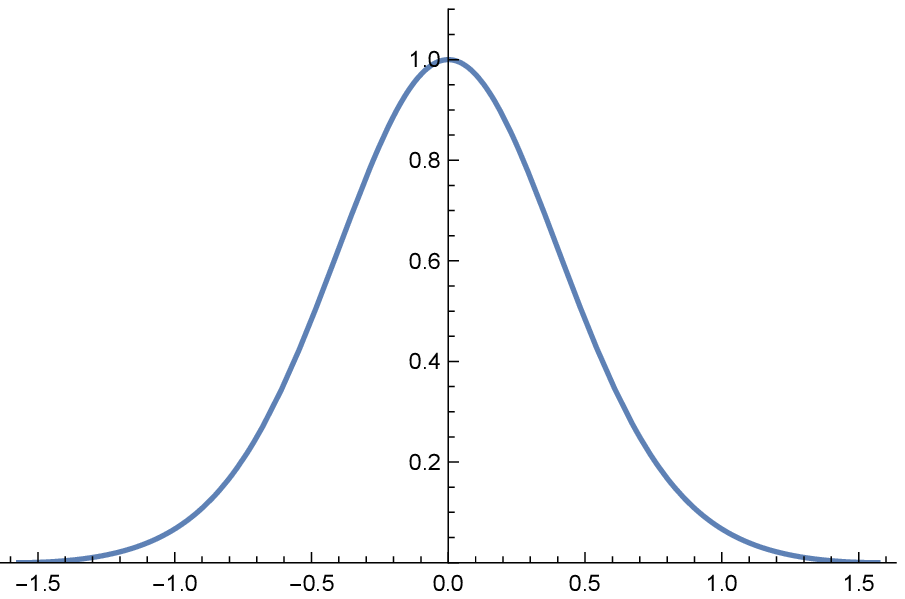} \\    
      {\footnotesize $\kappa = 4$}
      & \includegraphics[width=0.29\textwidth]{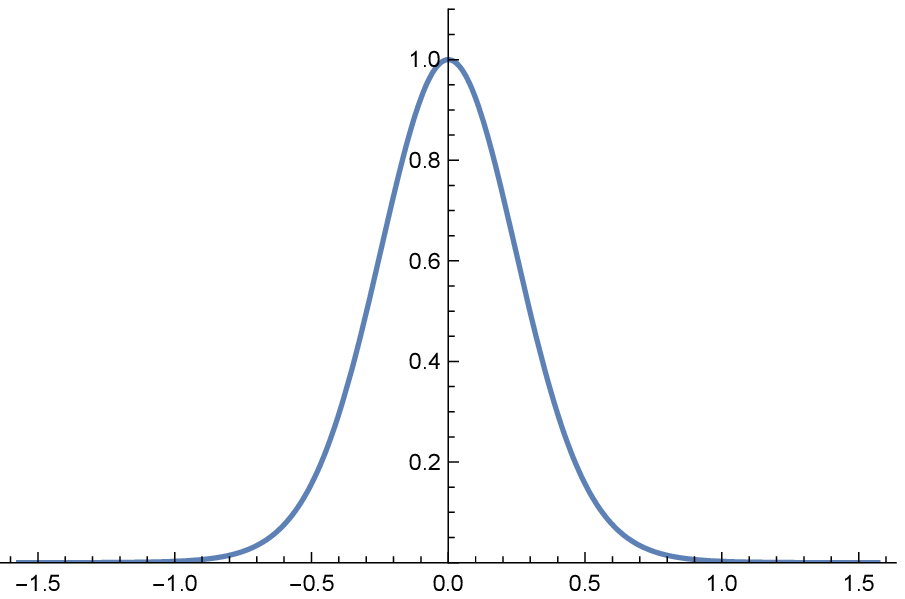}
      & \includegraphics[width=0.29\textwidth]{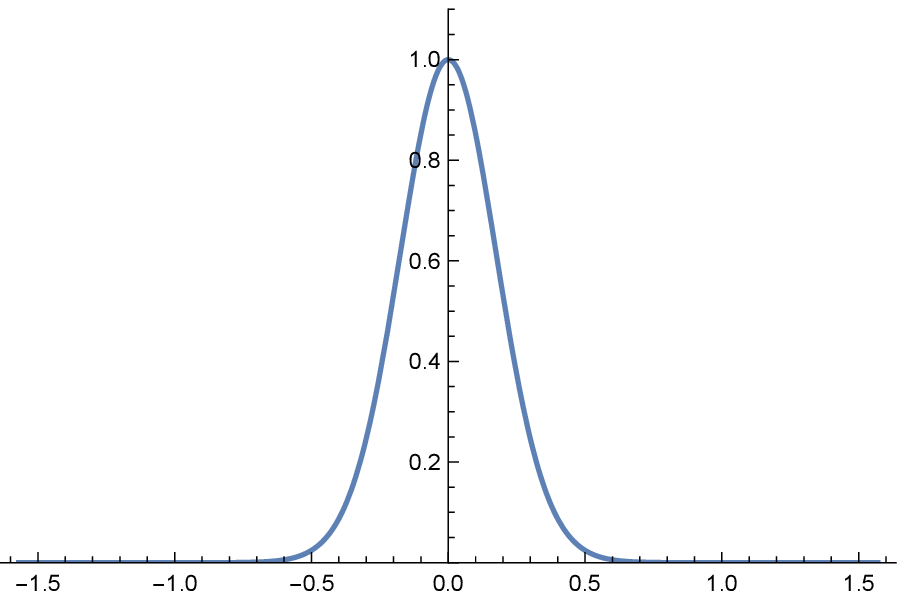}
      & \includegraphics[width=0.29\textwidth]{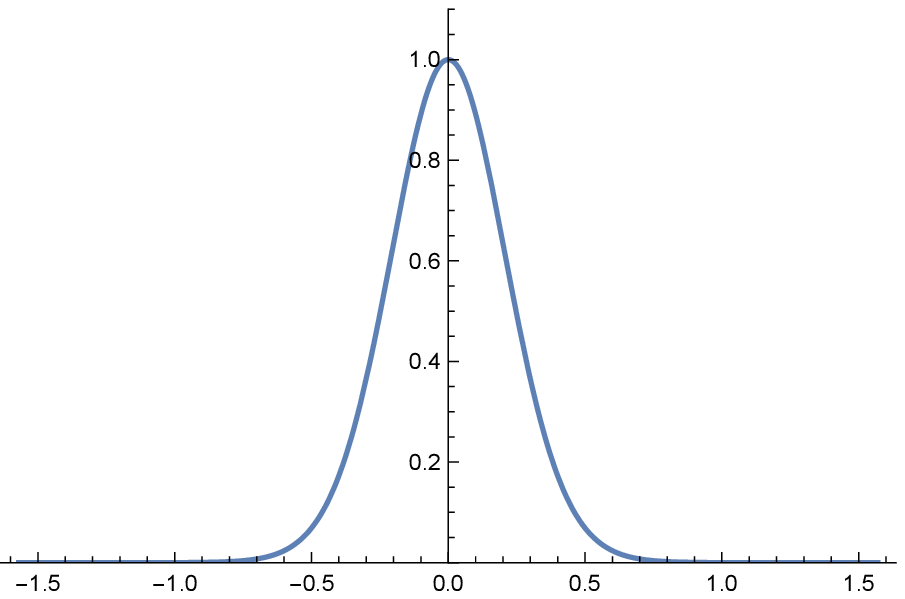} \\    
      {\footnotesize $\kappa = 8$}
      & \includegraphics[width=0.29\textwidth]{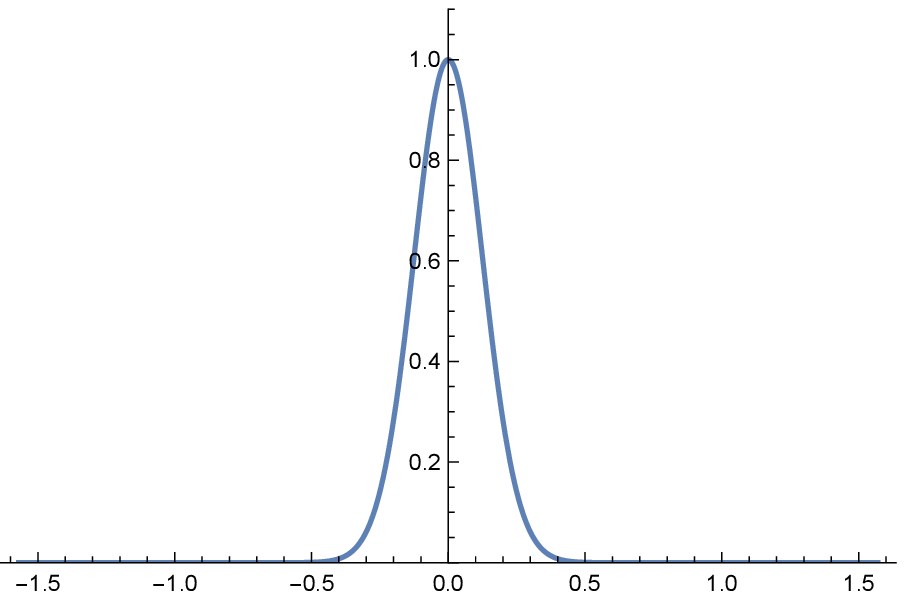}
      & \includegraphics[width=0.29\textwidth]{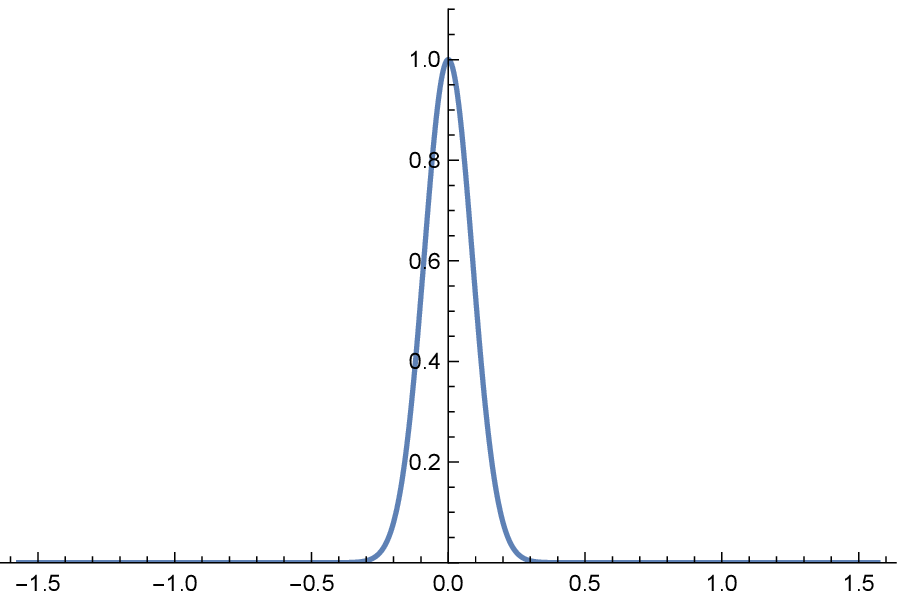}
      & \includegraphics[width=0.29\textwidth]{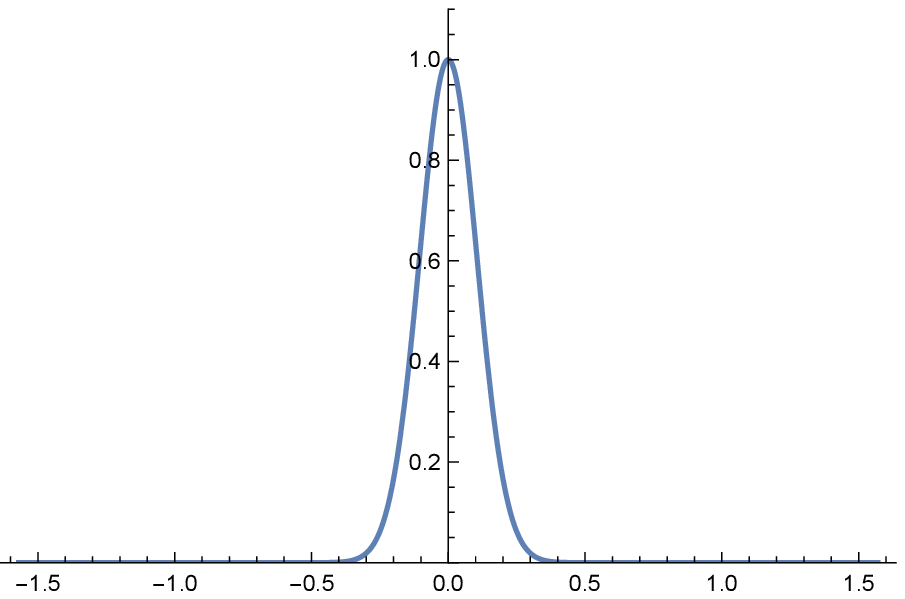} \\    
     \end{tabular}
  \end{center}
  \caption{Graphs of the orientation selectivity for {\em purely spatial
    models\/} of (left column) simple cells in terms of first-order
    directional derivatives of affine Gaussian kernels, (middle column) simple
    cells in terms of second-order directional derivatives of affine Gaussian
    kernels and (right column) complex cells in terms of directional
    quasi-quadrature measures that combine the first- and second-order
    simple cell responses in a Euclidean way for $C = 1/\sqrt{2}$
    and $\lambda = 1$,
    and shown for different values of the ratio
    $\kappa$ between the spatial scale parameters in the vertical
    {\em vs.\/}\ the horizontal directions. In contrast to the
    previous results in Figure~\protect\ref{fig-ori-sel-spat-anal},
    these curves show the orientation selectivity curves arising
    {\em when not adapting the angular frequency of the probing
      sine wave for each image orientation\/}.
    (top row) Results for $\kappa = 1$.
    (second row) Results for $\kappa = 2$.
    (third row) Results for $\kappa = 4$.
    (bottom row) Results for $\kappa = 8$.
    (Horizontal axes: orientation $\theta \in [-\pi/2, \pi/2]$.
     Vertical axes: Amplitude of the receptive field response relative
     to the maximum response obtained for $\theta = 0$.)}
  \label{fig-ori-sel-spat-anal-mod}
\end{figure*}

\begin{figure*}
  \begin{align}
    \begin{split}
      \label{eq-spat-compl-cell-mod}
      {\cal Q}_{0,\spat,\norm} L =
      \frac{e^{\frac{1}{\sqrt{2}}} \sqrt{\cos ^2\theta \, e^{-\sqrt{2}
            \left(\kappa ^2 \sin  ^2\theta+\cos ^2\theta\right)}
          \sqrt{\sin ^2(\beta +\omega  x_1 \cos \theta+\omega  x_2
            \sin \theta)} \sqrt{\cos ^4\theta \sin ^2(\beta+\omega
            x_1 \cos \theta+\omega  x_2 \sin \theta)+\cos ^2\theta}}}{\sqrt{\sqrt{\sin ^2(\beta +\omega  x_1)} \sqrt{\sin ^2(\beta +\omega x_1)+1}}}
     \end{split}
  \end{align}
  \caption{The expression for the oriented spatial quasi-quadrature measure
                 ${\cal Q}_{0,\spat,\norm} L$ in the purely spatial
                 model (\ref{eq-quasi-quad-dir-pure-spat-anal})
                 of a complex cell, when applied to a sine wave pattern
                 of the form (\ref{eq-sine-wave-model-spat-anal}),
                 for $\omega = \sqrt[4]{2} \, \lambda/\sigma_1$,
                 {\em when specifically not adapting the angular frequency of the
                   probing sine wave for each image orientation $\theta$.}}
   \label{fig-eq-spat-compl-cell-mod}
\end{figure*}

\begin{figure*}
  \begin{equation}
    \label{eq-resultant-2nd-order-simple-not-adapt}    
    R_{\simple,2} =
    \frac{\left(\left(\kappa ^2-1\right) \lambda ^2-2\right) \,
      _0\tilde{F}_1\left(2;\frac{1}{16} \left(\kappa ^2-1\right)^2 \lambda ^4\right)+4
      I_0\left(\frac{1}{2} \left(\kappa ^2-1\right) \lambda ^2\right)}{4
      \left(I_0\left(\frac{1}{2} \left(\kappa ^2-1\right) \lambda
          ^2\right)+I_1\left(\frac{1}{2} \left(\kappa ^2-1\right) \lambda ^2\right)\right)}
  \end{equation}
  \caption{The expression for the resultant of orientation
    selectivity curve for the second-order idealized model of simple
    cells, {\em when specifically not adapting the angular frequency of the
      probing sine wave for each image orientation $\theta$.}
    The function $_0\tilde{F}_1(a; z)$ is the hypergeometric function
    $\operatorname{Hypergeometric0F1Regularized}[a, z]$ in
    Mathematica, and the function $I_v(z)$ denotes the modified Bessel
    function of the first kind.}
  \label{fig-eq-resultant-2nd-order-simple-not-adapt}
\end{figure*}

\begin{figure*}[hbtp]
  \begin{center}
    \begin{tabular}{cc}
      {\em\footnotesize First-order simple cell\/}
      &       {\em\footnotesize Second-order simple cell\/} \\
      \includegraphics[width=0.30\textwidth]{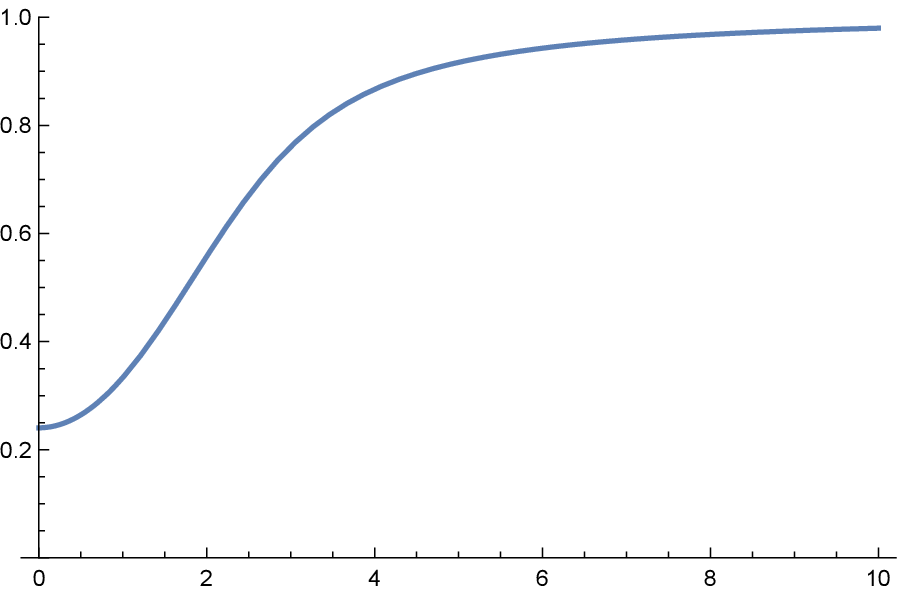}
      & \includegraphics[width=0.30\textwidth]{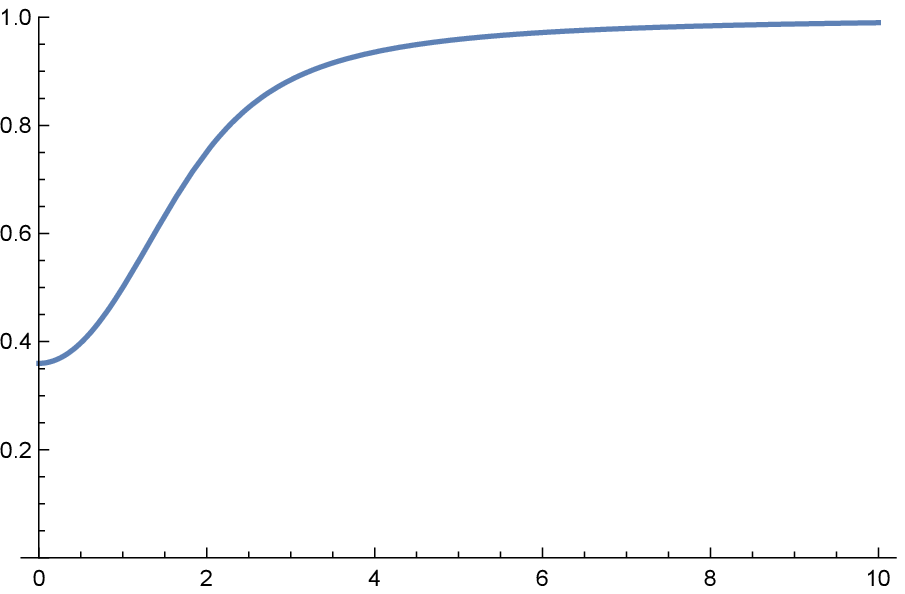}
    \end{tabular}
  \end{center}
  \caption{Graphs of the {\em resultant\/} $R$ of the orientation
    selectivity curves as function of the scale parameter ratio
    $\kappa$ for the different models of visual receptive fields based
    on affine Gaussian derivatives for $\lambda = 1$, {\em when specifically not
      adapting the angular frequency of the probing sine wave for each
      image orientation $\theta$.}
    (left) First-order model of simple cell. (right) Second-order
    model of simple cell.
    (Horizontal axes: scale parameter ratio $\kappa$. Vertical axes: resultant $R$.)}
  \label{fig-R-graphs-mod}
\end{figure*}

\subsection{Alternative self-similar selection of the parameters of
  purely spatial models of simple cells}
\label{sec-perturb-anal-self-sim-lambda}

A first possible generalization to consider of the above mentioned
theoretical analysis is by assuming that the angular frequency of the sine wave for
probing the receptive field is not necessarily selected from the angular frequency
$\hat{\omega}$ for which the receptive field assumes its maximum
value, but instead for some scalar constant $\lambda$ times this
frequency.

Thus, replacing $\hat{\omega}_{\varphi}$ according to
(\ref{eq-omega1-spat}) by
\begin{equation}
  \label{eq-omega1-spat-lambda}
  \hat{\omega}_{\varphi}
  = \frac{\lambda}{\sigma_1 \sqrt{\cos^2 \theta + \kappa^2 \sin^2 \theta}},
\end{equation}
and inserting this value into $A_{\varphi}(\theta, \omega;\; \sigma_1, \sigma_2)$
according to (\ref{eq-A-varphi}),
and similarly replacing  $\hat{\omega}_{\varphi\varphi}$
according to (\ref{eq-omega2-spat}) by
\begin{equation}
  \label{eq-omega2-spat-lambda}  
  \hat{\omega}_{\varphi\varphi}
  = \frac{\sqrt{2} \, \lambda}{\sigma_1 \sqrt{\cos^2 \theta + \kappa^2 \sin^2 \theta}},
\end{equation}
and inserting this expression into
$A_{\varphi\varphi}(\theta, \omega;\; \sigma_1, \sigma_2)$
according to (\ref{eq-A-varphivarphi}), 
as well as normalizing the resulting orientation selectivity curves to
having their peak values equal to 1 for the preferred orientation $\theta = 0$, then
gives that the resulting first- and second-order orientation
selectivity curves will be independent of the angular frequency scaling factor $\lambda$, and
of the forms
\begin{align}
  \begin{split}
     A_{\varphi,\max,\norm}(\theta;\; \kappa)
  = \frac{\left| \cos \theta \right|}{\sqrt{\cos ^2 \theta + \kappa ^2 \sin ^2\theta}},
  \end{split}\\
  \begin{split}
      A_{\varphi\varphi,\max,\norm}(\theta;\; \kappa)
  = \frac{\cos^2 \theta}
             {\cos ^2 \theta + \kappa ^2 \sin ^2\theta}, 
  \end{split}
\end{align}
that is of similar forms as previously derived in
Equations~({\ref{eq-ori-sel-simple-1der}) and~(\ref{eq-ori-sel-simple-2der}).
Thus, for this this type of generalization of the probing model, also
the compact measures in terms of the resultant $R$ according
to~(\ref{eq-R-simple-cell-1}) and~(\ref{eq-R-simple-cell-2})
and the bandwidth $B$ according to~(\ref{eq-B-simple-cell-1})
and (\ref{eq-B-simple-cell-2}) will be the same.

In this respect, there should be a certain robustness in the shapes of
the orientation selectivity curves, as well as concerning the compact
resultant and bandwidth measures, with regards to perturbations of
the scheme used for quantifying the orientation selectivity properties
of the receptive fields.

\subsection{Accumulation of orientation selectivity curves without any
  preferred selection of preferred frequencies for the sine wave probes}
\label{sec-perturb-anal-without-freq-adapt}

A different type of generalization of the probing scheme for the receptive
fields would be if we would not perform any adjustment of the
angular frequency of the sine wave probe with regard to the
frequency selectivity properties of the receptive field for every
image orientation of the sine wave probe, but instead keeping the
angular frequency the same for all the image orientations.

For the purely spatial model of receptive fields, then, with $\sigma_2 = \kappa \, \sigma_1$,
as well as with $\omega_{\varphi} = \lambda/\sigma_1$ for the first-order simple cell
models and with $\omega_{\varphi\varphi} = \sqrt{2} \, \lambda/\sigma_1$
for the second-order simple cell models, as corresponding to the
angular frequencies $\omega$ that lead to the maximum response for the
preferred direction $\theta = 0$ for the first- and second-order
models of simple cells, respectively,
according to (\ref{eq-omega1-spat}) and (\ref{eq-omega2-spat}),
the amplitude expressions
$A_{\varphi}(\theta, \omega;\; \sigma_1, \sigma_2)$
according to (\ref{eq-A-varphi})
and
$A_{\varphi\varphi}(\theta, \omega;\; \sigma_1, \sigma_2)$
according to (\ref{eq-A-varphivarphi}) then assume the forms
\begin{align}
  \begin{split}
    \label{eq-varphi-pure-spat-non-adapt}
    A_{\varphi}(\theta\; \kappa, \lambda)
    & =e^{-\frac{1}{2} (\kappa^2-1) \, \lambda^2 \sin^2 \theta} \, \left|\cos \theta \right|,
  \end{split}\\
  \begin{split}
    \label{eq-varphiphi-pure-spat-non-adapt}    
    A_{\varphi\varphi}(\theta;\; \kappa, \lambda)
    & = e^{-(\kappa^2-1) \, \lambda^2 \sin^2 \theta} \, \cos^2\theta.
  \end{split} 
\end{align}
By combining the responses from the first- and the second-order models
of simple cells according to (\ref{eq-L0-pure-spat-anal}) and
(\ref{eq-L00-pure-spat-anal}), respectively, into an idealized model
of complex cells according to (\ref{eq-quasi-quad-dir-pure-spat-anal}),
for the geometric average
$\omega = \sqrt{\omega_{\varphi} \, \omega_{\varphi\varphi}} = \sqrt[4]{2} \, \lambda/\sigma_1$
of the preferred angular frequencies for the first- and second-order
models of simple cells, respectively, and additionally normalizing the
resulting orientation selectivity curve to having its peak value for
$\theta = 0$, we obtain an expression for the resulting idealized
orientation selectivity curve according to
Equation~(\ref{eq-spat-compl-cell-mod}) in Figure~\ref{fig-eq-spat-compl-cell-mod}.

Figure~\ref{fig-ori-sel-spat-anal-mod} shows graphs of the idealized
models for orientation selectivity curves obtained in this way
for the special choice of $\lambda = 1$.
As can be seen from the graphs, the qualitative properties of the
orientation selectivity curves become more narrow for larger values of
the scale parameter ratio $\kappa$. Additionally, larger values of the
complementary parmeter $\lambda$ will also lead to more narrow
orientation selectivity curves, while lower values of the
complementary parameter will lead to broader orientation selectivity
curves. By varying the parameter $\lambda$ in the model, we can hence
predict the effects on the orientation selectivity curve, when using
a fixed value of the angular frequency, compared to the situation when
the angular frequency of the sine probe would have been adapted to (only) the
preferred orientation $\theta = 0$ of the receptive field.

The shapes of the resulting
idealized orientation selectivity curves do, however, differ somewhat
from the idealized orientation selectivity curves in
Figure~\ref{fig-ori-sel-spat-anal}, in that the orientation
selectivity curves obtained when not adapting the angular frequency
for each image orientation lead to more narrow orientation selectivity
properties, compared to the situation when adapting the angular
frequency of the sine probe for each angular frequency.

\subsubsection{Resultant measures for the orientation selectivity
  curves}

By computing the resultant according to (\ref{eq-def-resultant}) for
these orientation selectivity curves, we do furthermore get the
resultant for the first-order model of simple cells as
\begin{equation}
  R_{\simple,1} =
  \frac{2 \, \sqrt{\frac{2}{\pi }} \, e^{-\frac{1}{2} \left(\kappa ^2-1\right) \lambda
   ^2}}{\sqrt{\kappa ^2-1} \, \lambda \, \erf \left(\frac{\sqrt{\kappa ^2-1} \lambda
   }{\sqrt{2}}\right)}-\frac{2}{\left(\kappa ^2-1\right) \lambda ^2}+1
\end{equation}
and the resultant for the second-order model of simple cells according
to Equation~(\ref{eq-resultant-2nd-order-simple-not-adapt}) in
Figure~\ref{fig-eq-resultant-2nd-order-simple-not-adapt}.

Figure~\ref{fig-R-graphs-mod} shows graphs of these curves for the
special case of $\lambda = 1$. As can be seen from these graphs, the
resulting values of the resultant due to the more narrow orientation
selectivity properties resulting from this way of probing the
receptive fields confined to larger values of the resultant,
compared to the previous results in Figure~\ref{fig-R-graphs}.

Computing explicit expressions for the resultant of the corresponding
idealized model of complex cells is, however, not manageable in closed
form. Neither is it manageable to computate of explicit expressions
in closed form for the bandwidth descriptors for either the idealized
models of simple or complex cells, when using this probing criterion.

\subsection{Alternative self-similar selection of the parameters of
  velocity-adapted spatio-temporal models of simple cells}
\label{sec-perturb-anal-self-sim-lambda-spat-temp}

Concerning corresponding generalizations for a joint spatio-temporal
analysis, a first generalization to consider is by assuming that
neither the angular frequency $\omega$ of the probing sine wave nor its image
velocity $u$ are chosen from the angular frequency $\hat{\omega}$
or the image velocity $\hat{u}$, for which the receptive field assumes
its maximum value, but instead for some constants $\lambda$ and
$\mu$ times these values.

Let us, hence, replace $\hat{\omega}_{\varphi}$ and $\hat{u}_{\varphi}$
according to (\ref{eq-omega1-vel-adapt}) and (\ref{eq-u1-vel-adapt}),
respectively, by
\begin{equation}
  \label{eq-omega1-vel-adapt-perturb}  
   \hat{\omega}_{\varphi} = \frac{\lambda}{\sigma_1 \sqrt{\cos^2 \theta + \kappa^2 \sin^2 \theta}},
\end{equation}
\begin{equation}
  \label{eq-u1-vel-adapt-perturb}    
    \hat{u}_{\varphi} =  \mu \, v \cos \theta,
\end{equation}
and insert these values into
$A_{\varphi}(\theta, u, \omega;\; \sigma_1, \sigma_2, \sigma_t)$
according to (\ref{eq-A-varphi-spat-temp-anal}),
and similarly replace
$\hat{\omega}_{\varphi\varphi}$ and $\hat{u}_{\varphi\varphi}$
according to (\ref{eq-omega2-vel-adapt}) and (\ref{eq-u2-vel-adapt}),
respectively, by
\begin{equation}
  \label{eq-omega2-vel-adapt-perturb}
   \hat{\omega}_{\varphi\varphi}
   = \frac{\sqrt{2} \, \lambda}{\sigma_1 \sqrt{\cos^2 \theta + \kappa^2 \sin^2 \theta}},
\end{equation}
\begin{equation}
  \label{eq-u2-vel-adapt-perturb}      
    \hat{u}_{\varphi\varphi} =  \mu \, v \cos \theta.
\end{equation}
and insert these expressions into
$A_{\varphi\varphi}(\theta, u, \omega;\; \sigma_1, \sigma_2, \sigma_t)$
according to (\ref{eq-A-varphivarphi-spat-temp-anal}), which then,
after normalizing the resulting orientation selectivity curves to
having their peak values equal to 1, leads to
\begin{multline}
  A_{\varphi,\max,\norm}(\theta;\; \sigma_1, \sigma_t, \kappa, \mu) = \\
  = \frac{\left|\cos \theta \right| \, \sqrt{(\mu -1)^2 \, \sigma_t^2 \, v^2+\sigma_1^2}}{\sqrt{\kappa
   ^2 \, \sigma_1^2 \sin ^2\theta+\cos ^2\theta \left((\mu -1)^2 \, \sigma_t^2
   \, v^2+\sigma_1^2\right)}},
\end{multline}
\begin{multline}
  A_{\varphi\varphi,\max,\norm}(\theta;\; \sigma_1, \sigma_t, \kappa, \mu) = \\
  = \frac{\cos ^2\theta \left((\mu -1)^2 \, \sigma_t^2
   \, v^2+\sigma_1^2\right)}{\kappa ^2 \, \sigma_1^2 \sin ^2\theta+\cos ^2(\theta
   ) \left((\mu -1)^2 \, \sigma_t^2 \, v^2+\sigma_1^2\right)}.
\end{multline}
Notably, these orientation
selectivity curves are, although independent of the frequency scaling
factor $\lambda$, for general values of the velocity scaling
factor $\mu$, in addition to a dependency on the scale
parameter ratio $\kappa$, also strongly dependent on both the
spatial and temporal scale parameters $\sigma_1$ and $\sigma_t$
as well as the velocity parameter $v$ in the receptive field model.

\begin{figure*}
  \begin{multline}
    \label{eq-ori-sel-curves-1-spattemp-nonadapt}
    A_{\varphi,\max,\norm}(\theta;\; \sigma_1, \sigma_t, \kappa, \lambda, \mu) = 
    \left| \cos \theta \right| \, \exp \left(\frac{\lambda ^2 \left(-\kappa ^2 \, \sigma_1^2 \sin
          ^2\theta-( \left| \cos \theta \right| -1) \left((1-2 \mu ) \,
          \sigma_t^2 \, v^2+  \left| \cos \theta \right| 
            \left(\sigma_1^2+\sigma_t^2 \, v^2\right)+\sigma_1^2\right)\right)}{2
        \left((\mu -1)^2 \, \sigma_t^2 \,
          v^2+\sigma_1^2\right)}\right) = \\
    = \left| \cos \theta \right| \times 
        e^{- \frac{1}{2} \, \lambda^2 \, \tilde{\kappa}^2 \sin^2 \theta} \times
        e^{ - \frac{\lambda^2 \, ( \left| \cos \theta \right| -1)
            \left((1-2 \mu ) \, \tilde{v}^2 +  \left| \cos \theta \right| 
            \left(1 +\tilde{v}^2\right)+1\right)}{2((\mu -1)^2 \,
          \tilde{v}^2+1)}},
  \end{multline}
  \begin{multline}
    \label{eq-ori-sel-curves-2-spattemp-nonadapt}    
    A_{\varphi\varphi,\max,\norm}(\theta;\; \sigma_1, \sigma_t, \kappa, \lambda, \mu) = 
    \cos ^2\theta \, \exp \left(\frac{\lambda ^2 \left(-\kappa ^2 \, \sigma_1^2 \sin
          ^2\theta-(\left| \cos \theta \right| -1) \left((1-2 \mu ) \,
            \sigma_t^2 \, v^2+  \left| \cos \theta \right| 
            \left(\sigma_1^2+\sigma_t^2 \, v^2\right)+\sigma_1^2\right)\right)}{
        \left((\mu -1)^2 \, \sigma_t^2 \,
          v^2+\sigma_1^2\right)}\right) = \\
        = \cos^2 \theta \times
        e^{- \lambda^2 \, \tilde{\kappa}^2 \, \sin^2 \theta} \times
        e^{ - \frac{\lambda^2 \, ( \left| \cos \theta \right| -1)
            \left((1-2 \mu ) \, \tilde{v}^2 +  \left| \cos \theta \right| 
            \left(1 +\tilde{v}^2\right)+1\right)}{((\mu -1)^2 \,
          \tilde{v}^2+1)}}.
 \end{multline}
  \caption{The expressions for the orientation selectivity curve for
    the first- and second-order velocity-adapted spatio-temporal models of simple cells,
    {\em when specifically not adapting the angular frequency of the
      probing sine wave for each image orientation $\theta$.}}
  \label{fig-eq-ori-sel-curves-12-spattemp-nonadapt}
\end{figure*}

If we restrict the velocity scaling factor $\mu$ to $\mu = 1$,
however, then these orientation selectivity curves reduce to the forms
previously derived when also selecting the angular frequency of the
sine wave probe to the value $\hat{\omega}$, that leads to the
strongest response for the receptive field, of the forms
\begin{align}
  \begin{split}
    A_{\varphi,\max,\norm}(\theta;\; \kappa) 
    & = \frac{\left|\cos \theta \right|}{\sqrt{\cos ^2\theta + \kappa ^2 \sin ^2\theta}},
  \end{split}\\
  \begin{split}
    A_{\varphi\varphi,\max,\norm}(\theta;\; \kappa) 
    & = \frac{\cos ^2\theta}{\cos ^2\theta + \kappa ^2 \sin ^2\theta}.
  \end{split}
\end{align}
In this way, for the special case of choosing the velocity scaling
factor as $\mu = 1$, this spatio-temporal analysis generalizes the
results of the purely spatial analysis in
Section~\ref{sec-perturb-anal-self-sim-lambda}.

For values of the velocity scaling factor $\mu \neq 1$, the previous
role of the square of the scale parameter ratio $\kappa^2$ in the
idealized shape of the orientation selectivity curve is, however, now
replaced by the following parameter
\begin{equation}
  \label{eq-def-kappa-tilde-spat-temp}
  \tilde{\kappa}^2
  = \frac{\kappa^2 \, \sigma_1^2}{(\mu -1)^2 \, \sigma_t^2 \,  v^2+\sigma_1^2}
  = \frac{\kappa^2}{1 + (\mu -1)^2 \, \left(\frac{\sigma_t \, v}{\sigma_1}\right)^2},
\end{equation}
where the dimensionless ratio
\begin{equation}
  \label{eq-def-v-tilde}
  \tilde{v} = \frac{\sigma_t}{\sigma_1} \, v
\end{equation}
is a scale-normalized velocity parameter of the receptive field,
such that the shapes of the orientation selectivity curves in the
spatio-temporal case then assume the forms
\begin{align}
  \begin{split}
    A_{\varphi,\max,\norm}(\theta;\; \tilde{\kappa}) 
    & = \frac{\left|\cos \theta \right|}{\sqrt{\cos ^2\theta + \tilde{\kappa} ^2 \sin ^2\theta}},
  \end{split}\\
  \begin{split}
    A_{\varphi\varphi,\max,\norm}(\theta;\; \tilde{\kappa}) 
    & = \frac{\cos ^2\theta}{\cos ^2\theta + \tilde{\kappa} ^2 \sin ^2\theta}.
  \end{split}
\end{align}
Thus, for non-matching values of the parameters of the sine
wave probe to the optimal values with regard to any specific receptive field, the shapes of
the orientation selectivity curves will be affected by the degree of
such a mismatch between the parameters of the probing sine wave in
relation to the parameters of the receptive field.

\subsection{Accumulation of orientation selectivity curves without any
  preferred selection of preferred frequencies or velocities for the sine wave probes}
\label{sec-perturb-anal-without-freq-adapt-spat-temp}

Alternatively, we can also consider a situation, when neither the
angular frequency or the image velocity of the sine wave probe is
adapted to the properties of the receptive field.

Let us therefore introduce an angular frequency scaling parameter $\lambda$
as well as a velocity scaling parameter $\mu$ normalized such that
the special values $\lambda = 1$ and $\mu = 1$ would correspond to the
same experimental situation as if the angular frequency $\omega$ as well as the
image velocity $u$ would be adapted to correspond to the maximum
possible value of the receptive field response for the preferred
orientation $\theta = 0$.

Then, for the first-order model of the receptive field, we
parameterize the angular frequency and the image velocity according to
(\ref{eq-omega1-vel-adapt}) and (\ref{eq-u1-vel-adapt})
\begin{align}
  \begin{split}
    \omega_{\varphi}
    = \frac{\lambda }{\sqrt{(\mu -1)^2 \, \sigma_t^2 \, v^2+\sigma_1^2}},
  \end{split}\\
  \begin{split}
    u_{\varphi} = \mu \, v,
  \end{split}
\end{align}
while we for the second-order model of the receptive field we
parameterize the angular frequency and the image velocity
according to (\ref{eq-omega2-vel-adapt}) and (\ref{eq-u2-vel-adapt})
\begin{align}
  \begin{split}
    \omega_{\varphi}
    = \frac{\sqrt{2} \, \lambda}{\sqrt{(\mu -1)^2 \, \sigma_t^2 \, v^2+\sigma_1^2}},
  \end{split}\\
  \begin{split}
    u_{\varphi} = \mu \, v,
  \end{split}
\end{align}
which, after insertion into
$A_{\varphi}(\theta, u, \omega;\; \sigma_1, \sigma_2, \sigma_t)$
according to
(\ref{eq-A-varphi-spat-temp-anal})
and
$A_{\varphi\varphi}(\theta, u, \omega;\; \sigma_1, \sigma_2, \sigma_t)$
according to
(\ref{eq-A-varphivarphi-spat-temp-anal}), 
leads to the expressions for the orientation
selectivity curves for the first- and second-order receptive fields
shown in Equations~(\ref{eq-ori-sel-curves-1-spattemp-nonadapt})
and (\ref{eq-ori-sel-curves-2-spattemp-nonadapt}) in
Figure~\ref{fig-eq-ori-sel-curves-12-spattemp-nonadapt},
if we also normalize these curves to having their peak values equal to 1.

From a more detailed examination of this expression, when reformulated
in terms of the transformed scale parameter ratio $\tilde{\kappa}$
according to (\ref{eq-def-kappa-tilde-spat-temp}) and the
dimensionless scale-normalized velocity parameter $\tilde{v}$
according to (\ref{eq-def-v-tilde}), we can not that the resulting
orientation selectivity curves constitue products of (i)~a cosine
function raised to the order of differentiation, (ii)~a negative exponential in terms of the
transformed scale parameter ratio $\tilde{\kappa}$ and (iii)~a
negative exponential in terms of the dimensionless scale-normalized
velocity parameter $\tilde{v}$, where the complementary frequency
scaling factor $\lambda$ and the complementary velocity scaling factor
$\mu$ are also dimensionless.

For this spatio-temporal analysis, the shape of the orientation selectivity curves will therefore
be strongly dependent on the relationships between the parameters of the probing sine
wave and the parameters of the receptive field. From such a perspective,
a conceptual advantage of the probing method used in the previous theoretical
analysis in Section~\ref{sec-anal-ori-sel} is therefore that, by adapting the
parameters of the sine wave probe to the their optimal values for each probing situation, we can
derive characteristic properties of the orientation selectivity
properties of the receptive field, that constitute fully characteristic
properties of that receptive field, and not depending on any otherwise
{\em a priori\/} unknown mismatch between the parameters of the sine
wave probe in relation to the parameters of the receptive field.

\section{Summary and discussion}
\label{sec-summ-disc}

We have presented an in-depth theoretical analysis of the orientation
selectivity properties for the spatial and spatio-temporal receptive fields
according to
the generalized Gaussian derivative model for visual receptive fields,
summarized in Sections~\ref{sec-spat-simple-cells}
and~\ref{sec-spat-temp-simpl-cells}.
This model for visual receptive fields has been previously derived in
an axiomatic manner, from symmetry properties of the environment, in
combination with requirements of internal consistency between image
representations over multiple spatial and temporal scales.
This model has notably also been
demonstrated to well capture the properties of biological simple cells in the
primary visual cortex. Building upon that theory for linear receptive
fields, we have also analysed the orientation selectivity for (some of
them new) non-linear models of complex cells,
summarized in Sections~\ref{sec-spat-model-comp-cells}
and~\ref{sec-spat-temp-compl-cells},
based on energy models that combine the output from such models for
simple cells for different orders of spatial differentiation.

Specifically, we have in Section~\ref{sec-anal-ori-sel}, with the
details concerning the analysis of the space-time separable
spatio-temp\-oral models in Appendix~\ref{app-anal-space-time-sep-rf},
analyzed how the orientation selectivity depends
on a scale parameter ratio $\kappa$, between the scale parameters in
the image orientations perpendicular to {\em vs.\/}\ parallel with the
preferred orientation of the receptive fields. Explicit expressions
for the resulting orientation selectivity curves have been derived, based on closed form
theoretical analysis, and it has been shown that, for all these models
of visual receptive fields, the degree of orientation selectivity
becomes more narrow with increasing values of the scale parameter ratio
$\kappa$. Additionally, we have in
Section~\ref{sec-rel-comp-quant-measures} derived closed-form
expressions for the resultant and the bandwidth of the orientation
selectivity curves, which can be used for interpreting previous
compact characterizations of neurophysiological measurements of
orientation selectivity to parameters in the idealized models for
visual receptive fields.

To compare the theoretical predictions obtained based on the affine
Gaussian derivative model for visual receptive fields, to what would
be obtained from basing the theoretical analysis on other types of
receptive field models, we have furthermore in
Section~\ref{app-ori-sel-anal-gabor}
presented a detailed orientation
selectivity analysis for an an affine Gabor model of visual receptive
fields. The results obtained from the affine Gabor model are consistent with
the results obtained from the affine Gaussian derivative model, in the
sense that the orientation selectivity becomes more narrow, as we
widen the receptive fields in the direction perpendicular to preferred
orientation of the receptive fields.

The results from the affine Gabor model do, on the other
hand, also differ from the results obtained from the affine Gaussian
derivative model, in the respect that the parameter space of the
affine Gabor model comprises one more degree of freedom, compared to
the parameter space of the affine Gaussian derivative
model. Specifically, a variability along that additional degree of
freedom does also
strongly affect the orientation selectivity of the
receptive fields according to the affine Gabor model. The connections between the degree of orientation selectivity and the elongation of the receptive fields
obtained from the affine Gaussian derivative model are in this respect
more specific than for the affine Gabor model. This points to
a both a qualitative similarity and a qualitative difference between
the affine Gaussian derivative model and the affine Gabor model.

We have also in Appendix~\ref{app-aff-transf-prop-gabor}
shown that a generalized version of the affine Gabor model also supports
affine covariance, as the affine Gaussian derivative model does.

Concerning extensions of the approach, we have in the present
treatment regarding receptive fields according to the generalized
Gaussian derivative model for visual receptive fields,
limited ourselves to receptive fields corresponding to only
first- and second-order derivatives over the spatial domain. Modelling results
by Young (\citeyear{You87-SV}) have, however, demonstrated that
receptive fields up to fourth order of spatial differentiation may be
present in the primary visual cortex. It is straightforward to
extend our analysis to third- and fourth-order directional
derivatives,
which would then give other closed-form expressions for
the orientation selectivity of the receptive fields
(and more narrow than the orientation selectivity for
first- and second-order derivatives). Models of complex
cells involving third- and fourth-order spatial derivatives could also
be formulated and be theoretically analyzed, although experimental
support for such extended models of complex cells may currently not be available.

A more conceptual way of extending the modelling work would also be to
incorporate a complementary spatial smoothing step in model of complex
cells, as used in (Lindeberg \citeyear{Lin20-JMIV} Section~5).
In the models for complex cells used in the theoretical analysis
in this paper, the responses of first- and second-order spatial derivative
based receptive fields have been throughout combined in a pointwise
manner, over both space and time. A more
general approach would, however, be to perform weighted integration of such pointwise
contributions over a wider support region over space and time, with
the size over the spatial image domain and the duration over the temporal
domain proportional to the local spatial and temporal scales at which
the spatial and temporal derivatives are computed.
In the present treatment, we have, however, not extended the complex
cell models in
that way, although it could be well motivated, mainly to simplify the
treatment, and to limit the complexity of the theoretical analysis.
If the orientation selectivity curves obtained from theoretical
analysis are to be fitted to data from actual neurophysiological
measurements, it does, however, seem advisable to complement the
models for complex cells by explicit spatial integration, as done in
Equation~(46) in (Lindeberg \citeyear{Lin20-JMIV}).

Concerning the spatio-temporal models of the receptive fields, we have
also limited ourselves to performing a non-causal temporal analysis,
where the temporal smoothing kernels are 1-D Gaussian kernels.
To perform a corresponding time-causal temporal analysis, based on
using the time-causal limit kernel (\ref{eq-time-caus-lim-kern})
for temporal smoothing, one can
perform a Fourier analysis to determine how the probing sine wave will
be affected, using the closed-form expression for the Fourier
transform of the time-causal limit kernel, in a similar way as in the
temporal scale selection analysis in
(Lindeberg \citeyear{Lin17-JMIV} Section~5.2).

Concerning the way that the orientation selectivity curves are
defined, we have in the above theoretical analysis for the affine
Gaussian derivative model throughout%
\footnote{In the orientation selectivity analysis for receptive fields
  according to the affine Gabor model
  in Appendix~\ref{app-ori-sel-anal-gabor}, we did, however, not optimize
  the angular frequency $\omega$ of the probing sine wave for each inclination angle
  $\theta$, because that was harder to accomplish, in terms of
  closed-form mathematical expressions.}
in Section~\ref{sec-anal-ori-sel} assumed that the
wavelength $\hat{\omega}$ (and for the spatio-temporal analysis also
the image velocity $\hat{u}$) of the probing sine wave is optimized for each
image orientation separately. Another possibility is to instead assume that the
wavelength is only optimized for the preferred orientation of the
receptive field only, and then held constant for all the other
inclination angles $\theta$.
Changing the probing method in such a way would then
change the shapes of the orientation selectivity curves, but could be
easily performed, based on the principles for theoretical analysis
outlined in the above treatment.

In Section~\ref{sec-perturb-anal} we have specifically presented
a complementary theoretical analysis of such alternative probing methods,
for both the purely spatial
models and the velocity-adapted spatio-temporal models of the
receptive fields, and demonstrated how such modelling may lead
to more narrow orientation selectivity curves, compared to the
mainly studied probing method, where the angular frequency of the sine
wave probe is adapted to each orientation of the sine wave stimuli.

One could possibly also consider extending the analysis to different
values of the scale normalization powers $\gamma$ and $\Gamma$,
than using the maximally scale invariant choices $\gamma = 1$
and $\Gamma = 0$ used for simplicity in this treatment. Then, however,
some type of post-normalization may, however, also be necessary, to
make it possible to appropriately compare receptive field responses
between multiple spatial and temporal scales, which otherwise are
perfectly comparable when using $\gamma = 1$ and $\Gamma = 0$.

The complementary analysis in
Sections~\ref{sec-perturb-anal-self-sim-lambda}
and~\ref{sec-perturb-anal-self-sim-lambda-spat-temp},
where the frequency of the probing sine
wave is not selected from the maximum over the angular frequency, but
instead from a constant $\lambda$ times that angular frequency value,
demonstrates that in an experimental situation where the angular
frequency is adapted to each image orientation, the resulting
normalized orientation selectivity curves are invariant under
variations of the parameter $\lambda$.
In the case when the angular frequency of the sine wave probe is
not adapted to each stimulus orientation,
as studied in Sections~\ref{sec-perturb-anal-without-freq-adapt}
and~\ref{sec-perturb-anal-without-freq-adapt-spat-temp},
the introduction of a
corresponding parameter $\lambda$ does, however, strongly influence
the shapes of the orientation selectivity curves, and then also in close
interaction with the scale parameter ratio $\kappa$.
Additionally, regarding the complementary analysis of velocity-adapted
spatio-temporal receptive fields in
Sections~\ref{sec-perturb-anal-self-sim-lambda-spat-temp}
and~\ref{sec-perturb-anal-without-freq-adapt-spat-temp},
a mismatch between the velocity of the probing sine wave in relation
to the preferred velocity of the receptive field may also strongly
affect the shapes of the orientation selectivity curves.

A take-home message from this complementary analysis is that the
shapes of the orientation selectivity curves may, in these respects, be
strongly dependent on the parameters of the sine wave stimuli, that are
used for probing the orientation selectivity properties of the
receptive fields, unless adapting the parameters of the sine
wave probe to maximize the response properties of the receptive field.

The proposed probing method in
Section~\ref{sec-anal-ori-sel}, where the angular frequency of the
sine wave probe is for each image orientation determined from the
angular frequency that maximizes the response over the angular
frequencies, and for velocity-adapted spatio-temporal receptives the
image velocity of the stimulus probe is also determined from the image
velocity that maximizes the response over the image velocities, is in
this respect very special, in the sense that the resulting orientation
selectivity do not risk being hampered by an otherwise possible mismatch between the
parameters of stimulus probe in relation to the inherent parameters of
the receptive field.

Concerning possible limitations of the approach, it should be
emphasized that the models for visual receptive fields in the
generalized Gaussian derivative model are highly idealized.
They have been derived from mathematical analysis based on idealized theoretical
assumptions, regarding symmetry properties of the environment, while
they have not been quantitatively adjusted to the receptive field shapes of
actual biological neurons. Hence, the results from the presented
theoretical analysis should be interpreted as such, as the results of
a maximally idealized model, and not with specific aims of providing a
numerically accurate representation of actual biological neurons.
The receptive field models also constitute pure feed-forward models
with no feed-back mechanisms, which are otherwise known to be
important in biological vision.

In view of such a background, it is, however, highly interesting to
see how well the derived orientation selectivity curves in
Figures~\ref{fig-ori-sel-spat-anal},
\ref{fig-ori-sel-veladapt-anal} and~\ref{fig-ori-sel-spat-temp-sep-anal}
reflect the qualitative shapes of the biologically established
orientation selectivity curves recorded in the primary visual cortex by Nauhaus {\em et al.\/}\ (\citeyear{NauBenCarRin09-Neuron}), as reflected in the qualitative comparison to those results in our companion paper (Lindeberg \citeyear{Lin24-arXiv-HypoElongVarRF}).

Additionally, it is interesting to note that corresponding closed-form
calculations of the resultant, for the orientation selectivity curves
for the idealized models of simple cells based on affine Gaussian derivatives,
lead to predictions that are qualitatively very similar to the
experimentally obtained histograms of the resultant for the
orientation selectivity curves, accumulated for
biological simple cells by Goris {\em et al.\/}\ (\citeyear{GorSimMov15-Neuron}),
if extended to models of simple cells in terms of affine Gaussian derivatives up
to order 4, see (Lindeberg \citeyear{Lin24-arXiv-HypoElongVarRF})
for further details.

It is also known that the distinction between simple and complex cells
may not be as distinct as proposed in the initial work by Hubel and
Wiesel, where there could instead be a scale of gradual transitions
between simple and complex cells,
see
Alonso and Martinez (\citeyear{AloMar98-NatNeurSci}),
Mechler and Ringach (\citeyear{MecRin02-VisRes})
and Li {\em et al.\/}\ (\citeyear{LiLiuChoZhaTao15-JNeuroSci})
for treatments with different views on this topic.
Considering that the forms of the
derived orientation selectivity are often similar for our models of
simple and complex cells (as summarized in
Table~\ref{tab-summ-ori-sel-diff-models}),
one may speculate that the presented results
could be relevant with respect to such a wider contextual background.

It may also be worth stating that we do not in any way exclude the possibility of
deriving corresponding prediction results for other possible models of
receptive fields, or for models of neural mechanisms that lead to the
formation of visual receptive fields. A motivation, for in this paper restricting the
analysis in this paper to the idealized receptive fields according to the
generalized Gaussian derivative model and to the idealized
receptive fields according to the Gabor model,
is that these models may be regarded the most commonly used
and well-established idealized functional models for linear receptive fields in
the primary visual cortex. These receptive field models do
specifically reasonably well reproduce the qualitative shape of
many receptive fields in the primary visual cortex, as have been established by
neurophysiological measurements. The formulation of possible Gabor
models for spatio-temporal receptive fields does, however, appear as a
more open topic. Therefore, we have here restricted the analysis of the affine Gabor
model to purely spatial receptive fields.

Considering also that the relationship between the orientation selectivity and the degree of elongation of the receptive fields is more direct for the affine-Gaussian-derivative-based model than for the affine Gabor model, an overall conclusion from this work is therefore that the affine-Gaussian-derivative-based model for visual receptive fields is conceptually much easier to analyze with regard to orientation selectivity properties than the affine Gabor model.

Based on the theoretical analysis presented in this paper, we propose
that very interesting work could be done to match the
output from measurements of the orientation selectivity properties of
biological receptive fields to theoretical models of their functional
properties.

\appendix

\section{Appendix}

\begin{figure*}[hbtp]
  \begin{center}
    \begin{tabular}{cccc}
      & {\em\footnotesize First-order first-order simple cell\/}
      &       {\em\footnotesize Second-order second-order simple cell\/}
      &       {\em\footnotesize Complex cell\/} \\
      {\footnotesize $\kappa = 1$}
      & \includegraphics[width=0.29\textwidth]{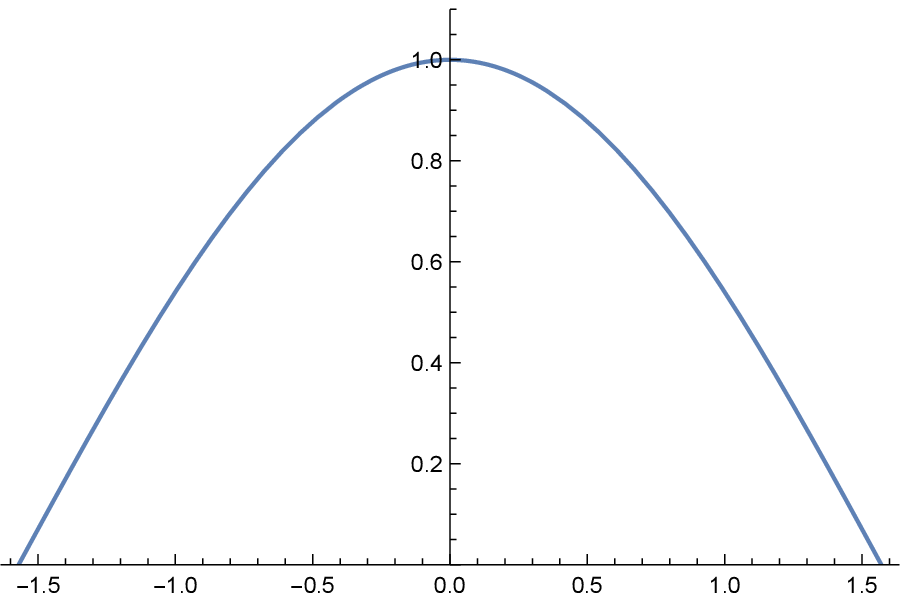}
      & \includegraphics[width=0.29\textwidth]{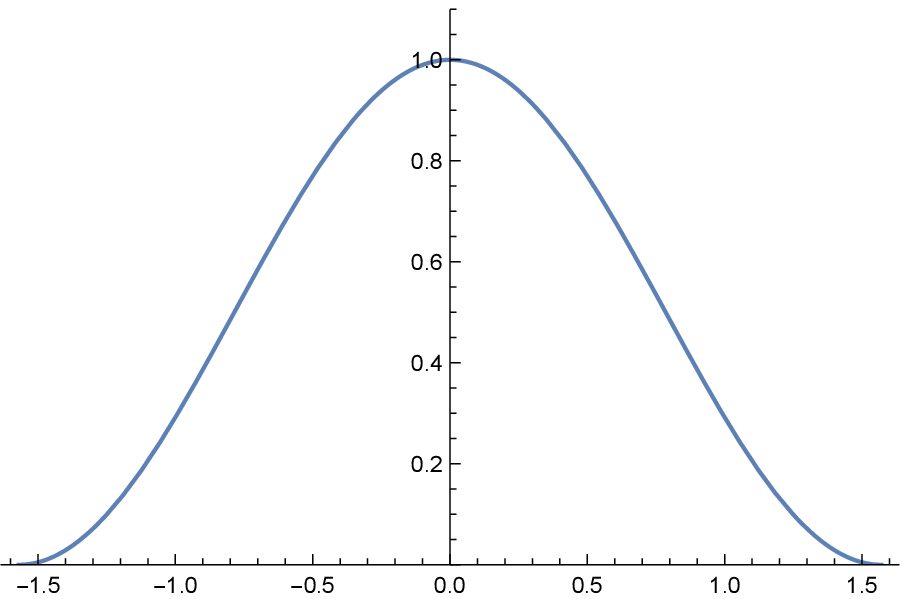}
      & \includegraphics[width=0.29\textwidth]{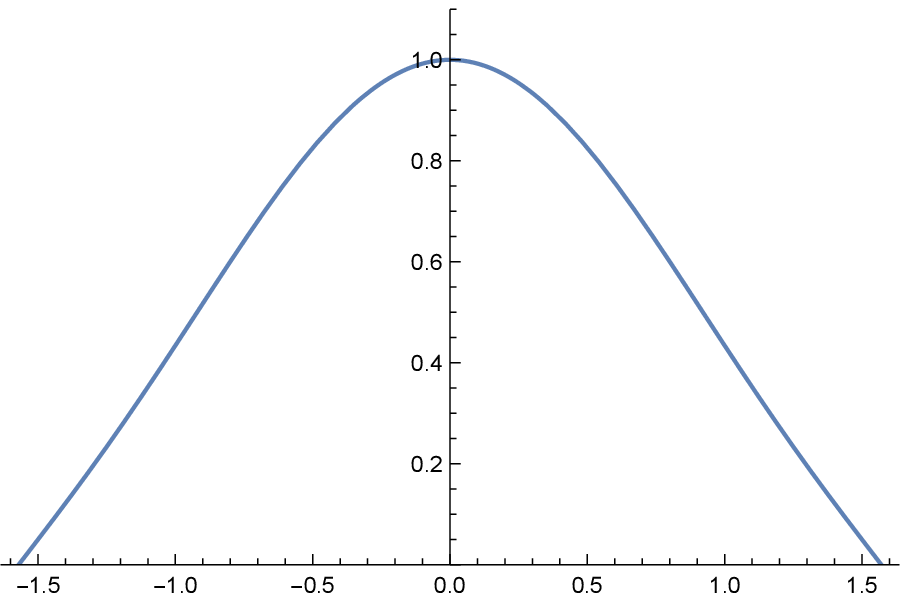} \\    
      {\footnotesize $\kappa = 2$}
      & \includegraphics[width=0.29\textwidth]{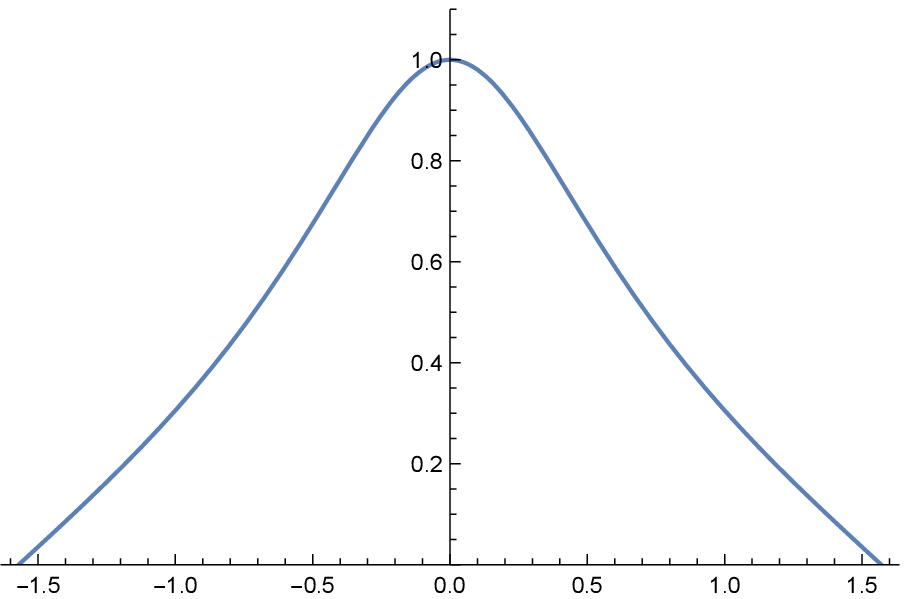}
      & \includegraphics[width=0.29\textwidth]{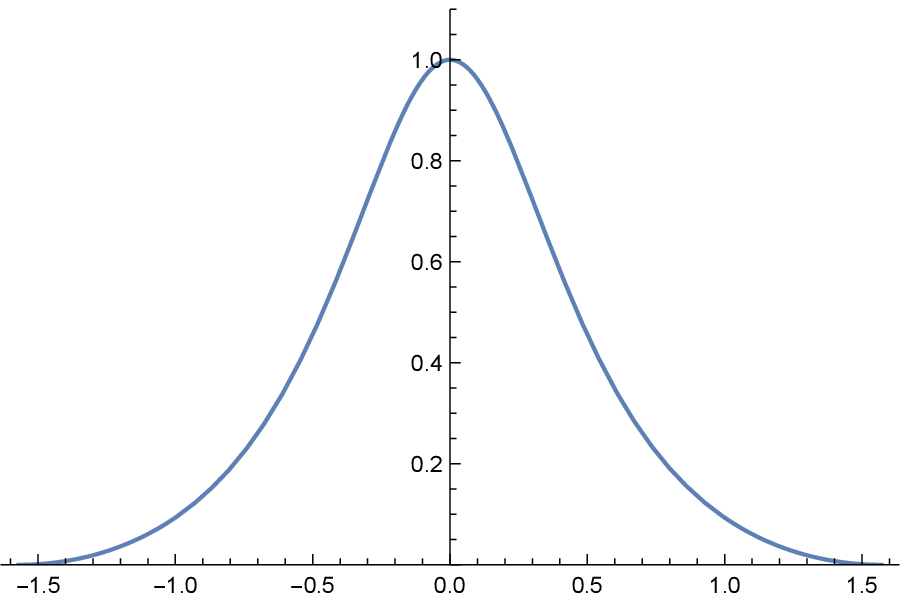}
      & \includegraphics[width=0.29\textwidth]{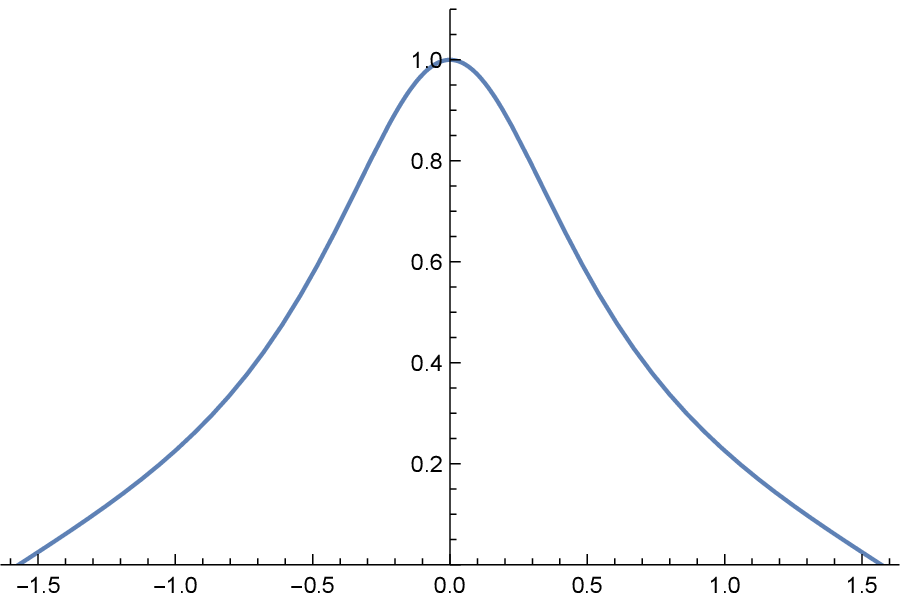} \\    
      {\footnotesize $\kappa = 4$}
      & \includegraphics[width=0.29\textwidth]{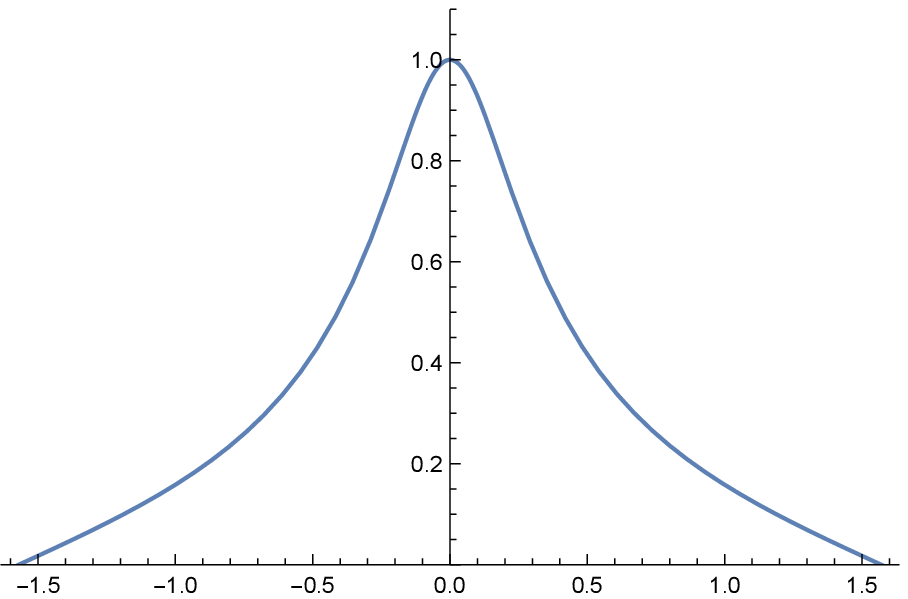}
      & \includegraphics[width=0.29\textwidth]{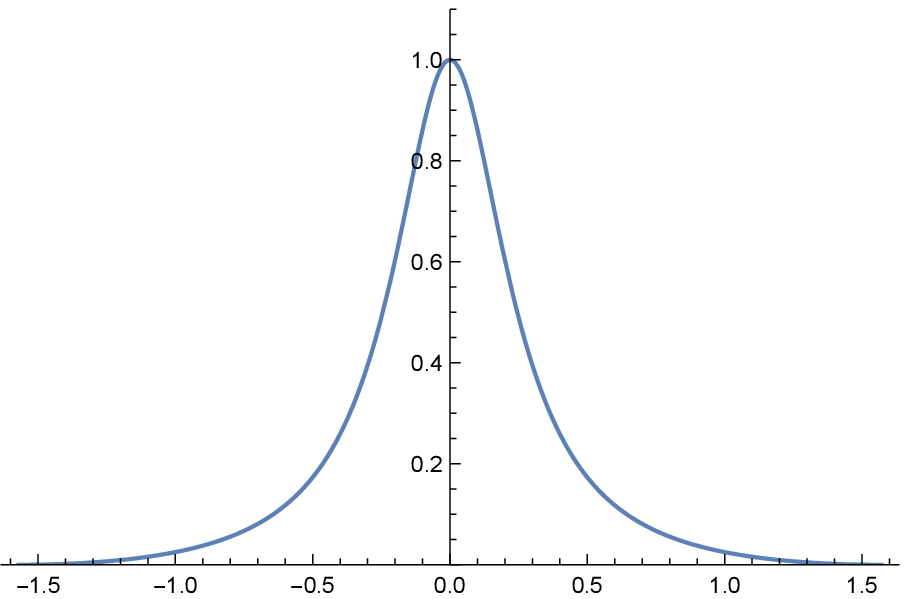}
      & \includegraphics[width=0.29\textwidth]{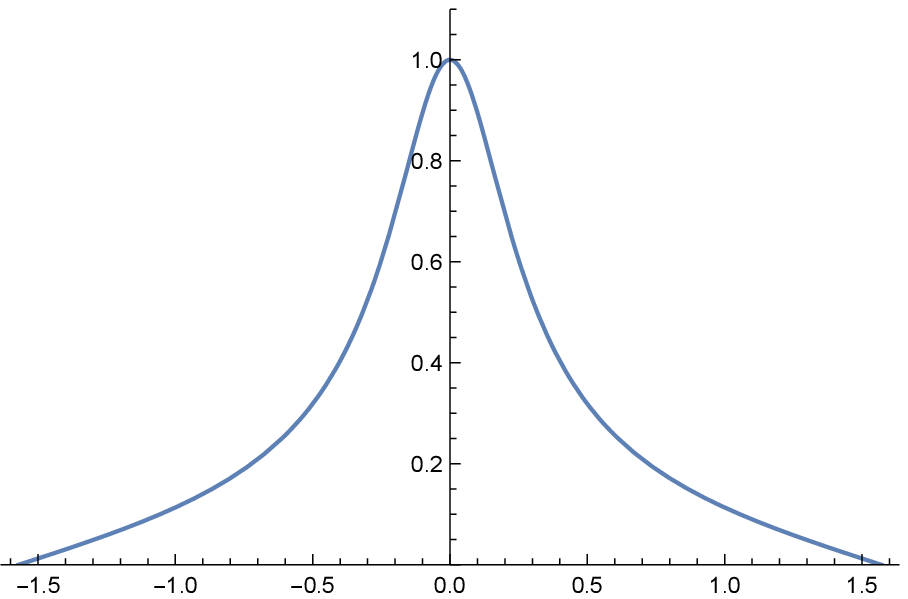} \\    
      {\footnotesize $\kappa = 8$}
      & \includegraphics[width=0.29\textwidth]{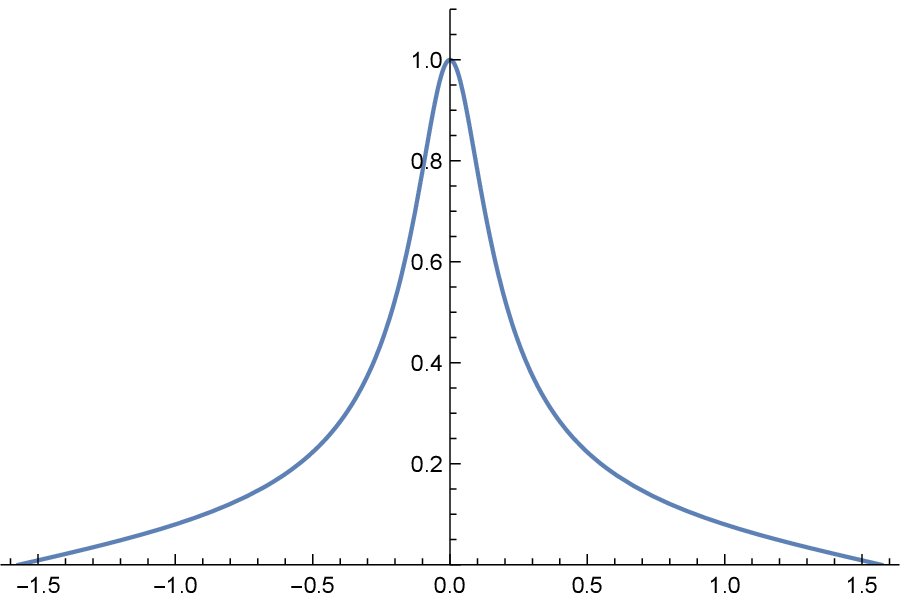}
      & \includegraphics[width=0.29\textwidth]{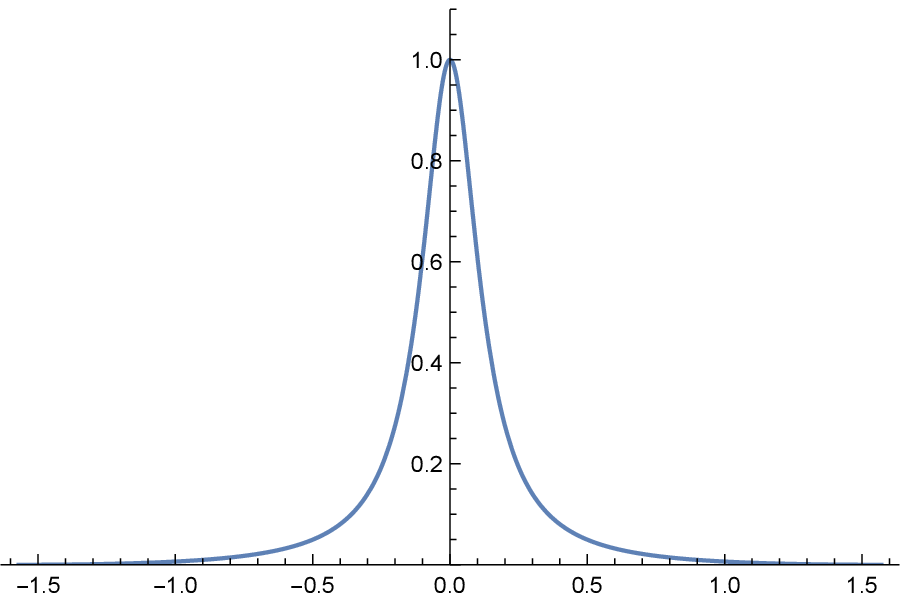}
      & \includegraphics[width=0.29\textwidth]{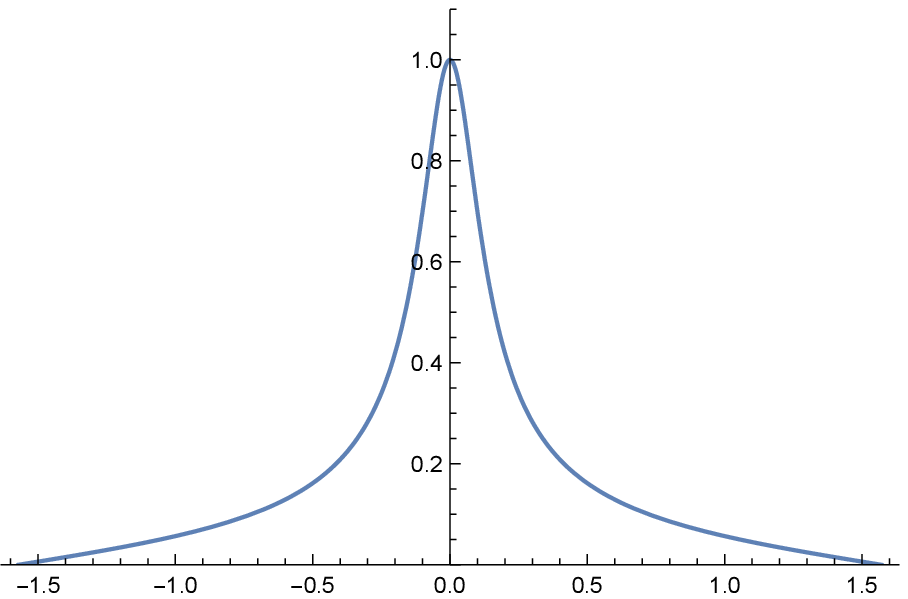} \\    
     \end{tabular}
  \end{center}
  \caption{Graphs of the orientation selectivity for {\em space-time separable
    spatio-temporal models\/} of (left column) simple cells in terms of first-order
    directional derivatives of affine Gaussian kernels combined with
    first-order temporal derivatives of temporal Gaussian kernels, (middle column) simple
    cells in terms of second-order directional derivatives of affine Gaussian
    kernels combined with second-order temporal derivatives of
    temporal Gaussian kernels
    and (right column) complex cells in terms of directional
    quasi-quadrature measures that combine the first- and second-order
    simple cell responses in a Euclidean way for $C_{\varphi} = 1/\sqrt{2}$,
    and $C_t = 1/\sqrt{2}$ shown for different values of the ratio
    $\kappa$ between the spatial scale parameters in the vertical
    {\em vs.\/}\ the horizontal directions. Observe how the degree of orientation
    selectivity varies strongly depending on the eccentricity
    $\epsilon = 1/\kappa$ of the receptive fields.
    (top row) Results for $\kappa = 1$.
    (second row) Results for $\kappa = 2$.
    (third row) Results for $\kappa = 4$.
    (bottom row) Results for $\kappa = 8$.
    (Horizontal axes: orientation $\theta \in [-\pi/2, \pi/2]$.
     Vertical axes: Amplitude of the receptive field response relative
     to the maximum response obtained for $\theta = 0$.)}
  \label{fig-ori-sel-spat-temp-sep-anal}
\end{figure*}

\subsection{Analysis for space-time separable models of
  spatio-temporal receptive fields}
\label{app-anal-space-time-sep-rf}

To investigate the directional selectivity of our spatio-temporal
models for simple and complex cells, we will analyze their response
properties to a moving sine wave of the form
\begin{multline}
  \label{eq-moving-sine-wave}
  f(x_1, x_2, t) = \\ 
  = \sin
     \left(
       \omega \cos (\theta) \, (x_1 - u_1 t) + \omega \sin (\theta) \, (x_2 -  u_2 t) + \beta
     \right),
\end{multline}
where we choose the velocity vector $(u_1, u_2)^T$ parallel to the inclination angle
$\theta$ of the grating according to $(u_1, u_2)^T = (u \cos \theta, u
\sin \theta)^T$, which, in turn, implies the form
\begin{multline}
  \label{eq-moving-sine-wave-spec}
  f(x_1, x_2, t) = 
  \sin
     \left(
       \omega \cos (\theta) \, x_1  + \omega \sin (\theta) \, x_2 -  u \, t + \beta
     \right).
\end{multline}
Let us initially perform such an analysis for space-time separable
models of simple and complex cells, in which the velocity vector
$v$ in the spatio-temporal receptive field models is set to zero.

For simplicity, we initially perform the theoretical analysis for
non-causal spatio-temporal receptive field models, where the temporal components are
given as scale-normalized temporal derivatives of 1-D temporal
Gaussian kernels.

In the following, we will name our models of spatio-temporal receptive
fields according to the orders of differentiation with respect to
space and time.

\subsubsection{First-order first-order simple cell}

Consider a space-time separable receptive field
corresponding to a {\em first-order\/} scale-normalized
Gaussian derivative with scale parameter $\sigma_1$ in the horizontal
$x_1$-direction, a zero-order Gaussian kernel with scale parameter
$\sigma_2$ in the vertical $x_2$-direction, and a {\em first-order\/}
scale-normalized Gaussian derivative with scale parameter $\sigma_t$
in the temporal direction, corresponding to $\varphi = 0$, $v = 0$,
$\Sigma _0= \diag(\sigma_1^2, \sigma_2^2)$,
$m = 1$ and $n = 1$ in (\ref{eq-spat-temp-RF-model-der-norm-caus}):
\begin{align}
  \begin{split}
    & T_{0,t,\norm}(x_1, x_2, t;\; \sigma_1, \sigma_2, \sigma_t) =
  \end{split}\nonumber\\
  \begin{split}
     & = \frac{\sigma_1 \sigma_t}{(2 \pi)^{3/2} \, \sigma_1 \sigma_2 \sigma_t} \,
            \partial_{x_1} \partial_{t}
              \left( e^{-x_1^2/2\sigma_1^2 - x_2^2/2 \sigma_2^2 - t^2/2\sigma_t^2} \right)
  \end{split}\nonumber\\
  \begin{split}
    & = \frac{x_1 t}{(2 \pi)^{3/2} \, \sigma_1^2 \sigma_2 \sigma_t^2} \,
               e^{-x_1^2/2\sigma_1^2 - x_2^2/2 \sigma_2^2 - t^2/2\sigma_t^2}.
  \end{split}
\end{align}
The corresponding receptive field response is then, after solving the
convolution integral in Mathematica,
\begin{align}
   \begin{split}
     L_{0,t,\norm}(x_1, x_2, t;\; \sigma_1, \sigma_2, \sigma_t) =
  \end{split}\nonumber\\
  \begin{split}
    & = \int_{\xi_1 = -\infty}^{\infty}  \int_{\xi_2 = -\infty}^{\infty} \int_{\zeta = -\infty}^{\infty}
             T_{0,t,\norm}(\xi_1, \xi_2, \zeta;\; \sigma_1, \sigma_2, \sigma_t)
  \end{split}\nonumber\\
  \begin{split}
    & \phantom{= = \int_{\xi_1 = -\infty}^{\infty}  \int_{\xi_2 = -\infty}^{\infty}}
             \times f(x_1 - \xi_1, x_2 - \xi_2, t - \zeta) \, d \xi_1 \xi_2 d\zeta
  \end{split}\nonumber\\
  \begin{split}
    & = \omega^2 \, \sigma_1 \, \sigma_t \, u \cos (\theta) \,
           e^{-\frac{1}{2} \omega^2 (\sigma_1^2 \cos^2 \theta + \sigma_2^2 \sin^2 \theta + \sigma_t^2 u^2)}
  \end{split}\nonumber\\
  \begin{split}
    \label{eq-L0t-space-time-sep-anal}
    & \phantom{= =}
           \times \sin
             \left(
                \omega \cos (\theta) \, x_1  + \omega \sin (\theta) \, x_2 -  u \, t + \beta
             \right),
   \end{split}         
\end{align}
{\em i.e.\/},\ a sine wave with amplitude
\begin{multline}
  A_{\varphi,t}(\theta, u, \omega;\; \sigma_1, \sigma_2, \sigma_t) = \\
  = \omega^2 \, \sigma_1 \, \sigma_t \, u \left| \cos \theta \right| \,
      e^{-\frac{1}{2} \omega^2 (\sigma_1^2 \cos^2 \theta + \sigma_2^2 \sin^2 \theta + \sigma_t^2 u^2)}.
\end{multline}
This expression first increases and then decreases with respect to
both the angular frequency $\omega$ and the velocity $u$ of the sine wave.
Selecting the values of $\hat{\omega}$ and $\hat{u}$ at which this expression
assumes its maximum over $\omega$ and $u$
\begin{equation}
  \hat{\omega}_{\varphi,t}
  = \frac{1}{\sigma_1 \sqrt{\cos^2 \theta + \kappa^2 \sin^2 \theta}},
\end{equation}
\begin{equation}
    \hat{u}_{\varphi,t}
    = \frac{\sigma_1}{\sigma_t} \sqrt{\cos^2 \theta + \kappa^2 \sin^2 \theta},
\end{equation}
and again reparameterizing the other spatial scale parameter
$\sigma_2$ as $\sigma_2 = \kappa \, \sigma_1$, gives
that the maximum amplitude measure over
spatial and temporal scales is
\begin{equation}
  A_{\varphi,t,\max}(\theta;\; \kappa)
  = \frac{\left| \cos \theta \right|}{e \sqrt{\cos ^2 \theta + \kappa ^2 \sin ^2\theta}}.
\end{equation}
Note that again this directional selectivity measure is independent of the
spatial scale parameter $\sigma_1$ as well as independent of the temporal scale
parameter $\sigma_t$, because of the
scale-invariant property of scale-normalized derivatives for
scale normalization power $\gamma = 1$. 

The left column in Figure~\ref{fig-ori-sel-spat-temp-sep-anal} shows
the result of plotting the measure $A_{\varphi,t,\max}(\theta;\; \kappa)$ of
the orientation selectivity as function of the inclination angle $\theta$
for a few values of the scale parameter ratio $\kappa$, with the values rescaled
such that the peak value for each graph is equal to 1. As we can see
from the graphs, as for the previous purely spatial model of the
receptive fields, the degree of orientation selectivity increases
strongly with the value of $\kappa$.

\subsubsection{First-order second-order simple cell}

Consider a space-time separable receptive field
corresponding to a {\em first-order\/} scale-normalized
Gaussian derivative with scale parameter $\sigma_1$ in the horizontal
$x_1$-direction, a zero-order Gaussian kernel with scale parameter
$\sigma_2$ in the vertical $x_2$-direction, and a {\em second-order\/}
scale-normalized Gaussian derivative with scale parameter $\sigma_t$
in the temporal direction, corresponding to $\varphi = 0$, $v = 0$,
$\Sigma_0 = \diag(\sigma_1^2, \sigma_2^2)$,
$m = 1$ and $n = 2$ in (\ref{eq-spat-temp-RF-model-der-norm-caus}):
\begin{align}
  \begin{split}
    & T_{0,tt,\norm}(x_1, x_2, t;\; \sigma_1, \sigma_2, \sigma_t) =
  \end{split}\nonumber\\
  \begin{split}
     & = \frac{\sigma_1 \sigma_t^2}{(2 \pi)^{3/2} \, \sigma_1 \sigma_2 \sigma_t} \,
            \partial_{x_1} \partial_{tt}
              \left( e^{-x_1^2/2\sigma_1^2 - x_2^2/2 \sigma_2^2 - t^2/2\sigma_t^2} \right)
  \end{split}\nonumber\\
  \begin{split}
    & = - \frac{x_1 (t^2 - \sigma_t^2)}{(2 \pi)^{3/2} \, \sigma_1^2 \sigma_2 \sigma_t^3} \,
               e^{-x_1^2/2\sigma_1^2 - x_2^2/2 \sigma_2^2 - t^2/2\sigma_t^2}.
  \end{split}
\end{align}
The corresponding receptive field response is then, after solving the
convolution integral in Mathematica,
\begin{align}
   \begin{split}
     L_{0,tt,\norm}(x_1, x_2, t;\; \sigma_1, \sigma_2, \sigma_t) =
  \end{split}\nonumber\\
  \begin{split}
    & = \int_{\xi_1 = -\infty}^{\infty}  \int_{\xi_2 = -\infty}^{\infty} \int_{\zeta = -\infty}^{\infty}
             T_{0,tt,\norm}(\xi_1, \xi_2, \zeta;\; \sigma_1, \sigma_2, \sigma_t)
  \end{split}\nonumber\\
  \begin{split}
    & \phantom{= = \int_{\xi_1 = -\infty}^{\infty}  \int_{\xi_2 = -\infty}^{\infty}}
             \times f(x_1 - \xi_1, x_2 - \xi_2, t - \zeta) \, d \xi_1 \xi_2 d\zeta
  \end{split}\nonumber\\
  \begin{split}
    & = - \omega^3 \, \sigma_1 \, \sigma_t^2 \, u^2 \cos (\theta) \,
           e^{-\frac{1}{2} \omega^2 (\sigma_1^2 \cos^2 \theta + \sigma_2^2 \sin^2 \theta + \sigma_t^2 u^2)}
  \end{split}\nonumber\\
  \begin{split}
    \label{eq-L0tt-space-time-sep-anal}
    & \phantom{= =}
           \times \cos
             \left(
                \omega \cos (\theta) \, x_1  + \omega \sin (\theta) \, x_2 -  u \, t + \beta
             \right),
   \end{split}         
\end{align}
{\em i.e.\/},\ a cosine wave with amplitude
\begin{multline}
  A_{\varphi,tt}(\theta, u, \omega;\; \sigma_1, \sigma_2, \sigma_t) = \\
  = \omega^3 \, \sigma_1 \, \sigma_t^2 \, u^2 \left| \cos \theta \right| \,
      e^{-\frac{1}{2} \omega^2 (\sigma_1^2 \cos^2 \theta + \sigma_2^2 \sin^2 \theta + \sigma_t^2 u^2)}.
\end{multline}
This entity assumes its maximum over the angular frequency $\omega$ and the
image velocity $u$ at
\begin{equation}
  \hat{\omega}_{\varphi,tt}
  = \frac{1}{\sigma_1 \sqrt{\cos^2 \theta + \kappa^2 \sin^2 \theta}},
\end{equation}
\begin{equation}
    \hat{u}_{\varphi,tt}
    = \frac{\sqrt{2} \sigma_1}{\sigma_t} \sqrt{\cos^2 \theta + \kappa^2 \sin^2 \theta},
\end{equation}
and again reparameterizing the other spatial scale parameter
$\sigma_2$ as $\sigma_2 = \kappa \, \sigma_1$, gives
that the maximum amplitude measure over
spatial and temporal scales is
\begin{equation}
  A_{\varphi,tt,\max}(\theta;\; \kappa) 
  = \frac{2 \left| \cos \theta \right|}{e^{3/2} \sqrt{\cos ^2 \theta + \kappa ^2 \sin ^2\theta}},
\end{equation}
{\em i.e.\/}, of a similar form as the previous measure 
$A_{\varphi,tt,\max}(\theta;\; \kappa)$, while being multiplied by
another constant.

\subsubsection{Second-order first-order simple cell}

Consider a space-time separable receptive field
corresponding to a {\em second-order\/} scale-normalized
Gaussian derivative with scale parameter $\sigma_1$ in the horizontal
$x_1$-direction, a zero-order Gaussian kernel with scale parameter
$\sigma_2$ in the vertical $x_2$-direction, and a {\em first-order\/}
scale-normalized Gaussian derivative with scale parameter $\sigma_t$
in the temporal direction, corresponding to $\varphi = 0$, $v = 0$,
$\Sigma_0 = \diag(\sigma_1^2, \sigma_2^2)$,
$m = 2$ and $n = 1$ in (\ref{eq-spat-temp-RF-model-der-norm-caus}):
\begin{align}
  \begin{split}
    & T_{00,t,\norm}(x_1, x_2, t;\; \sigma_1, \sigma_2, \sigma_t) =
  \end{split}\nonumber\\
  \begin{split}
     & = \frac{\sigma_1^2 \sigma_t}{(2 \pi)^{3/2} \, \sigma_1 \sigma_2 \sigma_t} \,
            \partial_{x_1 x_1} \partial_{t}
              \left( e^{-x_1^2/2\sigma_1^2 - x_2^2/2 \sigma_2^2 - t^2/2\sigma_t^2} \right)
  \end{split}\nonumber\\
  \begin{split}
    & = - \frac{(x_1^2 - \sigma_1^2) t}{(2 \pi)^{3/2} \, \sigma_1^3 \sigma_2 \sigma_t^2} \,
               e^{-x_1^2/2\sigma_1^2 - x_2^2/2 \sigma_2^2 - t^2/2\sigma_t^2}.
  \end{split}
\end{align}
The corresponding receptive field response is then, after solving the
convolution integral in Mathematica,
\begin{align}
   \begin{split}
     L_{00,t,\norm}(x_1, x_2, t;\; \sigma_1, \sigma_2, \sigma_t) =
  \end{split}\nonumber\\
  \begin{split}
    & = \int_{\xi_1 = -\infty}^{\infty}  \int_{\xi_2 = -\infty}^{\infty} \int_{\zeta = -\infty}^{\infty}
             T_{00,t,\norm}(\xi_1, \xi_2, \zeta;\; \sigma_1, \sigma_2, \sigma_t)
  \end{split}\nonumber\\
  \begin{split}
    & \phantom{= = \int_{\xi_1 = -\infty}^{\infty}  \int_{\xi_2 = -\infty}^{\infty}}
             \times f(x_1 - \xi_1, x_2 - \xi_2, t - \zeta) \, d \xi_1 \xi_2 d\zeta
  \end{split}\nonumber\\
  \begin{split}
    & = -\omega^3 \, \sigma_1^2 \, \sigma_t \, u \cos^2 (\theta) \,
           e^{-\frac{1}{2} \omega^2 (\sigma_1^2 \cos^2 \theta + \sigma_2^2 \sin^2 \theta + \sigma_t^2 u^2)}
  \end{split}\nonumber\\
  \begin{split}
    \label{eq-L00t-space-time-sep-anal}
    & \phantom{= =}
           \times \cos
             \left(
                \omega \cos (\theta) \, x_1  + \omega \sin (\theta) \, x_2 -  u \, t + \beta
             \right),
   \end{split}         
\end{align}
{\em i.e.\/},\ a cosine wave with amplitude
\begin{multline}
  A_{\varphi\varphi,t}(\theta, u, \omega;\; \sigma_1, \sigma_2, \sigma_t) = \\
  = \omega^3 \, \sigma_1^2 \, \sigma_t \, u \cos^2 \theta \,
      e^{-\frac{1}{2} \omega^2 (\sigma_1^2 \cos^2 \theta + \sigma_2^2 \sin^2 \theta + \sigma_t^2 u^2)}.
\end{multline}
This entity assumes its maximum over spatial scale $\sigma_1$ and over
temporal scale $\sigma_t$ at
\begin{equation}
   \hat{\omega}_{\varphi\varphi,t}
   = \frac{\sqrt{2}}{\sigma_1 \sqrt{\cos^2 \theta + \kappa^2 \sin^2 \theta}},
\end{equation}
\begin{equation}
    \hat{u}_{\varphi\varphi,t}
    = \frac{\sigma_1}{\sqrt{2}\sigma_t} \sqrt{\cos^2 \theta + \kappa^2 \sin^2 \theta},
\end{equation}
and again reparameterizing the other spatial scale parameter
$\sigma_2$ as $\sigma_2 = \kappa \, \sigma_1$, gives
that the maximum amplitude measure over
spatial and temporal scales is
\begin{equation}
  A_{\varphi\varphi,t,\max}(\theta;\; \kappa)
  = \frac{2 \cos^2 \theta}
             {e^{3/2} \left( \cos^2 \theta + \kappa ^2 \sin^2\theta \right)}.
\end{equation}

\subsubsection{Second-order second-order simple cell}

Consider a space-time separable receptive field
corresponding to a {\em second-order\/} scale-normalized
Gaussian derivative with scale parameter $\sigma_1$ in the horizontal
$x_1$-direction, a zero-order Gaussian kernel with scale parameter
$\sigma_2$ in the vertical $x_2$-direction, and a {\em second-order\/}
scale-normalized Gaussian derivative with scale parameter $\sigma_t$
in the temporal direction, corresponding to $\varphi = 0$, $v = 0$,
$\Sigma_0 = \diag(\sigma_1^2, \sigma_2^2)$,
$m = 2$ and $n = 2$ in (\ref{eq-spat-temp-RF-model-der-norm-caus}):
\begin{align}
  \begin{split}
    & T_{00,tt,\norm}(x_1, x_2, t;\; \sigma_1, \sigma_2, \sigma_t) =
  \end{split}\nonumber\\
  \begin{split}
     & = \frac{\sigma_1^2 \sigma_t^2}{(2 \pi)^{3/2} \, \sigma_1 \sigma_2 \sigma_t} \,
            \partial_{x_1 x_1} \partial_{tt}
              \left( e^{-x_1^2/2\sigma_1^2 - x_2^2/2 \sigma_2^2 - t^2/2\sigma_t^2} \right)
  \end{split}\nonumber\\
  \begin{split}
    & = \frac{(x_1^2 - \sigma_1^2) (t^2 - \sigma_t^2)}{(2 \pi)^{3/2} \, \sigma_1^3 \sigma_2 \sigma_t^3} \,
               e^{-x_1^2/2\sigma_1^2 - x_2^2/2 \sigma_2^2 - t^2/2\sigma_t^2}.
  \end{split}
\end{align}
The corresponding receptive field response is then, after solving the
convolution integral in Mathematica,
\begin{align}
   \begin{split}
     L_{00,tt,\norm}(x_1, x_2, t;\; \sigma_1, \sigma_2, \sigma_t) =
  \end{split}\nonumber\\
  \begin{split}
    & = \int_{\xi_1 = -\infty}^{\infty}  \int_{\xi_2 = -\infty}^{\infty} \int_{\zeta = -\infty}^{\infty}
             T_{00,tt,\norm}(\xi_1, \xi_2, \zeta;\; \sigma_1, \sigma_2, \sigma_t)
  \end{split}\nonumber\\
  \begin{split}
    & \phantom{= = \int_{\xi_1 = -\infty}^{\infty}  \int_{\xi_2 = -\infty}^{\infty}}
             \times f(x_1 - \xi_1, x_2 - \xi_2, t - \zeta) \, d \xi_1 \xi_2 d\zeta
  \end{split}\nonumber\\
  \begin{split}
    & = \omega^4 \, \sigma_1^2 \, \sigma_t^2 \, u^2 \cos^2 (\theta) \,
           e^{-\frac{1}{2} \omega^2 (\sigma_1^2 \cos^2 \theta + \sigma_2^2 \sin^2 \theta + \sigma_t^2 u^2)}
  \end{split}\nonumber\\
  \begin{split}
    \label{eq-L00tt-space-time-sep-anal}
    & \phantom{= =}
           \times \sin
             \left(
                \omega \cos (\theta) \, x_1  + \omega \sin (\theta) \, x_2 -  u \, t + \beta
             \right),
   \end{split}         
\end{align}
{\em i.e.\/},\ a sine wave with amplitude
\begin{multline}
  A_{\varphi\varphi,tt}(\theta, u, \omega;\; \sigma_1, \sigma_2, \sigma_t) = \\
  = \omega^4 \, \sigma_1^2 \, \sigma_t^2 \, u^2 \cos^2 \theta \,
      e^{-\frac{1}{2} \omega^2 (\sigma_1^2 \cos^2 \theta + \sigma_2^2 \sin^2 \theta + \sigma_t^2 u^2)}.
\end{multline}
This entity assumes its maximum over spatial scale $\sigma_1$ and over
temporal scale $\sigma_t$ at
\begin{equation}
   \hat{\omega}_{\varphi\varphi,tt}
   = \frac{\sqrt{2}}{\sigma_1 \sqrt{\cos^2 \theta + \kappa^2 \sin^2 \theta}},
\end{equation}
\begin{equation}
    \hat{u}_{\varphi\varphi,tt}
    = \frac{\sigma_1}{\sigma_t} \sqrt{\cos^2 \theta + \kappa^2 \sin^2 \theta},
\end{equation}
and again reparameterizing the other spatial scale parameter
$\sigma_2$ as $\sigma_2 = \kappa \, \sigma_1$, gives
that the maximum amplitude measure over
spatial and temporal scales is
\begin{equation}
  A_{\varphi\varphi,tt,\max}(\theta;\; \kappa)
  = \frac{4 \cos^2 \theta}
             {e^2 \left( \cos^2 \theta + \kappa^2 \sin^2 \theta \right)}.
\end{equation}
The middle column in Figure~\ref{fig-ori-sel-spat-temp-sep-anal}
shows the result of plotting the measure $A_{\varphi\varphi,tt,\max}(\theta;\; \kappa)$
of the orientation selectivity as function of the inclination angle
$\theta$ for a few values of the scale parameter ratio $\kappa$, with the values rescaled
such that the peak value is equal to 1. Again, the
degree of orientation selectivity increases strongly with the value of
$\kappa$, as for the first-order first-order model of a simple cell.

\subsubsection{Complex cell}

To model the spatio-temporal response of a complex cell according to
the directional sensitive spatio-temporal quasi-quadrature measure
(\ref{eq-spat-temp-RF-model-der-norm-caus}) based on space-time
separable spatio-temporal receptive fields,
we combine the responses of
the first-order first-order simple cell (\ref{eq-L0t-space-time-sep-anal}),
the first-order second-order cell (\ref{eq-L0tt-space-time-sep-anal}),
the second-order first-order simple cell (\ref{eq-L00t-space-time-sep-anal})
and the second-order second order cell (\ref{eq-L00tt-space-time-sep-anal})
for $v = 0$, $\Gamma_{\varphi} = 0$ and $\Gamma_t = 0$
according to 
\begin{align}
  \begin{split}
    & ({\cal Q}_{0,\sep,\norm} L)^2
  \end{split}\nonumber\\
  \begin{split}  
    & = L_{0,t,\norm}^2 + C_{\varphi} \, L_{00,t,norm}^2 +
 \end{split}\nonumber\\
  \begin{split}  
  \label{eq-quasi-quad-dir-spat-temp-sep-anal}
    & \phantom{=}
    + C_t \left( L_{0,tt,\norm}^2 + C_{\varphi} \, L_{00,tt,\norm}^2 \right).
  \end{split}
\end{align}
Selecting the angular frequency $\hat{\omega}$ as the geometric
average of the angular frequencies where the above spatio-temporal simple
cell models assume their maximum amplitude responses over spatial scales
\begin{equation}
  \hat{\omega}_{\cal Q}
  = \sqrt{\hat{\omega}_{\varphi,t}  \, \hat{\omega}_{\varphi,tt} \,
               \hat{\omega}_{\varphi\varphi,t} \, \hat{\omega}_{\varphi\varphi,tt}}
  = \frac{\sqrt[4]{2}}{\sigma_1 \sqrt{\cos^2 \theta + \kappa^2 \sin^2 \theta}},
\end{equation}
and selecting the image velocity $\hat{u}$ of the sine wave as the geometric
average of the image velocities where the above spatio-temporal simple
cell models assume their maximum amplitude responses over image
velocities
\begin{equation}
  \hat{u}_{\cal Q}
  = \sqrt{\hat{u}_{\varphi,t}  \, \hat{u}_{\varphi,tt} \,
               \hat{u}_{\varphi\varphi,t} \, \hat{u}_{\varphi\varphi,tt}}
  = \frac{\sigma_1}{\sigma_t} \sqrt{\cos^2 \theta + \kappa^2 \sin^2 \theta},
\end{equation}
as well as choosing the spatial and temporal weighting factors
$C_{\varphi}$ and $C_t$ between first- and second-order information as
$C_{\varphi} = 1/\sqrt{2}$ and $C_t = 1/\sqrt{2}$ according to
(Lindeberg \citeyear{Lin18-SIIMS}),
then implies that the spatio-temporal quasi-quadrature measure assumes
the form
\begin{equation}
  \label{eq-sep-compl-cell}
  A_{{\cal Q},\sep}(\theta;\; \kappa)
  = \frac{e^{-\sqrt{2}} \left| \cos \theta \right| \,
              \sqrt{2 + \kappa^2 + (2 - \kappa^2) \cos 2 \theta}}
              {\cos ^2\theta + \kappa ^2 \sin^2\theta}.
\end{equation}
Note that this expression is independent of both the spatial scale
parameter $\sigma_1$ and the temporal scale parameter $\sigma_t$,
because of the scale-invariant properties of scale-normalized
derivatives for scale normalization parameter $\gamma = 1$.
Moreover, this expression is also independent of the phase of the
signal, as determined by the spatial coordinates $x_1$ and $x_2$,
the time moment $t$ and the phase angle $\beta$.

The right column in Figure~\ref{fig-ori-sel-spat-temp-sep-anal} shows the
result of plotting the measure
$A_{{\cal Q},\sep}(\theta;\; \kappa)$ of the orientation
selectivity as function of the inclination angle $\theta$ for a few
values of the scale parameter ratio $\kappa$, with the values rescaled
such that the peak value for each graph is equal to 1. As can be seen from the
graphs, the degree of orientation selectivity increases strongly with
the value of $\kappa$ also for this spatio-temporal model of a complex cell, and in a
qualitatively similar way as for the simple cell models, regarding
both the purely spatial as well as the joint spatio-temporal models of the
simple cells.

\subsection{Affine transformation property for receptive fields
  according to a generalized affine Gabor model for simple cells}
\label{app-aff-transf-prop-gabor}

Given any positive definite spatial covariance matrix $\Sigma$ and any
angular frequency vector $\omega = (\omega_1, \omega_2)^T$, consider
a generalization of the affine Gabor model
(\ref{eq-aff-gabor-mod-even}) to a vector-valued affine
Gabor function of the form
\begin{equation}
  \label{def-genuine-aff-gabor-func}
  G(x;\; \Sigma, \omega) = g(x;\; \Sigma, \omega) \, e^{i \omega^T x},
\end{equation}
where $g(x;\; \Sigma, \omega)$ denotes a (here) two-dimensional
Gaussian kernel
\begin{equation}
  g(x;\; \Sigma)
  = \frac{1}{2 \pi \sqrt{\det \Sigma}} \, e^{-x^T \Sigma^{-1} x/2}.
\end{equation}
In the case when the angular frequency vector $\omega$ is parallel to
one of the eigenvectors of the spatial covariance matrix $\Sigma$,
this form of affine Gabor function spans a similar variability as the
previously treated affine Gabor model in
(\ref{eq-aff-gabor-mod-even}).
To make it possible to define a representation that is closed under
general affine transformations (affine covariance), we do, however,
here, relax the assumption that the orientation of the angular
frequency vector $\omega$ should be related to the dominant
orientations defined by the eigendirections of the
spatial covariance matrix $\Sigma$.

Let us now assume that we have two spatial images $f(x)$ and $f'(x')$
that are related according to a spatial affine transformation,
such that
\begin{equation}
  f'(x') = f(x)
\end{equation}
for
\begin{equation}
  \label{eq-def-aff-transf-gabor-cov-prop}
  x' = A \, x.
\end{equation}
Let us furthermore define affine Gabor representations over the two spatial
domains according to
\begin{align}
  \begin{split}
    L(x;\, \Sigma, \omega)
    & = (G(\cdot;\; \Sigma, \omega) * f(\cdot))(x;\; \Sigma, \omega)
  \end{split}\nonumber\\
  \begin{split}
    \label{eq-def-gabor-transf-conv-int1-proof}
    & = \int_{\xi \in \bbbr^2} G(\xi;\; \Sigma, \omega) \, f(x - \xi) \, d\xi,
   \end{split}\\
  \begin{split}
    L'(x';\, \Sigma', \omega')
    & = (G(\cdot;\; \Sigma', \omega') * f'(\cdot))(x';\; \Sigma', \omega')
  \end{split}\nonumber\\
  \begin{split}
    \label{eq-def-gabor-transf-conv-int2-proof}    
    & = \int_{\xi' \in \bbbr^2} G(\xi';\; \Sigma', \omega') \, f(x' - \xi') \, d\xi'.
   \end{split}
\end{align}
We would now like to investigate if the affine Gabor representations over the
two domains could be related to each other. To express such a relation,
we will perform the change of variables
\begin{equation}
  \label{eq-def-aff-transf-gabor-cov-prop-proof}
  \xi' = A \, \xi,
\end{equation}
which gives
\begin{equation}
  d\xi' = |\det A| \, d\xi,
\end{equation}
in the convolution integrals.

\medskip
\noindent
{\bf Step I:}
Let us first consider how the Gaussian function $g'(x';\, \Sigma')$ in
the Gabor function $G(x';\; \Sigma', \omega')$ transforms under the
change of variables. If we, in analogy with the transformation
property of the spatial covariance matrix in the affine Gaussian
derivative model for visual receptive fields
(Equation~(28) in Lindeberg \citeyear{Lin23-FrontCompNeuroSci})
assume that
\begin{equation}
  \Sigma' = A \, \Sigma \, A^T,
\end{equation}
which gives
\begin{equation}
  {\Sigma' }^{-1}= A^{-T} \, \Sigma^{-1} \, A^{-1}
\end{equation}
and
\begin{equation}
  \det \Sigma' = | \det A |^2 \, \det \Sigma,
\end{equation}
then we obtain
\begin{align}
  \begin{split}
    g(\xi';\; \Sigma')
    = \frac{1}{2 \pi \sqrt{\det \Sigma'}} \, e^{-{\xi'}^T {\Sigma'}^{-1} \xi'/2}
  \end{split}\nonumber\\
   \begin{split}
     = \frac{1}{2 \pi | \det A | \sqrt{\det \Sigma}} \,
     e^{-{A \xi}^T A^{-T} \, \Sigma^{-1} \, A^{-1} A \xi/2}
  \end{split}\\
  \begin{split}
    \label{eq-transf-prop-aff-gauss-gabor-proof}
     = \frac{1}{| \det A |} \, \frac{1}{2 \pi \sqrt{\det \Sigma}} \,
     e^{-\xi^T\, \Sigma^{-1} \xi/2}
    = \frac{1}{| \det A |} \, g(\xi;\; \Sigma).
  \end{split}
\end{align}

\medskip
\noindent
{\bf Step II:} If we further set
\begin{equation}
  \omega' = A^{-T} \omega,
\end{equation}
and combine with the transformation property
(\ref{eq-transf-prop-aff-gauss-gabor-proof}) of the affine Gaussian
kernel, then we have that the Gabor functions over the two domains are
related according to
\begin{align}
  \begin{split}
    G(\xi';\; \Sigma', \omega')
    = g(\xi';\; \Sigma') \, e^{i {\omega'}^T \xi'}
    = \frac{1}{| \det A |}  \, g(\xi;\; \Sigma) \, e^{i (A^{-T} \omega)^T A \xi}
  \end{split}\nonumber\\
  \begin{split}
    \label{eq-transf-prop-gabor-func}
    = \frac{1}{| \det A |}  \, g(\xi;\; \Sigma) \, e^{i \omega^T \xi}
    = \frac{1}{| \det A |} \, G(\xi;\; \Sigma, \omega).
  \end{split}
\end{align}

\medskip
\noindent
{\bf Step III:} Applied to the convolution integrals
(\ref{eq-def-gabor-transf-conv-int1-proof}) and
(\ref{eq-def-gabor-transf-conv-int2-proof}), the transformation
property (\ref{eq-transf-prop-gabor-func}) of the genuinely affine
Gabor function (\ref{def-genuine-aff-gabor-func})
thus implies that
\begin{align}
 \begin{split}
    L'(x';\, \Sigma', \omega')
     = (G(\cdot;\; \Sigma', \omega') * f'(\cdot))(x';\; \Sigma', \omega')
  \end{split}\nonumber\\
  \begin{split}
     = \int_{\xi' \in \bbbr^2} G(\xi';\; \Sigma', \omega') \, f(x' - \xi') \, d\xi'.
   \end{split}\nonumber\\
 \begin{split}
     = \int_{\xi \in \bbbr^2} \frac{1}{| \det A |}  \, G(\xi;\; \Sigma, \omega) \,
     f(x - \xi) \, | \det A | \, d\xi,
   \end{split}\nonumber\\
  \begin{split}
     = (G(\cdot;\; \Sigma, \omega) * f(\cdot))(x;\; \Sigma, \omega) = L(x;\, \Sigma, \omega).
  \end{split}
\end{align}

\medskip
\noindent
{\bf Summary of main result:} To conclude, we have shown that if two
images $f(x)$ and $f'(x')$ are related according to an affine transformation
\begin{equation}
  f'(x') = f(x)
\end{equation}
where
\begin{equation}
  x' = A \, x,
\end{equation}
then provided that affine Gabor representations are defined over the
two spatial domains according to
\begin{align}
  \begin{split}
    L(x;\, \Sigma, \omega)
    & = (G(\cdot;\; \Sigma, \omega) * f(\cdot))(x;\; \Sigma, \omega),
  \end{split}\\
  \begin{split}
    L'(x';\, \Sigma', \omega')
    & = (G(\cdot;\; \Sigma', \omega') * f'(\cdot))(x';\; \Sigma', \omega'),
  \end{split}
\end{align}
it then follows that these (genuinely) affine Gabor representations are
equal under a similar spatial affine transformation
\begin{equation}
  L'(x';\, \Sigma', \omega') = L(x;\, \Sigma, \omega),
\end{equation}
provided that the values of the receptive field parameters
$\Sigma$, $\Sigma'$, $\omega$ and $\omega'$ are appropriately matched
according to
\begin{align}
  \begin{split}
     \Sigma' = A \, \Sigma \, A^T,
   \end{split}\\
  \begin{split}
      \omega' = A^{-T} \omega.
   \end{split}
\end{align}
This affine transformation property constitutes a genuine affine
covariance property of the (here) generalized family of affine Gabor functions
according to
\begin{equation}
  G(x;\; \Sigma, \omega) = g(x;\; \Sigma, \omega) \, e^{i \omega^T x}.
\end{equation}
In this context, it should, however, be noted that restricted affine
Gabor model (\ref{eq-aff-gabor-mod-even}), where the angular frequency
vector is restricted to be parallel with one of the eigendirections of
the spatial covariance matrix $\Sigma$, cannot, however, be expected
to be covariant over the complete group of spatial affine
transformations. That restricted affine Gabor model will, however, be
covariant under similarity transformations (uniform scaling
transformations and spatial rotations) as well as to stretching
transformations in the orientation of the angular frequency vector.

\bibliographystyle{abbrvnat}

{\footnotesize
\bibliography{defs,tlmac}

\begin{thebibliography}{115}
\providecommand{\natexlab}[1]{#1}
\providecommand{\url}[1]{\texttt{#1}}
\expandafter\ifx\csname urlstyle\endcsname\relax
  \providecommand{\doi}[1]{doi: #1}\else
  \providecommand{\doi}{doi: \begingroup \urlstyle{rm}\Url}\fi

\bibitem[Abballe and Asari(2022)]{AbbAsa22-PONE}
L.~Abballe and H.~Asari.
\newblock Natural image statistics for mouse vision.
\newblock \emph{PLoS ONE}, 17\penalty0 (1):\penalty0 e0262763, 2022.

\bibitem[Adelson and Bergen(1985)]{AdeBer85-JOSA}
E.~Adelson and J.~Bergen.
\newblock Spatiotemporal energy models for the perception of motion.
\newblock \emph{Journal of Optical Society of America}, A~2:\penalty0 284--299,
  1985.

\bibitem[Albright(1984)]{Alb84-JNeuroPhys}
T.~D. Albright.
\newblock Direction and orientation selectivity of neurons in visual area {MT}
  of the macaque.
\newblock \emph{Journal of Neurophysiology}, 52\penalty0 (6):\penalty0
  1106--1130, 1984.

\bibitem[Almasi et~al.(2020)Almasi, Meffin, Cloherty, Wong, Yunzab, and
  Ibbotson]{AlmMefCloWonYunIbb20-CerCort}
A.~Almasi, H.~Meffin, S.~L. Cloherty, Y.~Wong, M.~Yunzab, and M.~R. Ibbotson.
\newblock Mechanisms of feature selectivity and invariance in primary visual
  cortex.
\newblock \emph{Cerebral Cortex}, 30\penalty0 (9):\penalty0 5067--5087, 2020.

\bibitem[Alonso and Martinez(1998)]{AloMar98-NatNeurSci}
J.-M. Alonso and L.~M. Martinez.
\newblock Functional connectivity between simple cells and complex cells in cat
  striate cortex.
\newblock \emph{Nature Neuroscience}, 1\penalty0 (5):\penalty0 395--403, 1998.

\bibitem[Berkes and Wiskott(2005)]{BerWis05-JVis}
P.~Berkes and L.~Wiskott.
\newblock Slow feature analysis yields a rich repertoire of complex cell
  properties.
\newblock \emph{Journal of Vision}, 5\penalty0 (6):\penalty0 579--602, 2005.

\bibitem[Blasdel(1992)]{Bla92-JNeuroSci}
G.~G. Blasdel.
\newblock Orientation selectivity, preference and continuity in monkey striate
  cortex.
\newblock \emph{Journal of Neuroscience}, 12\penalty0 (8):\penalty0 3139--3161,
  1992.

\bibitem[Bonhoeffer and Grinvald(1991)]{BonGri91-Nature}
T.~Bonhoeffer and A.~Grinvald.
\newblock Iso-orientation domains in cat visual cortex are arranged in
  pinwheel-like patterns.
\newblock \emph{Nature}, 353:\penalty0 429--431, 1991.

\bibitem[Bracewell(1999)]{Bra99}
R.~N. Bracewell.
\newblock \emph{The Fourier Transform and its Applications}.
\newblock McGraw-Hill, New York, 1999.
\newblock 3rd edition.

\bibitem[Burge(2020)]{Bur20-AnnRevVisSci}
J.~Burge.
\newblock Image-computable ideal observers for tasks with natural stimuli.
\newblock \emph{Annual Review of Vision Science}, 6:\penalty0 491--517, 2020.

\bibitem[Carandini(2006)]{Car06-JPhys}
M.~Carandini.
\newblock What simple and complex cells compute.
\newblock \emph{The Journal of Physiology}, 577\penalty0 (2):\penalty0
  463--466, 2006.

\bibitem[Carandini and Ringach(1997)]{CarRin97-VisRes}
M.~Carandini and D.~L. Ringach.
\newblock Predictions of a recurrent model of orientation selectivity.
\newblock \emph{Vision Research}, 37\penalty0 (21):\penalty0 3061--3071, 1997.

\bibitem[Cogno and Mato(2015)]{GonMat15-FrontNeurCirc}
S.~G. Cogno and G.~Mato.
\newblock The effect of synaptic plasticity on orientation selectivity in a
  balanced model of primary visual cortex.
\newblock \emph{Frontiers in Neural Circuits}, 9:\penalty0 42, 2015.

\bibitem[Conway and Livingstone(2006)]{ConLiv06-JNeurSci}
B.~R. Conway and M.~S. Livingstone.
\newblock Spatial and temporal properties of cone signals in alert macaque
  primary visual cortex.
\newblock \emph{Journal of Neuroscience}, 26\penalty0 (42):\penalty0
  10826--10846, 2006.

\bibitem[De and Horwitz(2021)]{DeHor21-JNPhys}
A.~De and G.~D. Horwitz.
\newblock Spatial receptive field structure of double-opponent cells in macaque
  {V1}.
\newblock \emph{Journal of Neurophysiology}, 125\penalty0 (3):\penalty0
  843--857, 2021.

\bibitem[DeAngelis and Anzai(2004)]{deAngAnz04-VisNeuroSci}
G.~C. DeAngelis and A.~Anzai.
\newblock A modern view of the classical receptive field: Linear and non-linear
  spatio-temporal processing by {V1} neurons.
\newblock In L.~M. Chalupa and J.~S. Werner, editors, \emph{The Visual
  Neurosciences}, volume~1, pages 704--719. MIT Press, 2004.

\bibitem[DeAngelis et~al.(1995)DeAngelis, Ohzawa, and
  Freeman]{DeAngOhzFre95-TINS}
G.~C. DeAngelis, I.~Ohzawa, and R.~D. Freeman.
\newblock Receptive field dynamics in the central visual pathways.
\newblock \emph{Trends in Neuroscience}, 18\penalty0 (10):\penalty0 451--457,
  1995.

\bibitem[Einh{\"a}user et~al.(2002)Einh{\"a}user, Kayser, K{\"o}nig, and
  K{\"o}rding]{EinKayKoeKoe02-EurJNeurSci}
W.~Einh{\"a}user, C.~Kayser, P.~K{\"o}nig, and K.~P. K{\"o}rding.
\newblock Learning the invariance properties of complex cells from their
  responses to natural stimuli.
\newblock \emph{European Journal of Neuroscience}, 15\penalty0 (3):\penalty0
  475--486, 2002.

\bibitem[Emerson et~al.(1987)Emerson, Citron, Vaughn, and
  Klein]{EmeCitVauKle87-JNeuroPhys}
R.~C. Emerson, M.~C. Citron, W.~J. Vaughn, and S.~A. Klein.
\newblock Nonlinear directionally selective subunits in complex cells of cat
  striate cortex.
\newblock \emph{Journal of Neurophysiology}, 58\penalty0 (1):\penalty0 33--65,
  1987.

\bibitem[Fang et~al.(2022)Fang, Cai, and Lu]{FanCaiLu-PNAS}
C.~Fang, X.~Cai, and H.~D. Lu.
\newblock Orientation anisotropies in macaque visual areas.
\newblock \emph{Proceedings of the National Academy of Sciences}, 119\penalty0
  (15):\penalty0 e2113407119, 2022.

\bibitem[Ferster and Miller(2000)]{FerMil00-AnnRevNeuroSci}
D.~Ferster and K.~D. Miller.
\newblock Neural mechanisms of orientation selectivity in the visual cortex.
\newblock \emph{Annual Review of Neuroscience}, 23\penalty0 (1):\penalty0
  441--471, 2000.

\bibitem[Franciosini et~al.(2019)Franciosini, Boutin, and
  Perrinet]{FraBouPer19-AnnCompNeurSciMeet}
A.~Franciosini, V.~Boutin, and L.~Perrinet.
\newblock Modelling complex cells of early visual cortex using predictive
  coding.
\newblock In \emph{Proc.\ 28th Annual Computational Neuroscience Meeting},
  2019.
\newblock Available from
  https://laurentperrinet.github.io/publication/franciosini-perrinet-19-cns/franciosini-perrinet-19-cns.pdf.

\bibitem[Geisler(2011)]{Gei11-VisRes}
W.~S. Geisler.
\newblock Contributions of ideal observer theory to vision research.
\newblock \emph{Vision Research}, 51\penalty0 (7):\penalty0 771--781, 2011.

\bibitem[Georgeson et~al.(2007)Georgeson, May, Freeman, and
  Hesse]{GeoMayFreHes07-JVis}
M.~A. Georgeson, K.~A. May, T.~C.~A. Freeman, and G.~S. Hesse.
\newblock From filters to features: {S}cale-space analysis of edge and blur
  coding in human vision.
\newblock \emph{Journal of Vision}, 7\penalty0 (13):\penalty0 7.1--21, 2007.

\bibitem[Ghodrati et~al.(2017)Ghodrati, Khaligh-Razavi, and
  Lehky]{GhoKhaLeh17-ProNeurobiol}
M.~Ghodrati, S.-M. Khaligh-Razavi, and S.~R. Lehky.
\newblock Towards building a more complex view of the lateral geniculate
  nucleus: {R}ecent advances in understanding its role.
\newblock \emph{Progress in Neurobiology}, 156:\penalty0 214--255, 2017.

\bibitem[Goris et~al.(2015)Goris, Simoncelli, and Movshon]{GorSimMov15-Neuron}
R.~L.~T. Goris, E.~P. Simoncelli, and J.~A. Movshon.
\newblock Origin and function of tuning diversity in {M}acaque visual cortex.
\newblock \emph{Neuron}, 88\penalty0 (4):\penalty0 819--831, 2015.

\bibitem[Hansard and Horaud(2011)]{HanHor11-NeurComp}
M.~Hansard and R.~Horaud.
\newblock A differential model of the complex cell.
\newblock \emph{Neural Computation}, 23\penalty0 (9):\penalty0 2324--2357,
  2011.

\bibitem[Hansel and van Vreeswijk(2012)]{HanVre12-JNeuroSci}
D.~Hansel and C.~van Vreeswijk.
\newblock The mechanism of orientation selectivity in primary visual cortex
  without a functional map.
\newblock \emph{Journal of Neuroscience}, 32\penalty0 (12):\penalty0
  4049--4064, 2012.

\bibitem[Hansen and Neumann(2008)]{HanNeu09-JVis}
T.~Hansen and H.~Neumann.
\newblock A recurrent model of contour integration in primary visual cortex.
\newblock \emph{Journal of Vision}, 8\penalty0 (8):\penalty0 8.1--25, 2008.

\bibitem[Heeger(1992)]{Hee92-VisNeuroSci}
D.~J. Heeger.
\newblock Normalization of cell responses in cat striate cortex.
\newblock \emph{Visual Neuroscience}, 9:\penalty0 181--197, 1992.

\bibitem[Hesse and Georgeson(2005)]{HesGeo05-VisRes}
G.~S. Hesse and M.~A. Georgeson.
\newblock Edges and bars: where do people see features in 1-{D} images?
\newblock \emph{Vision Research}, 45\penalty0 (4):\penalty0 507--525, 2005.

\bibitem[Hubel and Wiesel(1959)]{HubWie59-Phys}
D.~H. Hubel and T.~N. Wiesel.
\newblock Receptive fields of single neurones in the cat's striate cortex.
\newblock \emph{J Physiol}, 147:\penalty0 226--238, 1959.

\bibitem[Hubel and Wiesel(1962)]{HubWie62-Phys}
D.~H. Hubel and T.~N. Wiesel.
\newblock Receptive fields, binocular interaction and functional architecture
  in the cat's visual cortex.
\newblock \emph{J Physiol}, 160:\penalty0 106--154, 1962.

\bibitem[Hubel and Wiesel(1968)]{HubWie68-JPhys}
D.~H. Hubel and T.~N. Wiesel.
\newblock Receptive fields and functional architecture of monkey striate
  cortex.
\newblock \emph{The Journal of Physiology}, 195\penalty0 (1):\penalty0
  215--243, 1968.

\bibitem[Hubel and Wiesel(2005)]{HubWie05-book}
D.~H. Hubel and T.~N. Wiesel.
\newblock \emph{Brain and Visual Perception: {T}he Story of a 25-Year
  Collaboration}.
\newblock Oxford University Press, 2005.

\bibitem[Johnson et~al.(2008)Johnson, Hawken, and
  Shapley]{JohHawSha08-JNeuroSci}
E.~N. Johnson, M.~J. Hawken, and R.~Shapley.
\newblock The orientation selectivity of color-responsive neurons in {M}acaque
  {V1}.
\newblock \emph{The Journal of Neuroscience}, 28\penalty0 (32):\penalty0
  8096--8106, 2008.

\bibitem[Jones and Palmer(1987{\natexlab{a}})]{JonPal87a}
J.~Jones and L.~Palmer.
\newblock The two-dimensional spatial structure of simple receptive fields in
  cat striate cortex.
\newblock \emph{J. of Neurophysiology}, 58:\penalty0 1187--1211,
  1987{\natexlab{a}}.

\bibitem[Jones and Palmer(1987{\natexlab{b}})]{JonPal87b}
J.~Jones and L.~Palmer.
\newblock An evaluation of the two-dimensional {G}abor filter model of simple
  receptive fields in cat striate cortex.
\newblock \emph{J. of Neurophysiology}, 58:\penalty0 1233--1258,
  1987{\natexlab{b}}.

\bibitem[Jung et~al.(2022)Jung, Almasi, Sun, Yunzab, Cloherty, Bauquier,
  Renfree, Meffin, and Ibbotson]{JunAlmSunYunCloBauRenMefIbb22-SciAdv}
Y.~J. Jung, A.~Almasi, S.~H. Sun, M.~Yunzab, S.~L. Cloherty, S.~H. Bauquier,
  M.~Renfree, H.~Meffin, and M.~R. Ibbotson.
\newblock Orientation pinwheels in primary visual cortex of a highly visual
  marsupial.
\newblock \emph{Science Advances}, 8\penalty0 (39):\penalty0 eabn0954, 2022.

\bibitem[Koch et~al.(2016)Koch, Jin, Alonso, and Zaidi]{KocJinAloZai16-NatComm}
E.~Koch, J.~Jin, J.~M. Alonso, and Q.~Zaidi.
\newblock Functional implications of orientation maps in primary visual cortex.
\newblock \emph{Nature Communications}, 7\penalty0 (1):\penalty0 13529, 2016.

\bibitem[Koenderink(1984)]{Koe84}
J.~J. Koenderink.
\newblock The structure of images.
\newblock \emph{Biological Cybernetics}, 50\penalty0 (5):\penalty0 363--370,
  1984.

\bibitem[Koenderink and {van Doorn}(1987)]{KoeDoo87-BC}
J.~J. Koenderink and A.~J. {van Doorn}.
\newblock Representation of local geometry in the visual system.
\newblock \emph{Biological Cybernetics}, 55\penalty0 (6):\penalty0 367--375,
  1987.

\bibitem[Koenderink and {van Doorn}(1990)]{KoeDoo90-BC}
J.~J. Koenderink and A.~J. {van Doorn}.
\newblock Receptive field families.
\newblock \emph{Biological Cybernetics}, 63:\penalty0 291--298, 1990.

\bibitem[Koenderink and {van Doorn}(1992)]{KoeDoo92-PAMI}
J.~J. Koenderink and A.~J. {van Doorn}.
\newblock Generic neighborhood operators.
\newblock \emph{IEEE Transactions on Pattern Analysis and Machine
  Intelligence}, 14\penalty0 (6):\penalty0 597--605, Jun. 1992.

\bibitem[Kording et~al.(2004)Kording, Kayser, Einh{\"a}user, and
  Konig]{KorKayWinKon04-JNeuroPhys}
K.~P. Kording, C.~Kayser, W.~Einh{\"a}user, and P.~Konig.
\newblock How are complex cell properties adapted to the statistics of natural
  stimuli?
\newblock \emph{Journal of Neurophysiology}, 91\penalty0 (1):\penalty0
  206--212, 2004.

\bibitem[Kremkow et~al.(2016)Kremkow, Jin, Wang, and
  Alonso]{KreJinWanAlo16-Nature}
J.~Kremkow, J.~Jin, Y.~Wang, and J.~M. Alonso.
\newblock Principles underlying sensory map topography in primary visual
  cortex.
\newblock \emph{Nature}, 533\penalty0 (7601):\penalty0 52--57, 2016.

\bibitem[Kristensen and Sandberg(2021)]{KriSan21-SciRep}
D.~G. Kristensen and K.~Sandberg.
\newblock Population receptive fields of human primary visual cortex organised
  as dc-balanced bandpass filters.
\newblock \emph{Scientific Reports}, 11\penalty0 (1):\penalty0 22423, 2021.

\bibitem[Lampl et~al.(2001)Lampl, Anderson, Gillespie, and
  Ferster]{LamAndGilFer01-Neuron}
I.~Lampl, J.~S. Anderson, D.~C. Gillespie, and D.~Ferster.
\newblock Prediction of orientation selectivity from receptive field
  architecture in simple cells of cat visual cortex.
\newblock \emph{Neuron}, 30\penalty0 (1):\penalty0 263--274, 2001.

\bibitem[Li et~al.(2015)Li, Liu, Chou, Zhang, and
  Tao]{LiLiuChoZhaTao15-JNeuroSci}
Y.-T. Li, B.-H. Liu, X.-L. Chou, L.~I. Zhang, and H.~W. Tao.
\newblock Synaptic basis for differential orientation selectivity between
  complex and simple cells in mouse visual cortex.
\newblock \emph{Journal of Neuroscience}, 35\penalty0 (31):\penalty0
  11081--11093, 2015.

\bibitem[Lian et~al.(2021)Lian, Almasi, Grayden, Kameneva, Burkitt, and
  Meffin]{LiaAlmGraKamBurMef21-PLOSCompBiol}
Y.~Lian, A.~Almasi, D.~B. Grayden, T.~Kameneva, A.~N. Burkitt, and H.~Meffin.
\newblock Learning receptive field properties of complex cells in {V1}.
\newblock \emph{PLoS Computational Biology}, 17\penalty0 (3):\penalty0
  e1007957, 2021.

\bibitem[Lindeberg(1998)]{Lin97-IJCV}
T.~Lindeberg.
\newblock Feature detection with automatic scale selection.
\newblock \emph{International Journal of Computer Vision}, 30\penalty0
  (2):\penalty0 77--116, 1998.

\bibitem[Lindeberg(2011)]{Lin10-JMIV}
T.~Lindeberg.
\newblock Generalized {G}aussian scale-space axiomatics comprising linear
  scale-space, affine scale-space and spatio-temporal scale-space.
\newblock \emph{Journal of Mathematical Imaging and Vision}, 40\penalty0
  (1):\penalty0 36--81, 2011.

\bibitem[Lindeberg(2013)]{Lin13-BICY}
T.~Lindeberg.
\newblock A computational theory of visual receptive fields.
\newblock \emph{Biological Cybernetics}, 107\penalty0 (6):\penalty0 589--635,
  2013.

\bibitem[Lindeberg(2016)]{Lin16-JMIV}
T.~Lindeberg.
\newblock Time-causal and time-recursive spatio-temporal receptive fields.
\newblock \emph{Journal of Mathematical Imaging and Vision}, 55\penalty0
  (1):\penalty0 50--88, 2016.

\bibitem[Lindeberg(2017)]{Lin17-JMIV}
T.~Lindeberg.
\newblock Temporal scale selection in time-causal scale space.
\newblock \emph{Journal of Mathematical Imaging and Vision}, 58\penalty0
  (1):\penalty0 57--101, 2017.

\bibitem[Lindeberg(2018)]{Lin18-SIIMS}
T.~Lindeberg.
\newblock Dense scale selection over space, time and space-time.
\newblock \emph{SIAM Journal on Imaging Sciences}, 11\penalty0 (1):\penalty0
  407--441, 2018.

\bibitem[Lindeberg(2020)]{Lin20-JMIV}
T.~Lindeberg.
\newblock Provably scale-covariant continuous hierarchical networks based on
  scale-normalized differential expressions coupled in cascade.
\newblock \emph{Journal of Mathematical Imaging and Vision}, 62\penalty0
  (1):\penalty0 120--148, 2020.

\bibitem[Lindeberg(2021)]{Lin21-Heliyon}
T.~Lindeberg.
\newblock Normative theory of visual receptive fields.
\newblock \emph{Heliyon}, 7\penalty0 (1):\penalty0 e05897:1--20, 2021.
\newblock \doi{10.1016/j.heliyon.2021.e05897}.

\bibitem[Lindeberg(2023{\natexlab{a}})]{Lin23-BICY}
T.~Lindeberg.
\newblock A time-causal and time-recursive scale-covariant scale-space
  representation of temporal signals and past time.
\newblock \emph{Biological Cybernetics}, 117\penalty0 (1--2):\penalty0 21--59,
  2023{\natexlab{a}}.

\bibitem[Lindeberg(2023{\natexlab{b}})]{Lin23-FrontCompNeuroSci}
T.~Lindeberg.
\newblock Covariance properties under natural image transformations for the
  generalized {G}aussian derivative model for visual receptive fields.
\newblock \emph{Frontiers in Computational Neuroscience}, 17:\penalty0
  1189949:1--23, 2023{\natexlab{b}}.

\bibitem[Lindeberg(2024{\natexlab{a}})]{Lin24-arXiv-HypoElongVarRF}
T.~Lindeberg.
\newblock Do the receptive fields in the primary visual cortex span a
  variability over the degree of elongation of the receptive fields?
\newblock \emph{arXiv preprint arXiv:2404.04858}, 2024{\natexlab{a}}.

\bibitem[Lindeberg(2024{\natexlab{b}})]{Lin24-arXiv-UnifiedJointCovProps}
T.~Lindeberg.
\newblock Unified theory for joint covariance properties under geometric image
  transformations for spatio-temporal receptive fields according to the
  generalized {G}aussian derivative model for visual receptive fields.
\newblock \emph{arXiv preprint arXiv:2311.10543}, 2024{\natexlab{b}}.

\bibitem[Lindeberg and G{\aa}rding(1997)]{LG96-IVC}
T.~Lindeberg and J.~G{\aa}rding.
\newblock Shape-adapted smoothing in estimation of 3-{D} shape cues from affine
  distortions of local 2-{D} structure.
\newblock \emph{Image and Vision Computing}, 15\penalty0 (6):\penalty0
  415--434, 1997.

\bibitem[Lowe(2000)]{Low00-BIO}
D.~G. Lowe.
\newblock Towards a computational model for object recognition in {IT} cortex.
\newblock In \emph{Biologically Motivated Computer Vision}, volume 1811 of
  \emph{Springer LNCS}, pages 20--31. Springer, 2000.

\bibitem[Maldonado et~al.(1997)Maldonado, Godecke, Gray, and
  Bonhoeffer]{MalGodGraBon97-Science}
P.~E. Maldonado, I.~Godecke, C.~M. Gray, and T.~Bonhoeffer.
\newblock Orientation selectivity in pinwheel centers in cat striate cortex.
\newblock \emph{Science}, 276\penalty0 (5318):\penalty0 1551--1555, 1997.

\bibitem[Marcelja(1980)]{Mar80-JOSA}
S.~Marcelja.
\newblock Mathematical description of the responses of simple cortical cells.
\newblock \emph{Journal of Optical Society of America}, 70\penalty0
  (11):\penalty0 1297--1300, 1980.

\bibitem[Mardia(1972)]{Mar72}
K.~V. Mardia.
\newblock \emph{Statistics of Directional Data}.
\newblock Academic Press, London, 1972.

\bibitem[Martinez and Alonso(2001)]{MarAlo01-Neuron}
L.~M. Martinez and J.-M. Alonso.
\newblock Construction of complex receptive fields in cat primary visual
  cortex.
\newblock \emph{Neuron}, 32\penalty0 (3):\penalty0 515--525, 2001.

\bibitem[May and Georgeson(2007)]{MayGeo05-VisRes}
K.~A. May and M.~A. Georgeson.
\newblock Blurred edges look faint, and faint edges look sharp: {T}he effect of
  a gradient threshold in a multi-scale edge coding model.
\newblock \emph{Vision Research}, 47\penalty0 (13):\penalty0 1705--1720, 2007.

\bibitem[Mazurek et~al.(2014)Mazurek, Kager, and van
  Hooser]{MazKagHoo14-FrontNeurCirc}
M.~Mazurek, M.~Kager, and S.~D. van Hooser.
\newblock Robust quantification of orientation selectivity and direction
  selectivity.
\newblock \emph{Frontiers in Neural Circuits}, 8:\penalty0 92, 2014.

\bibitem[Mechler and Ringach(2002)]{MecRin02-VisRes}
F.~Mechler and D.~L. Ringach.
\newblock On the classification of simple and complex cells.
\newblock \emph{Vision Research}, 42\penalty0 (8):\penalty0 1017--1033, 2002.

\bibitem[Merkt et~al.(2019)Merkt, Sch{\"u}{\ss}ler, and
  Rotter]{MerSchRot19-PLOSCompBiol}
B.~Merkt, F.~Sch{\"u}{\ss}ler, and S.~Rotter.
\newblock Propagation of orientation selectivity in a spiking network model of
  layered primary visual cortex.
\newblock \emph{{PLoS} Computational Biology}, 15\penalty0 (7):\penalty0
  e1007080, 2019.

\bibitem[Merolla and Boahn(2004)]{MerBoa04-NIPS}
P.~Merolla and K.~Boahn.
\newblock A recurrent model of orientation maps with simple and complex cells.
\newblock In \emph{Advances in Neural Information Processing Systems (NIPS
  2004)}, pages 995--1002, 2004.

\bibitem[Moldakarimov et~al.(2014)Moldakarimov, Bazhenov, and
  Sejnowski]{MolBazSej14-PLOSCompBiol}
S.~Moldakarimov, M.~Bazhenov, and T.~J. Sejnowski.
\newblock Top-down inputs enhance orientation selectivity in neurons of the
  primary visual cortex during perceptual learning.
\newblock \emph{{PLoS} Computational Biology}, 10\penalty0 (8):\penalty0
  e1003770, 2014.

\bibitem[Movshon et~al.(1978)Movshon, Thompson, and
  Tolhurst]{MovThoTol78-JPhys}
J.~A. Movshon, E.~D. Thompson, and D.~J. Tolhurst.
\newblock Receptive field organization of complex cells in the cat's striate
  cortex.
\newblock \emph{The Journal of Physiology}, 283\penalty0 (1):\penalty0 79--99,
  1978.

\bibitem[Najafian et~al.(2022)Najafian, Koch, Teh, Jin, Rahimi-Nasrabadi,
  Zaidi, Kremkow, and Alonso]{NajKocTehJinRahZaiKreAlo22-NatureComm}
S.~Najafian, E.~Koch, K.~L. Teh, J.~Jin, H.~Rahimi-Nasrabadi, Q.~Zaidi,
  J.~Kremkow, and J.-M. Alonso.
\newblock A theory of cortical map formation in the visual brain.
\newblock \emph{Nature Communications}, 13\penalty0 (1):\penalty0 2303, 2022.

\bibitem[Nauhaus et~al.(2008)Nauhaus, Benucci, Carandini, and
  Ringach]{NauBenCarRin09-Neuron}
I.~Nauhaus, A.~Benucci, M.~Carandini, and D.~L. Ringach.
\newblock Neuronal selectivity and local map structure in visual cortex.
\newblock \emph{Neuron}, 57\penalty0 (5):\penalty0 673--679, 2008.

\bibitem[Nguyen and Freeman(2019)]{NguFre19-PLOSCompBiol}
G.~Nguyen and A.~W. Freeman.
\newblock A model for the origin and development of visual orientation
  selectivity.
\newblock \emph{{PLoS} Computational Biology}, 15\penalty0 (7):\penalty0
  e1007254, 2019.

\bibitem[Oleskiw et~al.(2024)Oleskiw, Lieber, Simoncelli, and
  Movshon]{OleLieSimMov23-bioRxiv}
T.~D. Oleskiw, J.~D. Lieber, E.~P. Simoncelli, and J.~A. Movshon.
\newblock Foundations of visual form selectivity for neurons in macaque {V1}
  and {V2}.
\newblock \emph{bioRxiv}, 2024.03.04.583307, 2024.

\bibitem[Pattadkal et~al.(2018)Pattadkal, Mato, van Vreeswijk, Priebe, and
  Hansel]{PatMatVrePriHan18-CellRep}
J.~J. Pattadkal, G.~Mato, C.~van Vreeswijk, N.~J. Priebe, and D.~Hansel.
\newblock Emergent orientation selectivity from random networks in mouse visual
  cortex.
\newblock \emph{Cell Reports}, 24\penalty0 (8):\penalty0 2042--2050, 2018.

\bibitem[Pei et~al.(2016)Pei, Gao, Hao, Qiao, and
  Ai]{PeiGaoHaoQiaAi16-NeurRegen}
Z.-J. Pei, G.-X. Gao, B.~Hao, Q.-L. Qiao, and H.-J. Ai.
\newblock A cascade model of information processing and encoding for retinal
  prosthesis.
\newblock \emph{Neural Regeneration Research}, 11\penalty0 (4):\penalty0 646,
  2016.

\bibitem[Porat and Zeevi(1988)]{PorZee88-PAMI}
M.~Porat and Y.~Y. Zeevi.
\newblock The generalized {G}abor scheme of image representation in biological
  and machine vision.
\newblock \emph{IEEE Transactions on Pattern Analysis and Machine
  Intelligence}, 10\penalty0 (4):\penalty0 452--468, 1988.

\bibitem[Priebe(2016)]{Pri16-AnnRevVisSci}
N.~J. Priebe.
\newblock Mechanisms of orientation selectivity in the primary visual cortex.
\newblock \emph{Annual Review of Vision Science}, 2:\penalty0 85--107, 2016.

\bibitem[Ringach(2002)]{Rin01-JNeuroPhys}
D.~L. Ringach.
\newblock Spatial structure and symmetry of simple-cell receptive fields in
  macaque primary visual cortex.
\newblock \emph{Journal of Neurophysiology}, 88:\penalty0 455--463, 2002.

\bibitem[Ringach(2004)]{Rin04-JPhys}
D.~L. Ringach.
\newblock Mapping receptive fields in primary visual cortex.
\newblock \emph{Journal of Physiology}, 558\penalty0 (3):\penalty0 717--728,
  2004.

\bibitem[Ringach et~al.(2002)Ringach, Shapley, and
  Hawken]{RinShaHaw03-JNeurSci}
D.~L. Ringach, R.~M. Shapley, and M.~J. Hawken.
\newblock {O}rientation selectivity in macaque {V1}: {D}iversity and laminar
  dependence.
\newblock \emph{Journal of Neuroscience}, 22\penalty0 (13):\penalty0
  5639--5651, 2002.

\bibitem[Rose and Blakemore(1974)]{RosBla74-ExpBrainRes}
D.~Rose and C.~Blakemore.
\newblock An analysis of orientation selectivity in the cat's visual cortex.
\newblock \emph{Experimental Brain Research}, 20:\penalty0 1--17, 1974.

\bibitem[Ruslim et~al.(2023)Ruslim, Burkitt, and Lian]{RusBurLia23-bioRxiv}
M.~A. Ruslim, A.~N. Burkitt, and Y.~Lian.
\newblock Learning spatio-temporal {V1} cells from diverse {LGN} inputs.
\newblock \emph{bioRxiv}, pages 2023--11, 2023.

\bibitem[Rust et~al.(2005)Rust, Schwartz, Movshon, and
  Simoncelli]{RusSchMovSim05-Neuron}
N.~C. Rust, O.~Schwartz, J.~A. Movshon, and E.~P. Simoncelli.
\newblock Spatiotemporal elements of macaque {V}1 receptive fields.
\newblock \emph{Neuron}, 46\penalty0 (6):\penalty0 945--956, 2005.

\bibitem[Sadeh and Rotter(2014)]{SadRot14-BICY}
S.~Sadeh and S.~Rotter.
\newblock Statistics and geometry of orientation selectivity in primary visual
  cortex.
\newblock \emph{Biological Cybernetics}, 108:\penalty0 631--653, 2014.

\bibitem[Sasaki et~al.(2015)Sasaki, Kimura, Ninomiya, Tabuchi, Tanaka, Fukui,
  Asada, Arai, Inagaki, Nakazono, Baba, Daisuke, Nishimoto, Sanada, Tani,
  Imamura, Tanaka, and
  Ohzawa]{SakKimNimTabTanFukAsaAraInaNakBabDaiNisSanTanImaTanOhz15-SciRep}
K.~S. Sasaki, R.~Kimura, T.~Ninomiya, Y.~Tabuchi, H.~Tanaka, M.~Fukui, Y.~C.
  Asada, T.~Arai, M.~Inagaki, T.~Nakazono, M.~Baba, K.~Daisuke, S.~Nishimoto,
  T.~M. Sanada, T.~Tani, K.~Imamura, S.~Tanaka, and I.~Ohzawa.
\newblock Supranormal orientation selectivity of visual neurons in
  orientation-restricted animals.
\newblock \emph{Scientific Reports}, 5\penalty0 (1):\penalty0 16712, 2015.

\bibitem[Schiller et~al.(1976)Schiller, Finlay, and
  Volman]{SchFinVol76-JNeuroPhys}
P.~H. Schiller, B.~L. Finlay, and S.~F. Volman.
\newblock {Q}uantitative studies of single-cell properties in monkey striate
  cortex. {II}. {O}rientation specificity and ocular dominance.
\newblock \emph{Journal of Neurophysiology}, 39\penalty0 (6):\penalty0
  1320--1333, 1976.

\bibitem[Scholl et~al.(2013)Scholl, Tan, Corey, and
  Priebe]{SchTanCorPri13-JNeurSci}
B.~Scholl, A.~Y.~Y. Tan, J.~Corey, and N.~J. Priebe.
\newblock Emergence of orientation selectivity in the mammalian visual pathway.
\newblock \emph{Journal of Neuroscience}, 33\penalty0 (26):\penalty0
  10616--10624, 2013.

\bibitem[Seri{\`e}s et~al.(2004)Seri{\`e}s, Latham, and
  Pouget]{SerLatPou04-NatNeuroSci}
P.~Seri{\`e}s, P.~E. Latham, and A.~Pouget.
\newblock Tuning curve sharpening for orientation selectivity: coding
  efficiency and the impact of correlations.
\newblock \emph{Nature Neuroscience}, 7\penalty0 (10):\penalty0 1129--1135,
  2004.

\bibitem[Serre and Riesenhuber(2004)]{SerRie04-AIMemo}
T.~Serre and M.~Riesenhuber.
\newblock Realistic modeling of simple and complex cell tuning in the {HMAX}
  model, and implications for invariant object recognition in cortex.
\newblock Technical Report AI Memo 2004-017, MIT Computer Science and Artifical
  Intelligence Laboratory, 2004.

\bibitem[Shapley et~al.(2003)Shapley, Hawken, and Ringach]{ShaHawRin03-Neuron}
R.~Shapley, M.~Hawken, and D.~L. Ringach.
\newblock Dynamics of orientation selectivity in the primary visual cortex and
  the importance of cortical inhibition.
\newblock \emph{Neuron}, 38\penalty0 (5):\penalty0 689--699, 2003.

\bibitem[Sharpee(2013)]{Sha13-AnnRevNeurSci}
T.~O. Sharpee.
\newblock Computational identification of receptive fields.
\newblock \emph{Annual Review of Neuroscience}, 36:\penalty0 103--120, 2013.

\bibitem[Somers et~al.(1995)Somers, Nelson, and Sur]{SomNelSur95-JNeuroSci}
D.~C. Somers, S.~B. Nelson, and M.~Sur.
\newblock An emergent model of orientation selectivity in cat visual cortical
  simple cells.
\newblock \emph{Journal of Neuroscience}, 15\penalty0 (8):\penalty0 5448--5465,
  1995.

\bibitem[Sompolinsky and Shapley(1997)]{SomSha97-CurrOpNeuroBio}
H.~Sompolinsky and R.~Shapley.
\newblock New perspectives on the mechanisms for orientation selectivity.
\newblock \emph{Current Opinion in Neurobiology}, 7\penalty0 (4):\penalty0
  514--522, 1997.

\bibitem[Touryan et~al.(2002)Touryan, Lau, and Dan]{TouLauDan02-JNeuroSci}
J.~Touryan, B.~Lau, and Y.~Dan.
\newblock Isolation of relevant visual features from random stimuli for
  cortical complex cells.
\newblock \emph{Journal of Neuroscience}, 22\penalty0 (24):\penalty0
  10811--10818, 2002.

\bibitem[Touryan et~al.(2005)Touryan, Felsen, and Dan]{TouFelDan05-Neuron}
J.~Touryan, G.~Felsen, and Y.~Dan.
\newblock Spatial structure of complex cell receptive fields measured with
  natural images.
\newblock \emph{Neuron}, 45\penalty0 (5):\penalty0 781--791, 2005.

\bibitem[Valois et~al.(2000)Valois, Cottaris, Mahon, Elfer, and
  Wilson]{ValCotMahElfWil00-VR}
R.~L.~D. Valois, N.~P. Cottaris, L.~E. Mahon, S.~D. Elfer, and J.~A. Wilson.
\newblock Spatial and temporal receptive fields of geniculate and cortical
  cells and directional selectivity.
\newblock \emph{Vision Research}, 40\penalty0 (2):\penalty0 3685--3702, 2000.

\bibitem[van Kleef et~al.(2010)van Kleef, , Cloherty, and
  Ibbotson]{KleCloIbb10-JPhys}
J.~P. van Kleef, , S.~L. Cloherty, and M.~R. Ibbotson.
\newblock Complex cell receptive fields: evidence for a hierarchical mechanism.
\newblock \emph{The Journal of Physiology}, 588\penalty0 (18):\penalty0
  3457--3470, 2010.

\bibitem[Vita et~al.(2024)Vita, Orsi, Stanko, Clark, and
  Tiriac]{BitOrsStaClaTir24-bioRxiv}
D.~J. Vita, F.~S. Orsi, N.~G. Stanko, N.~A. Clark, and A.~Tiriac.
\newblock Development and organization of the retinal orientation selectivity
  map.
\newblock \emph{bioRxiv:2024.03.27.585774}, 2024.

\bibitem[Walker et~al.(2019)Walker, Sinz, Cobos, Muhammad, Froudarakis, Fahey,
  Ecker, Reimer, Pitkow, and
  Tolias]{WalSinCobMuhFroFahEckReiPitTol19-NatNeurSci}
E.~Y. Walker, F.~H. Sinz, E.~Cobos, T.~Muhammad, E.~Froudarakis, P.~G. Fahey,
  A.~S. Ecker, J.~Reimer, X.~Pitkow, and A.~S. Tolias.
\newblock Inception loops discover what excites neurons most using deep
  predictive models.
\newblock \emph{Nature Neuroscience}, 22\penalty0 (12):\penalty0 2060--2065,
  2019.

\bibitem[Wallis and Georgeson(2009)]{WalGeo09-VisRes}
S.~A. Wallis and M.~A. Georgeson.
\newblock Mach edges: Local features predicted by 3rd derivative spatial
  filtering.
\newblock \emph{Vision Research}, 49\penalty0 (14):\penalty0 1886--1893, 2009.

\bibitem[Wang et~al.(2024)Wang, Dey, Lagos, Behnam, Callaway, and
  Stafford]{WanDeyLagBehCalSta24-CellRep}
H.~Wang, O.~Dey, W.~N. Lagos, N.~Behnam, E.~M. Callaway, and B.~K. Stafford.
\newblock Parallel pathways carrying direction-and orientation-selective
  retinal signals to layer 4 of the mouse visual cortex.
\newblock \emph{Cell Reports}, 43\penalty0 (3), 2024.

\bibitem[Wang and Spratling(2016)]{WanSpra16-CognComp}
Q.~Wang and M.~W. Spratling.
\newblock Contour detection in colour images using a neurophysiologically
  inspired model.
\newblock \emph{Cognitive Computation}, 8\penalty0 (6):\penalty0 1027--1035,
  2016.

\bibitem[Watkins and Berkley(1974)]{WatBer73-ExpBrainRes}
D.~W. Watkins and M.~A. Berkley.
\newblock The orientation selectivity of single neurons in cat striate cortex.
\newblock \emph{Experimental Brain Research}, 19:\penalty0 433--446, 1974.

\bibitem[Wei et~al.(2022)Wei, Merkt, and Rotter]{WeiMerRot22-bioRxiv}
W.~Wei, B.~Merkt, and S.~Rotter.
\newblock A theory of orientation selectivity emerging from randomly sampling
  the visual field.
\newblock \emph{bioRxiv}, pages 2022--07, 2022.

\bibitem[Wendt and Faul(2024)]{WenFay24-JVis}
G.~Wendt and F.~Faul.
\newblock Binocular luster elicited by isoluminant chromatic stimuli relies on
  mechanisms similar to those in the achromatic case.
\newblock \emph{Journal of Vision}, 24\penalty0 (3):\penalty0 7--7, 2024.

\bibitem[Yedjour and Yedjour(2024)]{YedYed24-CognNeurDyn}
H.~Yedjour and D.~Yedjour.
\newblock A spatiotemporal energy model based on spiking neurons for human
  motion perception.
\newblock \emph{Cognitive Neurodynamics}, pages 1--15, 2024.

\bibitem[Young(1987)]{You87-SV}
R.~A. Young.
\newblock The {G}aussian derivative model for spatial vision: {I}. {R}etinal
  mechanisms.
\newblock \emph{Spatial Vision}, 2\penalty0 (4):\penalty0 273--293, 1987.

\bibitem[Young and Lesperance(2001)]{YouLes01-SV}
R.~A. Young and R.~M. Lesperance.
\newblock The {G}aussian derivative model for spatio-temporal vision: {II}.
  {C}ortical data.
\newblock \emph{Spatial Vision}, 14\penalty0 (3, 4):\penalty0 321--389, 2001.

\bibitem[Young et~al.(2001)Young, Lesperance, and Meyer]{YouLesMey01-SV}
R.~A. Young, R.~M. Lesperance, and W.~W. Meyer.
\newblock The {G}aussian derivative model for spatio-temporal vision: {I}.
  {C}ortical model.
\newblock \emph{Spatial Vision}, 14\penalty0 (3, 4):\penalty0 261--319, 2001.

\end{thebibliography}
}

\end{document}